\documentstyle[12pt]{report}
\newlength{\aivwidth}   
\newlength{\tmpwidth}   

\newcommand{\phrd}[1]{Phys.\ Rev.\ {\bf D#1}}

\newcommand{\phrl}[1]{Phys.\ Rev.\ Lett.\ {\bf #1}}
\newcommand{\nphb}[1]{Nucl.\ Phys.\ {\bf B#1}}
\newcommand{\nph}[1]{Nucl.\ Phys.\ {\bf #1}}
\newcommand{\phlb}[1]{Phys.\ Lett.\ {\bf B#1}}

\newcommand{\zphc}[1]{Z.\ Phys.\ {\bf C#1}}

\newcommand{\ijmpha}[1]{Int.\ J.\ Mod.\ Phys.\ {\bf A#1}}

\newcommand{\phrp}[1]{Phys.\ Rep.\ {\bf #1}}

\newcommand{\be}{\begin{equation}}
\newcommand{\ee}{\end{equation}}
\newcommand{\bea}{\begin{eqnarray}}
\newcommand{\eea}{\end{eqnarray}}
\newcommand{\ba}{\begin{array}}
\newcommand{\ea}{\end{array}}
\newcommand{\eref}[1]{(\ref{#1})}
\newcommand{\nn}{\nonumber\\}

\newcommand{\dfun}{{\cal D}}
\newcommand{\Df}{{\cal D}}
\newcommand{\lag}{{\cal L}}

\newcommand{\dx}{d^4x\,}
\newcommand{\ep}{\epsilon}
\newcommand{\vp}{\varphi}
\newcommand{\pa}{\partial}
\newcommand{\suu}{$\rm SU(2)\times U(1)$}
\newcommand{\rx}{$\rm R_\xi$}
\newcommand{\cs}{|_{\scriptstyle \phi_a=0}}
\newcommand{\pcs}{|_{\scriptstyle \phi_a^{(1)}=0}}
\newcommand{\Q}{{\cal Q}}
\renewcommand{\P}{{\cal P}}
\newcommand{\bq}{\bar{q}}
\newcommand{\bp}{\bar{p}}
\newcommand{\Det}{\,\mbox{Det}\,}

\newcommand{\lib}{\bar{\lag}_I}
\newcommand{\dnn}{\delta(x^0-y^0)}
\newcommand{\vx}{{\bf x}}
\newcommand{\vy}{{\bf y}}
\newcommand{\dd}{\delta^3(\vx-\vy)}

\begin{document}

\title{Equivalence of Hamiltonian and Lagrangian\\Path Integral Quantization}
\author{Carsten Grosse-Knetter\thanks{E-Mail:
knetter@physw.uni-bielefeld.de}\\Dissertation}
\date{Universit\"at Bielefeld\\Fakult\"at f\"ur Physik\\November 1993\\[5mm]
BI-TP 93/56\\hep-ph/9311259}

\maketitle

\setcounter{page}{0}
This thesis is essentially a combination of the author's following
publications:
\thispagestyle{empty}
\begin{itemize}
\item C. Grosse-Knetter and R. K\"ogerler, {\em Unitary Gauge, Stueckelberg
Formalism and Gauge Invariant Models
for Effective Lagrangians\/}, \phrd{48}, 2865 (1993)
\item C. Grosse-Knetter, {\em Hamiltonian Quantization of Effective
Lagrangians with Massive Vector Fields\/}, \phrd{48}, 2854 (1993)
\item C. Grosse-Knetter, {\em Effective Lagrangians with Higher Derivatives\/},
Bielefeld Preprint BI-TP 93/29 (1993), hep-ph/9306321
\item C. Grosse-Knetter, {\em Equivalence of Hamiltonian and Lagrangian Path
Integral Quantization: Effective Gauge Theories\/},
Bielefeld Preprint BI-TP 93/40 (1993), hep-ph/9308201, to be published in
\phrd{}
\end{itemize}

\tableofcontents

\chapter{Introduction}
Effective Lagrangians containing massless or massive vector
fields with arbitrary
(non--Yang--Mills) self couplings are investigated very intensively in
the literature in order
to parametrize possible deviations of
the self interactions of the electroweak gauge bosons
 $W^\pm$, $Z$ and $\gamma$ \cite{nongauge,nonlin,gauge}
and of the gluons \cite{gluons} from the standard
model predictions with respect to experimental tests of these couplings.
In \cite{nongauge,nonlin,gauge,gluons}
it is always implicitely assumed that
the Feynman rules, which are the basis for calculations of
$S$-Matrix elements and cross sections, can directly be
obtained from the effective Lagrangian,
i.e., the quadratic terms in the Lagrangian
yield the propagators and the
cubic, quartic, etc. terms yield the vertices in the
standard manner. This simple
quantization rule is known as {\em Matthews's theorem\/}\footnote{Matthews
himself proved this
theorem only for the very simple case of interaction terms with at most
first powers of derivatives and he used the canonical quantization formalism
instead of the PI formalism \cite{mat}.}   \cite{mat}.
Within the framework of the
the Feynman path intergral (PI) formalism \cite{feyn,fapo,books}
(where the Feynman rules follow from
the generating functional) this theorem  can be reformulated as follows:
\begin{quote}{\it
Given a Lagrangian ${\cal L}_{}$
with an arbitrary interaction term,
the corresponding generating functional can
be written as a Lagrangian PI
\begin{equation} Z[J]=\int{\cal D}\Phi\,\exp\left\{
i\int d^4x\, [{\cal L}_{quant}
+J\Phi]\right\} \label{lpi}\end{equation}
(where $\Phi$ is a shorthand notation
for all fields in ${\cal L}_{quant}$). If
${\cal L}_{}$ has no gauge freedom,
the quantized
Lagrangian ${\cal L}_{quant}$ occurring in the
PI is identical to the primordial one
\begin{equation} {\cal L}_{quant}=
{\cal L}_{}. \label{eq}\end{equation}
If ${\cal L}_{}$ has a gauge freedom, the generating functional\/
{\rm(\ref{lpi})} is the same as the one obtained in
the Faddeev--Popov (FP) formalism\/
{\rm\cite{fapo,books}} (in an arbitrary gauge)
with the quantized Lagrangian
\begin{equation} {\cal L}_{quant}=
{\cal L}_{}+{\cal L}_{g.f.}+
{\cal L}_{ghost}, \label{fp} \end{equation}
which contains an additional gauge-fixing (g.f.)\ term and a ghost term.}
\end{quote}
The generating functional \eref{lpi} with \eref{eq} or \eref{fp}
is very convenient for practical calculations because it is
manifestly Lorentz invariant
(if a covariant gauge is choosen), it does not involve
the generalized momenta corresponding to
the fields and, as mentioned above, it
directly implies the Feynman rules; for this reason it is used in all
practical calculations \cite{nongauge,nonlin,gauge,gluons}.
However, it is well known that,
in general, quantization has to be based on
the more elaborate Hamiltonian PI formalism \cite{fad,bedu,sen,const,gity}.
The naive Lagrangian PI formalism \cite{feyn,fapo,books},
where (\ref{lpi}) with
(\ref{eq}) is taken as the ansatz
for the generating functional, can a priori only be applied to quantize
physical systems without derivative couplings and without
constraints. Thus, to prove
Matthews's theorem, one has
to derive the Lagrangian PI (\ref{lpi}) with
(\ref{eq}) or (\ref{fp}) from
the Hamiltonian PI, i.e.\ one has to
show that the correct Hamiltonian PI formalism
and the naive Lagrangian PI formalism are equivalent.
For this reason, I will not use the historical designation ``Matthews's
theorem'' for the above statement but the name {\em Hamilton--Lagrange
equivalence theorem (HLE theorem),\/} which is more adequate to its
modern PI formulation.

The HLE theorem was proven by Bernard and Duncan \cite{bedu}
for effective interactions of scalar fields
(without higher derivatives of the fields), i.e.\
for physical sytems which do not involve constraints.
Vector fields and fermion fields, however,
are subject to constraints.
Thus, in order to derive the HLE theorem for the general case,
one has to take into account the formalism of quantization
of constrained systems, which goes back to Dirac
\cite{dirac} and was formulated within the
path integral formalism by
Faddeev \cite{fad} (for first-class constrained, i.e.\
gauge invariant, systems)
and by Senjanovic \cite{sen} (for second-class constrained, i.e.\
gauge noninvariant, systems). Besides, in these works
the equivalence of Hamiltonian and Lagrangian PI quantization
was shown for Yang--Mills theories \cite{fad} and for
massive Yang--Mills theories \cite{sen} {\em without\/} additional
effective interaction terms.
Extensive treatises on Hamiltonian quantization of constrained systems can be
found in textbooks about this subject \cite{const,gity}.
In this thesis I will give a general proof of the HLE theorem\footnote{%
Recently several works about the equivalence of Hamiltonian and Lagrangian
PI quantization have been published \cite{bfvbv}. However, in \cite{bfvbv}
the equivalence of the (Hamiltonian) Batalin--Fradkin--Vilkovisky
(BFV) PI formalism \cite{bfv} to the
(Lagrangian) Batalin--Vilkovisky
(BV) PI formalism \cite{bv} is proven; both are very formal and
abstract formalisms, which are not directly connected to the usual
Hamiltonian \cite{fad,sen} or Lagrangian \cite{feyn,fapo} PI formalism.
My work is completely different from \cite{bfvbv} since I prove
the equivalence of the (Hamiltonian) Faddeev--Senjanovic PI \cite{fad,sen}
(which is the fundamental one because it can be derived from elemantary
dynamics) to the (Lagrangian) Feynman--Faddeev--Popov PI
\cite{feyn,fapo} (wich is the one used in all practical calculations
\cite{nongauge,nonlin,gauge,gluons}) {\em without\/} applying
the BFV \cite{bfv} or the BV \cite{bv} formalism.}
for arbitrary interactions of all physically important types
of particles, viz.\ scalars, fermions, massless and massive vector
bosons. I will also take into account the case of effective interactions
which involve higher derivatives of the fields.

Particular attention will be paid to
effective interactions of massive vector fields because
they are most interesting from the phenomenological and from the
theoretical point of view.
Three types of effective
theories with massive vector fields can be found in the
literature, namely
gauge noninvariant Lagrangians \cite{nongauge}, spontaneously broken
gauge theories (SBGTs) with a nonlinearly realized scalar sector and
without physical Higgs bosons \cite{nonlin}
(called gauged nonlinear $\sigma$-models or chiral Lagrangians)
and SBGTs with a
linearely realized scalar sector which contain (a) physical Higgs boson(s)
\cite{gauge}. In fact, the proof of the HLE theorem
for gauge noninvariant
Lagrangians can be extended to the case of gauge invariant Lagrangians
because, by applying a Stueckelberg transformation
\cite{stue,kugo}, each SBGT can be rewritten as a gauge
noninvariant model (after a nonlinear parametrization of the
scalar sector \cite{lezj,clt}
in the case of a linear Higgs model).
A Stueckelberg transformation
is a field transformation (that involves
derivatives of the fields) which results in removing all
unphysical scalar fields (pseudo-Goldstone fields)
from the Lagrangian.
The resulting Lagrangian is called the unitary gauge (U-gauge)
of the original SBGT. On the other hand, by using
the Stueckelberg formalism, an arbitrary
gauge noninvariant Lagrangian can be written as a (nonlinear) SBGT
\cite{bulo} and by
introduction of (a) physical Higgs boson(s) it can be extended to
a linear Higgs model.
In this thesis I will reformulate the Stueckelberg
formalism within the Hamiltonian formalism, thereby establishing the
physical equivalence of gauge invariant Lagrangians and
the corresponding gauge noninvariant U-gauge
Lagrangians. On this basis I will prove the HLE theorem for
SBGTs by first deriving the Faddeev--Popov PI for
the case of the U-gauge\footnote{It will be shown later
that for the case of this special gauge the Faddeev--Popov formalism yields
no explicit g.f.\ term and ghost
term.} and then using the equivalence of all gauges,
i.e.\ the independence of the $S$-matrix elements from the choice of
the gauge in the Faddeev--Popov formalism \cite{lezj,able},
in order to generalize this result to any other gauge.

A priori it is
not clear that two Lagrangians which are related by a Stueckelberg
transformation are equivalent, since such a transformation is not a simple
point transformation
because it involves derivatives of the unphysical scalar
fields; however, within the Hamiltonian formalism this
equivalence can be properly shown. In Hamiltonian framework
no more ``Stueckelberg transformation''
is made, instead, one passes from the gauge
noninvariant (second-class constrained) system to the
gauge invariant (first-class constrained)
system by a phase space enlargement
followed by an application of the constraints in order to convert
the Hamiltonian and the constraints themselves.

In distinction to massive vector fields,
massless vector fields necessarily have to be understood as gauge
fields. A Lagrangian with massless vector fields but gauge
noninvariant interactions of these has no physical meaning
because without a gauge-fixing term, which only becomes
introduced for gauge invariant
Lagrangians (within the Hamiltonian PI formalism as well
as within the
Lagrangian PI formalism), the operator occurring in the quadratic
part of the Lagrangian has no inverse and therefore it is
impossible to obtain a propagator for the vector fields.
I will present a general proof of the HLE theorem for
gauge theories with additional
arbitrary (non--Yang--Mills) self interactions of the gauge fields,
with arbitrary couplings of the gauge fields to scalar fields
and to fermion fields and with arbitrary interactions
among the scalar and fermion fields (which are all gauge invariant).
The proof also applies to the case of SBGTs,
i.e.\ gauge theories with massive
gauge fields, because one can assume
that the scalar fields that are coupled to the gauge fields have a
nonvanishing vacuum expectation value. In fact, this way one obtains a proof
of the HLE theorem for SBGTs alternative to the one announced above
which is more direct and which is not based on the Stueckelberg formalism.

Within the Hamiltonian PI formalism, a gauge
theory cannot be {\em directly\/} quantized
in the Lorentz-gauge or, for the case of SBGTs,
in the $\rm R_\xi$-gauge (which are the most convenient
gauges for practical calculations) because the
corresponding g.f.\ conditions
cannot be written as relations among
the fields and the conjugate fields alone, and
thus they are not g.f.\ conditions within the Hamiltonian framework.
Therefore, I will first derive the generating functional
\eref{lpi} with (\ref{fp})
in the Coulomb-gauge; due to the equivalence of all gauges
\cite{lezj,able} this result can then be rewritten in any other gauge.

When carrying out a complete analysis of
the extensions of the standard model, one necessarily has  to consider
effective interaction terms which depend on higher derivatives
\cite{phhide}. Therefore, I will also derive the HLE theorem for effective
Lagrangians with higher derivatives of the fields (effective higher-order
Lagrangians). However, theories described by
higher-order Lagrangians have quite unsatisfactory properties
\cite{ost,hide}, namely: there are
additional degrees of freedom, the energy is unbound from below,
the solutions of the equations of motion are not uniquely determined
by the
initial values of the fields and their first time derivatives and
the theory has no analytic limit for $\epsilon\to 0$
(where $\epsilon$ denotes the coupling constant
of the higher-order interaction term).
Clearly, these features are very undesirable
when dealing with effective Lagrangians in order to parametrize
small deviations from a renormalizable theory like the standard
model, in which no unphysical effects occur.

Fortunately however,
the abovementioned problems are absent if a
higher-order Lagrangian is considered to be an {\em effective\/}
one. This means, one assumes that there
exists a renormalizable theory with
heavy particles at an energy scale $\Lambda$
(``new physics''), and that the effective Lagrangian
parametrizes the effects of the ``new physics'' at an energy
scale lower than $\Lambda$ by expressing the contributions
of the heavy particles (which do not explicitely occur in it)
through nonrenormalizable effective interactions of the
light particles. Supposed that the renormalizable Lagrangian
describing the ``new physics'' does
not depend on higher derivatives,
it causes no unphysical effects and
therefore such effects also do not occur at the lower energy scale,
i.e.\ at the effective-Lagragian level.
Actually, I will show in this thesis that
in the first order of the effective coupling constant
$\epsilon$ (with $\epsilon\ll 1$)
all higher time derivatives can be eliminated
from the effective Lagrangian.
Higher powers of $\epsilon$ can be neglected
because an effective Lagrangian
is assumed to describe the effects of  well-behaved ``new physics'' in
the $O(\ep)$ approximation only; consequently
all ill-behaved effects (which do
not occur in the first order of $\epsilon$) become cancelled by
other $O(\epsilon^n)$ $(n>1)$ effects of the ``new physics''.

Each effective higher-order
Lagrangian can be reduced to a first-order Lagrangian because
one can apply the equations of motion (EOM) to
eliminate all higher time derivatives from the effective
interaction term (by neglecting higher powers of $\epsilon$).
This is a nontrivial statement because, in general, the EOM
must not be used to convert the Lagrangian. However, it was
shown in \cite{eom,pol}
that it is possible to find field transformations which have the
the same effect as the application
of the EOM to the effective interaction term
(in the first order of $\epsilon$). I will show that
these field transformations
are point transformations (and thus canonical
transformations) within the Hamiltonian formalism for higher-order Lagrangians
(Ostrogradsky formalism \cite{ost})
although they involve derivatives of the fields.
The reason for this is that within the Ostrogradsky formalism
the derivatives up to order $N-1$
are formally treated as independent coordinates if the
Lagrangian is of order $N$, and the order
$N$ of the Lagrangian can be chosen arbitrarily
 without affecting the physical content of the theory
\cite{gity} (as long as $N$ is
greater or equal to the order of the highest actually appearing
derivative).  This implies that the reduced first-order
Lagrangian which is obtained
from the primordial higher-order Lagrangian by applying the EOM
to its effective interaction term is
physically equivalent to this (at the classical and
at the quantum level).

In this thesis I will prove the HLE theorem for the case of
effective higher-order Lagrangians by reducing them to first-order
ones, as explained above,
and then applying the HLE theorem for effective Lagrangians with at
most first time derivatives of the fields.
Especially the treatment of fermion fields
can be simplified very much this way
because the EOM of these fields only depend on first
time derivatives. Therefore one can eliminate not only higher
but also first time derivatives of the fermion
fields from the effective interaction term
and thus it is sufficient to derive the HLE theorem
for the case of effective interactions
in which no time derivatives of these fields occur.
Furthermore, within the Ostrogradsky formalism,
the proof of the canonical equivalence of Lagrangians that are related by
a Stueckelberg transformation can be generalized to the higher-order case
and, besides, it can be simplified very much because, as mentioned above,
in this formalism a field transformation
which involves derivatives of the fields becomes a canonical
transformation.

In the Hamiltonian treatment of effective theories with
scalar fields in \cite{bedu} it is
assumed that the effective interaction term is proportional to a small
$\ep$ and in the subsequent procedure higher powers of $\ep$
are negelected because otherwise
it is not possible to find closed expressions for
the generalized velocities and the Hamiltonian in terms of the fields and the
generalized momenta (if there are higher than second
powers of the velocities in the Lagrangian). I will proceed similarly;
I will assume that the effective interactions, which are only the
{\em deviations\/} from the standard interactions (i.e.\ from the
Yang--Mills self-interactions of the vector fields, minimal gauge
couplings of these to the scalar and fermion fields, Yukawa couplings
of the scalars to the fermions
and derivative-free scalar self-interactions),
are proportional to a
coupling constant\footnote{In general, effective Lagrangians contain more
than one coupling constant in the nonstandard interaction terms. However, this
does not affect the results of this thesis because each effective coupling
constant $\ep_i$ can be written as $\ep_i=\ep g_i$ with $g_i\le 1$, where
$\ep \ll 1$ is the same for all $\ep_i$.}
$\epsilon$ with $\epsilon\ll 1$.
In the proof of the HLE theorem I will
only consider terms which are at most first order
in $\epsilon$, neglecting
higher powers\footnote{One should keep
in mind that the neglection of higher powers of $\ep$ is not a restriction
to the tree level because one can consider loops in which one vertex
follows from the effective interaction term ($\propto\ep$) while the other(s)
are standard (Yang--Mills) vertices.} of $\epsilon$. This treament is
justified when dealing with
phenomenologically motivated effective Lagrangians as in
\cite{nongauge,nonlin,gauge,gluons}
since these are considered in order to investigate the effects of
{\em small\/} deviations from the standard model and since an effective
Lagrangian only describes the $O(\ep)$-approximation of ``new physics''
(see above).

It will turn out that
the result (\ref{eq}) or (\ref{fp}) is only correct up to
additional quartically divergent terms, i.e.\ terms
proportional to $\delta^4(0)$.
According to \cite{bedu}
I will neglect $\delta^4(0)$-terms when
establishing the equivalence
of Hamiltonian and Lagrangian PI quantization
because they become zero in dimensional regularization \cite{bedu,leib}.

Throughout this thesis, I will
introduce the source terms in the PI {\em after\/}
all manipulations will
have been done. This is consistent because the source terms
for the ghost fields have to be introduced later, anyway. Actually, if the
source terms would be considered from the beginning, they would not
remain unchanged in the subsequent treatment. However, a change in the
source terms does not effect the $S$-matrix elements \cite{able}.

This thesis is organized as follows: In chapter~2 I show how the unitary
gauge of a spontaneously broken gauge theory  can be constructed
within the simple Lagrangian path integral formalism,
I give an introduction to the Stueckelberg formalism and I apply it in order to
show that each effective Lagrangian can be written in a gauge invariant form.
Chapter~3 contains a brief review of the dynamics and the
Hamiltonian path integral
quantization of constrained systems. In chapter~4 I prove the
Hamilton--Lagrange equivalence theorem (HLE theorem)
for effective interactions of massive vector fields, namely for
gauge noninvariant effective Lagrangians and for spontaneously broken gauge
theories (with linearly and with nonlinearly realized symmetry) and I
reformulate the Stueckelberg formalism within the Hamiltonian framework.
In chapter~5 I prove the equivalence of
Lagrangians which are related to each other by a field
transformation that involves derivatives,
I show that a higher-order effective Lagrangian can be reduced to a
first-order one by applying the equations of motion to the effective
interaction term and
I derive the HLE theorem for effective Lagrangians with higher derivatives of
the fields. In chapter~6 I present
a general proof of the HLE theorem
for effective gauge theories. {Chapter~7 contains the summary of
the results derived in this thesis.}

\chapter{The Unitary Gauge and the Stueckelberg Formalism}
\markboth{CHAPTER 2. UNITARY GAUGE AND STUECKELBERG FORMALISM}{}
In this chapter I will study the various approaches to the unitary gauge of a
spontaneously broken gauge theory within the simple {\em Lagrangian\/}
(Faddeev--Popov) PI formalism; i.e., I will show
how the unphysical scalar fields can be
removed  from such a theory by quantizing it in this
formalism.
I will give an introduction to the Stueckelberg formalism and I will
utilize it in order to connect gauge noninvariant
Lagrangians with massive vector fields (containing standard or nonstandard
self-interactions of these) with
(linearly or nonlinearly realized) SBGTs.

The treatment of this chapter is a bit aside from the main
point of this thesis,
namely {\em Hamiltonian\/} PI quantiztion, because here all considerations
are based on the naive
Lagrangian PI ansatz. However, the results obtained in this
chapter and the physical methods introduced here
(especially the Stueckelberg
formalism) are of importance for the Hamiltonian treatment of SBGTs in the
following chapters.

SBGTs contain unphysical degrees of
freedom, the pseudo-Goldstone scalars.
At the classical level, the unphysical
fields can be removed by means of a gauge
transformation, i.e., for given values
of the pseudo-Goldstone fields at
each space-time point there exists a gauge transformation (in which the
choice of the gauge parameters depends on these values)
that maps the unphysical fields identically to
zero. This gauge, which is
characterized by the fact that the Lagrangian
contains only ``physical fields'', is called the unitary gauge (U-gauge).

However, this naive definition of the U-gauge
cannot be applied in quantum physics which is best seen within the framework of
Feynman's path integral formalism \cite{feyn,fapo,books}, where
quantization is based on the generating functional \eref{lpi}.
Since the gauge transformation which removes the unphysical fields
is dependent on the values of these fields at each space-time point,
it cannot be applied to the
generating functional where a functional integration over {\em all\/}
values of the fields is done.
In other words there is not a ``universal'' gauge
transformation which maps arbitrary
pseudo-Goldstone fields equal to zero.

There are three (equivalent) ways of
constructing a gauge without unphysical
fields (i.e.\
without pseudo-Goldstone fields and without ghost fields) within the
Lagrangian PI formalism. I discuss them mainly
for the case of linearly and minimally realized
SBGTs (i.e.\ those which contain
physical Higgs scalars and
which are renormalizable). The case
of nonlinear and/or nonminimal theories
will also be treated at the end of this chapter. The construction of the
U-gauge within the correct Hamiltonian PI formalism will be
discussed in the chapters 4 and 5 of this thesis.

The first and most direct
procedure to construct  the U-gauge within the Faddeev--Popov
(FP) formalism
is simply to impose the gauge-fixing condition that the
pseudo-Goldstone fields be equal to zero
(which can be done because of the
existence of the abovementioned gauge transformation). The
resulting FP $\delta$-function
is used in order to integrate out the
unphysical scalar fields while the FP determinant can be
written as an exponential function
without introducing ghost fields, which yields a
quartically divergent
(i.e.\ proportional to $\delta^4(0)$)
nonpolynomial Higgs self-coupling term.
Alternatively, the FP determinant can be rewritten as a ghost term
with static ghost fields. All ghost loops are quartically divergent
and the ghost term can be replaced by the abovementioned
Higgs self-interaction term, which has the same effect on physical
matrix elements.

The second method is to construct the $\rm R_\xi$-gauge
\cite{lezj,able,fls,wein2} in which the unphysical fields are still
present but with masses proportional to the free parameter
$\sqrt{\xi}$, and then to take the limit $\xi\to\infty$. In this
limit, the unphysical fields
get infinite masses and decouple. However, the
ghost-ghost-scalar couplings get
infinite, too, with the consequence, that
the ghost term does not completely vanish; I will show that some of
the ghost loops vanish in this limit and the others become
quartically divergent. The contribution of the latter terms
yields again the quartically divergent Higgs self-interaction term.
The $\rm R_\xi$-limiting
procedure goes back to \cite{wein2,leya}. I will
apply this formalism to an SBGT with only partly broken symmetry,
which requires a modification of the $\rm R_\xi$-gauge, namely the
introduction of different parameters $\xi$ corresponding to the
massive and to the massless gauge fields.

The third way is most similar
to the classical treatment: the unphysical fields
are decoupled from the physical
ones by applying appropriate field transformations \cite{lezj}.
This procedure consists of two
subsequent field transformations; first the unphysical
scalars are paramatrized
nonlinearly and then they are decoupled and can be
integrated out. When making field transformations in the PI,
one also has to take into account
the Jacobian determinant which arises owing
to the change of the functional integration measure
\cite{lezj,sast}; in this case this yields
again the abovementioned Higgs self-interaction term.

Thus, all three methods lead to
a quantized U-gauge Lagrangian
which contains, in addition to the classical U-gauge Lagrangian,
an  extra nonpolynomial quartically  divergent Higgs self-interaction
term. The same term was derived by
quantizing the classical U-gauge Lagrangian
canonically \cite{leya,wein1}
where it emerges as a remnant of covariantization. In section~4.4
I will derive it within the Hamiltonian PI formalism.
However, the $\delta^4(0)$-term becomes zero if dimensional
regularization is applied \cite{bedu,leib}. Besides, it was shown in
\cite{apqu,jogl}, that this term cancels against other
quartic divergences arising from vector-boson loops, so that,
when summing over all Feynman diagrams, no quartic divergent terms
contribute to the $S$-matrix elements (within a renormalizable theory).
Due to the equivalence of all gauges \cite{lezj,able},
loop calculations can
either be carried out within the $\rm R_\xi$-gauge or within the
U-gauge. They seem to be simpler
in the U-gauge than in the $\rm R_\xi$-gauge
because there are less Feynman
diagrams due to the absence of unphysical fields.
On the other hand, the form of the
vector-boson propagator in this gauge
(which is proportional to the zeroth instead of
the inverse second power of the energy) induces higher
divergences in the single Feynman diagrams which are,
however, cancelled when summing over the different diagrams.
These higher divergences do not
occur from the beginning if one uses the $\rm R_\xi$-gauge.
Furthermore,
calculations within the U-gauge suffer from
ambiguities in the determination of the finite part
of an $S$-matrix element
\cite{fls,jawe}. Thus, the $\rm R_\xi$-gauge seems to be more
adequate for loop calculations than the U-gauge.

Therefore, for practical purposes it is not
so useful to remove the
unphysical degrees of freedom from an SBGT. Instead, one
should go the reverse way, i.e., when dealing with a
gauge noninvariant Lagrangian, one should introduce
unphysical fields in order to rewrite
this theory as an SBGT, which enables the choice of the
$\rm R_\xi$-gauge and simplifies loop calculations.
Actually,
the third of the abovementioned procedures can be
reversed, i.e.\ an
SBGT can be ``reconstructed'' from its
(gauge noninvariant) U-gauge Lagrangian. To do this,
scalar fields, which are initially
completely decoupled, are introduced
into the theory by multiplying an
(infinite) constant to the generating
functional which contains the
functional integration over these fields.
The unphysical scalars are then coupled to the physical fields by an
appropriate field
transformation. In the next
step unphysical and physical scalar fields together are
rewritten in a linearized form.
In \cite{clt} a formal proof was given that this way each tree unitary
Lagrangian can be written as an SBGT. I will
explicitly carry out this procedure for the case of the electroweak
standard model taking into account the
(Lagrangian) PI formalism.
This method of constructing
SBGTs by such ``field-enlarging transformations''
represents the non-Abelian version of the Stueckelberg formalism
\cite{stue,kugo,sots,dtt}, which
in its original form was studied only for
theories without physical Higgs bosons where it leads
to the problem of nonpolynomial
interactions and
nonrenormalizability (in non-Abelian theories). The existence
of physical scalars, however,
enables a linearization of the scalar sector, so
that renormalizable Stueckelberg models can be constructed.

By applying a Stueckelberg transformation to a simple massive
Yang--Mills theory (without physical scalars)
one obtains a gauged nonlinear $\sigma$-model. I will
briefly review
and summarize the three different approaches to such a
model discussed in the literature, namely the Stueckelberg formalism
\cite{kugo},
gauging a nonlinear $\sigma$-model \cite{bash}
(i.e.\ a model with a
global nonlinearly realized spontaneously broken symmetry) and taking the
limit $M_H\to\infty$ of a Higgs model \cite{apbe}.

In the electroweak phenomenology effective Lagrangians
with extra
non--Yang Mills vector-boson self-interactions are considered
(see e.g.\ \cite{nongauge}) in order
to parametrize possible deviations from the standard model. Applying a
Stueckelbeg transformation to such an effective Lagrangian allows to
rewrite it as an SBGT. This result has already been obtained in
\cite{bulo}, however without identifying the corresponding
transformation as a Stueckelberg transformation.
Within the resulting SBGT the
gauge group acts nonlinearly on the unphysical fields,
there are nonpolynomial interactions and they are nonrenormalizable.
However, the
gauge freedom enables the choice of the $\rm R_\xi$-gauge (where the
vector-boson propagators have a good high-energy behaviour)
in order to carry out loop calculations within an effective
theory, which shows that the loops
in such a theory do not diverge as severely as one would expect by naive
power counting. Furthermore I will show that each
effective Lagrangian with massive vector bosons
can even be extended to an SBGT with {\em linearly\/}
realized symmetry by introducing a
physical Higgs boson. This makes the loop
corrections even smaller.

Within this chapter renormalizable
SBGTs are discussed by taking the example of the
$\rm SU(2)\times U(1)$ standard model (SM) of the electroweak interaction
\cite{sm}
since it is of greatest phenomenological interest.
Similarly, in the discussion of effective Lagrangians
I restrict myself to
theories containing the electroweak vector bosons,
which leads again to $\rm SU(2)\times U(1)$ gauge invariance. However, the
results obtained in this chapter can easily be generalized to an arbitrary
gauge group.

The results of this chapter were obtained in collaboration with
R.~K\"ogerler. They have first been published in \cite{gkko}.


\section{Preliminaries and Notation}
In this section I introduce my notation of the electroweak standard model
and of
the Faddeev--Popov formalism. For a systematic treatment of these subjects,
the reader is referred to the original literature \cite{fapo,sm} and,
especially, to textbooks on quantum field theory and particle physics
(e.g.\ \cite{books}).

The SM gauge fields corresponding to
the gauge groups SU(2) and U(1) are $W_{\mu}^i$ $(i=1,2,3)$
and $B_{\mu}$, respectively.
For practical purposes the $W$ field
is parametrized in terms of a $2\times 2$ matrix:
\begin{equation} W_{\mu}\equiv\frac{1}{2}W_{\mu}^i\tau_i.\end{equation}
The matrix-valued field strength tensors are
\bea
W_{\mu\nu}&\equiv&\pa_\mu W_\nu-\pa_\nu W_\mu+ig[W_\mu,W_\nu],\nn
B_{\mu\nu}&\equiv&\pa_\mu B_\nu-\pa_\nu B_\mu.\label{fst}
\eea
The scalar fields $\tilde{h}$ and $\varphi_i$
are parametrized by means of
as a $2\times 2$ matrix as well:
\begin{equation}\Phi\equiv\frac{1}{\sqrt{2}}(
\tilde{h}{\bf{1}}+i\tau_i\varphi_i) .\label{higgslin}\end{equation}
Furthermore I consider
one fermionic doublet $\Psi$ (the generalization to more doubletts
works as usual) consisting of an up-type field $u$ and a down-type
field $d$ (quark or lepton)
\be
\Psi\equiv\left(\ba{c}u \\ d \ea \right),\qquad \Psi_{L,R}\equiv
\frac{1}{2}
(1 \mp \gamma_5)\Psi
\ee
with the fermion mass matrix
\begin{equation} M_f\equiv\left(
\begin{array}{cc}m_u&0\\0&m_d\end{array}\right).
\end{equation}
With the appropriate covariant derivatives
\bea
D_\mu \Phi &\equiv& \pa_\mu \Phi+ig W_\mu \Phi-\frac{i}{2}g'\Phi
B_\mu\tau_3,\nn
D_\mu\Psi_L&\equiv&\left(\pa_\mu+igW_\mu+\frac{i}{2}g'(B-L)B_\mu\right)
\Psi_L,\nn
D_\mu\Psi_R&\equiv&\left(\pa_\mu+\frac{i}{2}g'(\tau_3+B-L)B_\mu\right)
\Psi_R\label{cdsm}\eea
(with $B$ and $L$ being the baryon and lepton number of $\Psi$),
the SM Lagrangian takes the well known form \cite{books,sm}
\begin{eqnarray}{\cal L}_{SM}&=&-\frac{1}{2}
{\,\rm tr\,}(W^{\mu\nu}W_{\mu\nu})-
\frac{1}{4}(B^{\mu\nu}B_{\mu\nu})\nonumber\\ &&
+\frac{1}{2}{\,\rm tr\,}\left[(D^\mu\Phi)^\dagger
(D_\mu\Phi)\right]-
\frac{1}{2}\mu^2{\,\rm tr\,}
(\Phi\Phi^\dagger)-\frac{1}{4}\lambda{\,(\rm tr\,}
(\Phi\Phi^\dagger))^2\nonumber\\ &&
+i(\bar{\Psi}_L
\gamma_\mu D^\mu\Psi_L+\bar{\Psi}_R\gamma_\mu D^\mu\Psi_R)
-\frac{\sqrt{2}}{v}
(\bar{\Psi}_L\Phi M_f\Psi_R+\bar{\Psi}_R M_f\Phi^\dagger
\Psi_L)\label{linv}\end{eqnarray}
(with $\mu^2<0$ and $\lambda>0$).
${\cal L}_{SM}$ is invariant under the local
$\rm SU(2)\times U(1)$ gauge transformations
\begin{eqnarray} W_\mu&\to &S(x)W_\mu S^\dagger(x)
-\frac{i}{g} S(x)\partial_\mu S^\dagger(x)
,\nonumber\\ B_\mu&\to &B_\mu-\partial_\mu\beta(x),
\nonumber\\
\Phi& \to & S(x)\Phi\exp\left(-\frac{i}{2}g'\beta(x)\tau_3\right),
\nonumber\\
\Psi_L&\to & S(x)\exp\left(\frac{i}{2}g'
(B-L)\beta(x)\right)\Psi_L,\nonumber\\
\Psi_R& \to & \exp\left(\frac{i}{2}g'
(\tau_3+B-L)\beta (x)\right)\Psi_R
\label{gaugetrafo}\end{eqnarray}
with
\begin{equation}
 S(x)\equiv\exp\left(\frac{i}{2}g\alpha_i(x)\tau_i\right),\end{equation}
(where $g$ and $g'$ are the SU(2) and U(1) coupling constants,
respectively). $\alpha_i(x)$
and $\beta(x)$ denote the four gauge parameters.
The nonvanishing
vacuum expectation value (VEV) of the scalar
field $\Phi$ \eref{higgslin} is
\begin{equation} \langle \Phi\rangle_0=
\frac{v}{\sqrt{2}}{\bf 1}\quad\mbox{with}\quad
v=\sqrt{\frac{-\mu^2}{\lambda}}.\label{vev}\end{equation}
The fields \begin{equation} h\equiv\tilde{h}-v\label{higgsshift}
\end{equation}
and $\varphi_i$ have a
vanishing VEV. $h$ is the Higgs field and $\varphi_i$
are the pseudo-Goldstone fields.
The physical gauge-boson
fields $W^\pm_\mu$, $Z_\mu$ and $A_\mu$ (photon) are
the well known combinations of $W_{\mu}^i$ and $B_\mu$:
\bea
W^\pm_\mu&\equiv&\frac{1}{\sqrt{2}}\left(W^1_\mu\mp W^2_\mu\right),\nn
Z_\mu&\equiv& \cos\theta_W W^3_\mu-\sin\theta_W B_\mu,\nn
A_\mu&\equiv& \sin\theta_W W^3_\mu+\cos\theta_W B_\mu
\label{mix}\eea
(with $\theta_W$ being the Weinberg
angle defined by $\tan\theta_W\equiv\frac{g'}{g}$).

In order
to quantize the theory one usually starts from the Lagrangian PI ansatz
\cite{feyn,books}
\begin{equation} Z=\int{\cal D} W_{\mu}^i{\cal D}
B_\mu {\cal D} h{\cal D}\varphi_i {\cal D}
\Psi{\cal D}\bar{\Psi}\,
\exp\left(i\int d^4x\,
{\cal L}_{SM}\right). \label{pi}\end{equation}
$Z$ contains an infinite constant due to the gauge freedom.
This is removed within the Faddeev--Popov (FP) formalism \cite{fapo,books}
by imposing the general gauge-fixing conditions
\begin{equation}
F_a(W_{\mu}^i,B_\mu,\Phi)=C_a(x),\qquad a=1,\ldots ,4\label{gf}
\end{equation}
(where $C_a(x)$ are arbitrary funcions) and then rewriting $Z$
(after dropping the infinite constant) as
\be Z=\int{\cal D} W_{\mu}^i{\cal D} B_\mu
{\cal D} h{\cal D}\varphi_i
{\cal D}\Psi{\cal D}\bar{\Psi}\,
\delta^4[F_a-C_a(x)]{\,\rm Det\,}
\left(\frac{\delta F_a(x)}{\delta\alpha_b(y)}\right)
\exp\left(i\int d^4x\, {\cal L}_{SM}\right) \label{pigf}
\ee
($\alpha_a=(\alpha_i,\beta)$).
Since (\ref{pigf}) is independent of the $C_a$
\cite{lezj}, one can construct the weighted average over them
(with the weight functions
$\exp\left(-\frac{i}{2\xi_a}\int d^4x\, C_a^2\right)$, $\xi_a$
being a set of free parameters\footnote{Usually
all $\xi_a$ are taken to be
equal, but for my purposes I
allow also different $\xi_a$.\label{diffxi}}).
Then one expresses the FP determinant through  the
ghost fields $\eta_a^{}$, $\eta_a^\ast$ by using
\begin{equation} {\,\rm Det\,}
\left(\frac{\delta F_a(x)}{\delta\alpha_b(y)}\right)\propto
\int{\cal D}\eta^{}_a{\cal D}\eta^\ast_a\,
\exp\left(-i\int d^4x d^4y\, \eta^\ast_a(x)\frac{\delta F_a(x)}{\delta
\alpha_b(y)}\eta^{}_b(y)\right).\label{ghosts}\end{equation}
As a result, (\ref{pigf}) can be written as \cite{books,lezj,able,fls,wein2}
\begin{equation} Z=\int{\cal D} W_{\mu}^i{\cal D} B_\mu
{\cal D} h{\cal D}\varphi_i {\cal D}\Psi{\cal D}\bar{\Psi}
{\cal D}\eta^{}_a{\cal D}\eta^\ast_a\,
\exp\left(i\int d^4x \, {\cal L}_{quant}\right) \label{pieff}
\end{equation}
with the quantized Lagrangian
\begin{eqnarray} {\cal L}_{quant}&=&{\cal L}_{SM}-\frac{1}
{2\xi_a}F_a^2-\eta^\ast_a\frac{\pa
F_a}{\pa \alpha_b}\eta^{}_b
\nonumber\\&\equiv&{\cal L}_{SM}+{\cal L}_{g.f.}+
{\cal L}_{ghost},\label{leff}\end{eqnarray}
which contains an additional gauge-fixing term a ghost
term\footnote{Note that $\lag_{ghost}=-\eta^\ast_a\frac{\pa
F_a}{\pa \alpha_b}\eta^{}_b$ is a convenient shorthand notation for
$\lag_{ghost}=-\int d^4y\,\eta^\ast_a(x)\frac{\delta
F_a(x)}{\delta \alpha_b(y)}\eta^{}_b(y)$ (which is a local expression because
$\frac{\delta F_a(x)}{\delta \alpha_b(y)}$ is proportional to $\delta^4(x-y)$
or its derivatives). This notation will be used throughout this thesis.}.
Finally, source terms for all (physical and
unphysical) fields have to be introduced.

Later in this thesis I will also consider effective Lagrangians
with an additional arbitrary \suu -invariant interaction term $\lag_I$,
\be
\lag_{eff}=\lag_{SM}+\ep\lag_I.
\label{lageff}\ee
Obviously, the g.f.\ term and the ghost term in \eref{leff}, which only
depend on the choice of the g.f. conditions  \eref{gf},
remain unchanged if,
instead of $\lag_{SM}$ \eref{linv}, such a Lagrangian
is quantized; i.e.\ these terms are independent of the form of $\lag_I$.


\section{Derivation of the U-Gauge by Gauge Fixing}
In this section I explain the ``direct way'' of constructing the
U-gauge within the
FP formalism by setting the unphysical pseudo-Goldstone
fields  equal to zero from the beginning. This is done
by imposing the
gauge-fixing conditions
\bea
F_i&\equiv&\varphi_i=0,\qquad\qquad i=1,2,3,\nonumber\\
F_4&\equiv&\partial_\mu A^\mu=C(x)\label{ugf}.
\end{eqnarray}
This is a possible choice
of g.f.\ conditions since the $\varphi_i$ can be transformed
to zero by means of a gauge transformation.
The fourth g.f.\ condition is necessary
because the unbroken symmetry $\rm U(1)_{em}$
has to be fixed as well.
(\ref{pigf}) now takes the form
\be Z=\int{\cal D} W_{\mu}^i{\cal D} B_\mu {\cal D}
h{\cal D}\varphi_i{\cal D}\Psi{\cal D}\bar{\Psi}\,
\delta^3[\varphi_i]\delta[\partial_\mu A^\mu-C(x)]
{\,\rm Det\,}
\left(\frac{\delta F_a(x)}{\delta\alpha_b(y)}\right)
\exp\left(i\int d^4x\, {\cal L}_{SM}\right). \ee
Only the second $\delta$-function is treated
in the way explained above,
yielding an
appropriate ${\cal L}_{g.f.}$. The other one enables to
carry out the
${\cal D} \varphi_i$ integration with the result that all terms with
pseudo-Goldstone
fields in the SM Lagrangian  and in the ghost term
become equal to zero. Thus, the unphysical pseudo-Goldstone
bosons are removed from the
theory. The quantized Lagrangian becomes
\begin{equation} {\cal L}_{quant}={\cal L}_{SM}|_{\varphi_i=0}
-\frac{1}{2\xi_\gamma}(\partial_\mu
A^\mu)^2+{\cal L}_{ghost}|_{\varphi_i=0}\label{leffu}\end{equation}
where $\lag_{ghost}$ is given by \eref{leff}.
Except for ${\cal L}_{ghost}$ this is
identical to the classical U-gauge Lagrangian
with fixed $\rm U(1)_{em}$.

Let me now consider the ghost term
and show that in this gauge also the ghost fields can be removed
from the theory.
Corresponding to the gauge boson mixing \eref{mix} I define the parameters
\begin{eqnarray}
\alpha_Z&=&\cos\theta_W \alpha_3 -\sin\theta_W \beta,\nonumber\\
\alpha_\gamma&=&\sin\theta_W \alpha_3 +\cos\theta_W
\beta.\end{eqnarray}
{}From (\ref{gaugetrafo}) one finds the changes of the
$F_a$ \eref{ugf} under infinitesimal gauge transformations.
\begin{eqnarray} \delta F_1&=& \delta\varphi_1=
\frac{g}{2}(v+h)\delta\alpha_1+O(\varphi_i),\nonumber\\\delta F_2&=&
\delta\varphi_2=\frac{g}{2}(v+h)\delta\alpha_2+O(\varphi_i),
\nonumber\\\delta F_3&=& \delta\varphi_3=
\frac{g}{2\cos\theta_W}(v+h)\delta\alpha_Z+O(\varphi_i),\nonumber\\
\delta F_4&=& \delta(\partial_\mu A^\mu)=
-\Box\delta\alpha_\gamma+O(\varphi_i)+O(W_{\mu}^i)
.\label{infgauge}\end{eqnarray}
First, one can see that all terms which are proportional to the
pseudo-Goldstone fields $\varphi_i$ (denoted as $O(\vp_i)$)
yield vanishing contributions to the ghost terms after
integrating out the $\delta$-function, as explained above.
Second, $\eta_\gamma$ (the ghost belonging to the
electromagnetic gauge freedom)
is a physically inert field: it is not possible
to construct a Feynman diagram
with internal $\eta_\gamma$-lines, because
(\ref{infgauge}) only yields (besides
a kinetic term for $\eta_\gamma$)
vertices with outgoing $\eta_\gamma$-lines
(and incoming $\eta^\pm$-lines coupled to
$W^\pm$) but no vertices with an
incoming $\eta_\gamma$. Thus, the field
$\eta_\gamma$ can be integrated out.

After removing all redundant terms, one can express the resulting
FP determinant as
\begin{equation} {\,\rm Det\,}\left(\frac{\delta F_a(x)}
{\delta\alpha_b(y)}\right)={\,\rm Det\,}
\left(\frac{g}{2}(v+h)\, {\rm diag}
\left[1,1,\frac{1}{\cos\theta_W}\right] \delta^4(x-y)\right).
\label{detfp}\end{equation}
Since the argument of the determinant has the simple form $M_{ab}(x)
\delta^4(x-y)$, one  can
express the functional
determinant (``Det'') in terms of the ordinary one
(``det'') by using the relation \cite{sast}
\begin{equation} {\,\rm Det\,} (M_{ab}(x)\delta^4(x-y))=
\exp\left[\delta^4(0)\int d^4x\, \ln(\det M_{ab}(x))
\right].\label{det}\end{equation}
Thus one finds
\begin{equation}{\,\rm Det\,}\left(\frac{\delta F_a(x)}
{\delta\alpha_b(y)}\right)=\exp\left( i\int
dx\, \left(-3i\delta^4(0)
\ln\left(1+\frac{h}{v}\right)-i\delta^4(0)\ln\frac{g^3v^3}
{8\cos\theta_W}\right)\right) .\end{equation}
This means that in this case (in contrast to the
$\rm R_\xi$ gauge, where
the argument of the FP determinant contains derivatives of the
$\delta$-function and thus \eref{det} cannot be applied)
the FP determinant can be written as an exponential function
without introducing unphysical ghost fields. As a result
one finds (neglecting a constant and using $M_W=\frac{vg}{2}$)
the ghostless FP term to the Lagrangian:
\begin{equation}{\cal L}_{ghost}=-3i\delta^4(0)\ln\left
(1+\frac{g}{2M_W}h\right).\label{extra}\end{equation}
(\ref{leffu}) with \eref{extra} shows that ${\cal L}_{quant}$
(in the gauge defined
by (\ref{ugf}))
contains no unphysical fields, neither pseudo-Goldstone nor ghost
fields\footnote{\label{foot}If the
unbroken subgroup is non-Abelian the ghost fields belonging to this
subgroup are still present.
These can be removed by
choosing the axial gauge $t_\mu A_b^\mu=C_b(x)$
for the massless gauge bosons
instead of the Lorentz gauge as in (\ref{ugf}).}.
Instead, there is the extra term (\ref{extra}) describing a
quartically divergent
nonpolynomial Higgs self-interaction.

The extra term (\ref{extra}) can alternatively
be derived from the Feynman diagrams
obtained by applying \eref{ghosts} in order to
express the determinant (\ref{detfp}) in terms of
usual ghost fields. I am going now to present also
this derivation, since it makes the role
of the new interaction term more transparent.
In this procedure the ghost term
is (with $\eta^\pm=\frac{1}{\sqrt{2}}(\eta_1\mp i\eta_2)$)
\begin{eqnarray}
{\cal L}_{ghost}&=&-M_W\eta^{+\ast}\eta^+ -M_W\eta^{-\ast}\eta^-
-M_Z\eta^{\ast}_Z\eta^{}_Z
\nonumber\\ && -\frac{g}{2}\eta^{+\ast}\eta^+h -
\frac{g}{2}\eta^{-\ast}\eta^-h -\frac{g}{2\cos\theta_W}\eta_Z^\ast
\eta^{}_Z h .\label{ufp}\end{eqnarray}
There are no kinetic terms of the ghost fields, only
mass terms and couplings
to the Higgs boson. This means that the ghost propagators
are static ones, i.e. inverse masses.
Figure~\ref{feynrulesug} shows the Feynman rules
derived from (\ref{ufp}).
\begin{figure}
\vspace{5cm}
\caption[]{\label{feynrulesug}Feynman
rules obtained from the ghost term (\ref{ufp})
in the U-gauge. In all
figures the solid lines represent the Higgs lines and
the dotted ones the ghost lines.}
\end{figure}
Since the ghost fields exclusively couple to the Higgs boson,
they only contribute to Feynman diagrams with ghost loops
connected to an arbitrary
number of Higgs lines (Figure~\ref{ghostloop}),
which can be internal or external ones.
\begin{figure}
\vspace{7cm}
\caption[]{\label{ghostloop}Ghost loop connected to $N$
Higgs lines contributing to the
Feynman diagrams in the U-gauge.
The internal ghosts may be $\eta^\pm$ or
$\eta_Z$.}
\end{figure}
The Feynman rules (Fig.~\ref{feynrulesug}) imply that
the contribution of such a loop with $N$ ghost
propagators coupled to $N$ Higgs bosons to the amplitude
is (with a factor $(2\pi)^{-4}$ for
the loop and one $(-1)$ due to the Fermi
statistics of the ghosts)
\begin{equation} -\int \frac{d^4p}{(2\pi)^4}
\left(-\frac{g}{2M_W}\right)^N=-\delta^4(0)
\left(-\frac{g}{2M_W}\right)^N \label{higgsloopcont} \end{equation}
for an internal $\eta^\pm$ as well
as for an internal $\eta_Z$. One can see that
such a ghost loop effectively provides for a quartically
divergent
$N$-Higgs
self-coupling. Let me for a moment go to the one-loop level (where
the Higgs lines connected to the
ghost loop are ``tree lines'') and consider
all subdiagrams like Fig.~\ref{ghostloop}
with a fixed number $N$  of
Higgs lines. The sum of their contributions is
\begin{equation} -3\delta^4(0)(N-1)!
\left(-\frac{g}{2M_W}\right)^N. \label{vertexfactor}\end{equation}
The factor $3$ is due to the three types of
internal ghosts and the factor $(N-1)!$ is due to the $(N-1)!$ different
possibilities to attach the $N$ Higgs lines
to such a loop (as one can easily verify by
induction). Thus all the ghost loops with $N$ Higgs lines
together can be replaced by an
extra $N$-Higgs vertex (Figure~\ref{extravertex}) with the
quartically divergent
vertex factor (\ref{vertexfactor}).
\begin{figure}
\vspace{5cm}
\caption[]{\label{extravertex}Extra quartically
divergent $N$-Higgs-boson vertex.}
\end{figure}
Considering a combinatorical factor of
$1/N!$ due to the $N!$ different
possibilities to attach $N$ Higgs lines to
the such a vertex, all the extra $N$-Higgs
vertices (with all possible  values of
$N$) can be derived from the Lagrangian
\begin{equation}
{\cal L}_{extra}=3i\delta^4(0)\sum_{N=1}^\infty\left(\frac{1}{N}
\left(-\frac{g}{2M_W}\right)^Nh^N\right)=-3i\delta^4(0)\ln\left(
1+\frac{g}{2M_W}h\right),\label{extraalt}\end{equation}
which is identical to (\ref{extra}).

This one-loop derivation can easily be
generalized to all loop orders without changing the result:
If the Higgs lines  in Figs.~\ref{ghostloop}
and \ref{extravertex} are not ``tree lines'' but
connected to loops among themselves or to
other ghost loops or extra vertices of this type,
the only thing in the above discussion that changes is the
combinatorics.
However, the combinatorical factors of $(N-1)!$ for the
$N$-Higgs ghost-loop and
of $N!$ for the $N$-Higgs vertex change by the same extra
factor so that this cancels.
One can easily see that, no
matter how the Higgs lines are connected, for each
way to  attach $N$ Higgs lines to a loop like Fig.~\ref{ghostloop}
there are $N$ ways to
attach them to a vertex like Fig.~\ref{extravertex},
corresponding to the $N$ cyclic permutations.

This alternative derivation
of the extra term (\ref{extra}) to the quantized Lagrangian
(although it is more elaborate)
shows explicitly the meaning of (\ref{extra}):
$\delta^4(0)$ has to be interpreted as a quartically
divergent integral stemming
from
(\ref{higgsloopcont}) and
can be expressed in terms of a cut-off\footnote{Remember that $\delta^4(0)$
vanishes in dimensional regularization \cite{bedu,leib}.}
$\Lambda$ as
$\frac{\Lambda^4}{(2\pi)^4}$ and
the logarithm has to be evaluated in a power series
as in (\ref{extraalt}) which yields
a (nonpolynomial) self-interaction of an
arbitrary number of Higgs bosons.
One can  see that the unphysical
ghost fields can effectively be removed from the
theory by taking the U-gauge
but they do not completely decouple (as the
pseudo-Goldstone fields do): there remains the additional
interaction term (\ref{extra}) as a
remnant. However there are no more explicit ghost fields in this
term.

Remembering the last paragraph of the preceding section, one can easily
generalize this result to the quantization of an effective Lagrangian
with an additional arbitrary (but gauge invariant)
interaction term \eref{lageff}. The
quantized U-gauge Lagrangian turns out to be
\be
\lag_{quant}=\lag_{eff}\Big|_{\vp_i=0}-\frac{1}{2\xi_\gamma}
(\pa_\mu A^\mu)^2+\lag_{extra}\label{ugaugefp}\ee
with the same $\lag_{extra}$ \eref{extra}
as in the SM case (due to the independence
of the ghost term from the effective interaction term).
This result will be used when deriving the HLE theorem for effective SBGTs
later in this thesis.


\section[R${}_{\xi}$-Limiting Procedure]{R${}_{\bf \xi}$-Limiting Procedure}
The second approach
to the unitary gauge is to start from the SM quantized in the
$\rm R_\xi$-gauge and then to take the limit $\xi\to\infty$.
To construct this limit, one has to modify the
$\rm R_\xi$-gauge a bit because the photon propagator
\begin{equation}
i\frac{-g^{\mu\nu}+(1-\xi)\frac{\textstyle p^\mu p^\nu}{\textstyle
p^2}}{p^2} \label{gammaprop} \end{equation}
would become infinite for $\xi\to\infty$,
while the propagator of the massive gauge bosons
\begin{equation}  i\frac{-g^{\mu\nu}+(1-\xi)\frac{\textstyle
p^\mu p^\nu}{\textstyle p^2-\xi M_B^2}}{p^2-M_B^2}, \label{mgbp}
\end{equation}
($M_B=M_W$ or $M_Z$) remains finite in this limit.
I impose the usual $\rm R_\xi$ gauge-fixing
conditions \cite{books,lezj,able,fls,wein2}
and express them in terms of the mass
eigenstates $A^\mu$ and $Z^\mu$ of the neutral sector
(instead of $W_3^\mu$ and $B^\mu$):
\begin{eqnarray}  F_{1,2}&\equiv&\partial_\mu
W^\mu_{1,2}-\xi M_W\varphi_{1,2}=C_{1,2}(x),\nonumber\\
F_3&\equiv&\partial_\mu Z^\mu -\xi M_Z \varphi_3=C_3(x),\nonumber\\
F_4&\equiv&\partial_\mu A^\mu =C_4(x).\label{rgf}\end{eqnarray}
In order to obtain the
corresponding $\lag_{g.f.}$ I make use of the
possibility to introduce different parameters $\xi_a$
for each $F_a$ ($a=1,\ldots ,4$) in the g.f. term in \eref{leff}
(see footnote~\ref{diffxi} of this chapter). The
$\lag_{g.f.}$ most convenient for my purposes is
\begin{equation} {\cal L}_{g.f.}=-\frac{1}{2\xi}
\left[\sum_{i=1}^2(\partial_\mu W^\mu_i-\xi M_W
\varphi_i)^2+(\partial_\mu Z^\mu -\xi M_Z \varphi_3)^2\right]-
\frac{1}{2\xi_\gamma}(\partial_\mu A^\mu)^2\label{lgfr}\end{equation}
with {\em two\/} free parameters
$\xi$ and $\xi_\gamma$. The ghost term
depends only on $\xi$ since
(\ref{rgf}) depends only on $\xi$, thus the sole
difference between this gauge and the usual $\rm R_\xi$-gauge
is that $\xi$ in the photon
propagator is replaced by $\xi_\gamma$. Now one can take the limit
$\xi\to\infty$ \cite{wein2,leya}
(which does not affect physical
observables since these do not depend on $\xi$
\cite{lezj,able})
while the unbroken subgroup $\rm U(1)_{em}$
is fixed in an arbitrary
gauge specified by a finite $\xi_\gamma$
so that the photon propagator remains finite.

A complete list of Feynman rules for the $\rm R_\xi$-gauged
SM is, e.g., given in the book of Bailin and Love
\cite{books}. The $\xi$-dependent propagators and vertices are:
\begin{itemize}
\item The propagators of the massive gauge bosons \eref{mgbp},
which become Proca propagators of
massive spin-one particles for infinite $\xi$. Thus, the contribution
of those parts of the g.f.\ term, which are quadratic in the massive
gauge fields, vanishes in this limit.
\item The propagators of the
pseudo-Goldstone bosons and of the ghost fields
(except for $\eta_\gamma$, which is massless)
\begin{equation} i\frac{1}{p^2-\xi M_B^2}.\end{equation}
For $\xi\to\infty$, these particles
acquire infinite mass and their propagators
become zero. If
there would be no $\xi$-dependent vertices these particles
would completely decouple.
\item All
couplings of a (physical or unphysical) scalar to a ghost pair.
These are proportional to $\xi$ and become
infinite for $\xi\to\infty$.
\end{itemize}
Now I examine which Feynman
diagrams with lines corresponding to
unphysical fields do not vanish for $\xi\to\infty$.
Since the pseudo-Goldstone
and the $\eta^\pm,\eta_Z$ propagators behave as
$\xi^{-1}$ for large $\xi$ their number has to equal the number of
scalar--ghost--ghost vertices ($\propto\xi$) in
such diagrams. This means:
\begin{itemize}
\item All propagators $\propto\xi^{-1}$
have to be coupled to $\xi$-dependent
vertices at {\em both\/} ends, i.e.,
couplings of unphysical fields to
gauge fields or to fermion fields (which are independent of $\xi$)
do not contribute in this limit.
\item Only those $\xi$-dependent vertices yield
nonvanishing contributions
which couple to two $\xi$-dependent and
one $\xi$-independent propagator.
These are the $h\eta\eta^\ast$ and the
$\varphi^\pm\eta^{\pm\ast}\eta_\gamma$
vertices. However, the latter do not
contribute, since they exist only with
an incoming $\eta_\gamma$ but not with
an outgoing one, thus it is not possible
to construct ghost loops with
them.
\end{itemize}
As a consequence, all graphs with pseudo-Goldstone
lines or with $\eta_\gamma$-lines vanish for
$\xi\to\infty$. Thus,
the $\varphi_i$-fields and the ghost field $\eta_\gamma$
can be neglected altogether
in this limit. This means that the only nonvanishing diagrams containing
unphysical particles are those with
ghost loops that are exclusively coupled to Higgs bosons
(Fig.~\ref{ghostloop}), i.e. the same diagrams which contribute within the
alternative derivation of the U-gauge in the previous section.
The corresponding Feynman rules in the \rx -gauge are given in
Fig.~\ref{feynrx}.
\begin{figure}
\vspace{5cm}
\caption[]{\label{feynrx}Feynman
rules for Fig.~\ref{ghostloop} in the $\rm R_\xi$-gauge.}
\end{figure}

The contribution of such a loop with $N$ external Higgs lines
for $\xi\to\infty$ is (for internal $\eta^\pm$ as for $\eta_Z$)
\bea&& -\lim_{\xi\to\infty}\int\frac{d^4p}{(2\pi)^4}
\left(-\frac{i}{2}\xi gM_W\right)^N
\prod_{i=1}^N\frac{i}{p_i^2-\xi M_W^2}\nn&&
=-\int \frac{d^4p}{(2\pi)^4}\left(-\frac{g}{2M_W}
\right)^N=-\delta^4(0)
\left(-\frac{g}{2M_W}\right)^N.\label{looplim}\eea
(The momenta $p_i$ of the
internal ghosts do not have to be specified to
construct the limit.) This is identical to (\ref{higgsloopcont}); therefore
one can transfer the discussion of the previous section and finds for
the limit of the ghost term
\begin{equation}
\lim_{\xi\to\infty}{\cal L}_{ghost}=-3i\delta^4(0)\ln\left
(1+\frac{g}{2M_W}h\right),\end{equation}
which is again the extra term
(\ref{extra}).

Thus in the $\rm R_\xi$-gauge, after taking the limit $\xi\to\infty$,
the pseudo-Goldstone fields and even the massless ghost field
$\eta_\gamma$ (which has a $\xi$-independent
propagator) decouple
completely, while the effects of the
massive ghosts can be expressed through the (ghostless)
extra interaction term (\ref{extra}).
Therefore I obtain the same result
for the U-gauge Lagrangian as in the previous
section\footnote{Actually, it is a priori
not clear that the limit $\xi\to\infty$
can be taken {\em before\/}
the loop
integration in (\ref{looplim}). The
fact that the obtained result is identical
to that of the alternative derivations justifies  this treatment.}.


\section{Decoupling the Unphysical Scalars}
In this section I derive the U-gauge of the SM
by applying appropriate field
transformations to the gauge invariant Lagrangian (\ref{linv}). I start with
reparametrizing (point transforming)
the scalar sector of the theory, (\ref{higgslin}) with \eref{higgsshift},
nonlinearly as
\cite{lezj,higgs}:
\begin{equation}
\Phi=\frac{1}{\sqrt{2}}((v+h){\bf 1}+i\tau_i\varphi_i)=
\frac{1}{\sqrt{2}}(v+\rho)
\exp\left(i\frac{
\zeta_i\tau_i}{v}\right).\label{nonlin}\end{equation}
Here, $\rho$ is the new Higgs field and the $\zeta_i$ are the new
pseudo-Goldstone fields. One can see that in
this parametrization the Lagrangian
(\ref{linv}) contains nonpolynomial
interactions of the $\zeta_i$ to the gauge
bosons and to the fermions, which stem from expanding
the exponential
in the kinetic term of $\Phi$ and in the Yukawa
term. At the quantum level
this is not the whole story: The basis of quantization is not the Lagrangian
(\ref{linv}) but the PI (\ref{pi}).
Therefore one has to transform the integration measure in the PI, too,
which yields a functional Jacobian
determinant \cite{lezj,sast} according to
\begin{equation}
{\cal D} h {\cal D}\varphi_i ={\cal D}\rho{\cal D}\zeta_i\,
{\,\rm Det\,}\frac{\delta(h,\varphi_i)}{\delta(\rho ,\zeta_i)}.
\end{equation}
The explicit form of the field transformation (\ref{nonlin}) is
\begin{eqnarray} h&=&(v+\rho)\cos\tilde{\zeta}-v,\nonumber\\
\varphi_i&=&(v+\rho)\hat{\zeta_i}
\sin\tilde{\zeta}\label{nonlintrans}\end{eqnarray}
(with $\zeta\equiv\sqrt{
\zeta_1^2+\zeta_2^2+\zeta_3^2}$, $\hat{\zeta}_i\equiv\zeta_i/\zeta$
and $\tilde{\zeta}\equiv\zeta/v$).
One can again use (\ref{det}) in order
to express the functional Jacobian determinant in terms of the ordinary one,
wich is given by
\begin{equation}
\det\frac{\partial(h,\varphi_i)}{\partial(\rho ,\zeta_i)}=
(v+\rho)^3\frac{\sin^2\tilde{\zeta}}{v\zeta^2}. \end{equation}
With (\ref{det}) one finds that  the
change of the functional intergration measure  yields
the following extra term to the Lagrangian (after dropping a constant)
\begin{equation}
{\cal L}^\prime=-3i\delta^4(0)\ln\left(1+\frac{g}{2M_W}\rho\right)
-i\delta^4(0)\ln\left(\frac{\sin^2\tilde{\zeta}}
{\tilde{\zeta}^2}\right).
\label{twoextra}\end{equation}
The first term is again the
quartically divergent nonpolynomial Higgs
self-coupling term (\ref{extra}),
but there is also a quartically divergent nonpolynomial
self-interaction term of the nonlinearly realized
pseudo-Goldstone fields.

Introducing  the field combination
\begin{equation}
U\equiv\exp\left(i\frac{\zeta_i\tau_i}{v}\right), \label{u}
\end{equation}
one can deduce from (\ref{gaugetrafo}) the behaviour of $\rho$ and
$\zeta_i$
under $\rm SU(2)\times U(1)$ gauge transformations:
\begin{eqnarray} \rho&\to&\rho,\nonumber\\
U&\to&S(x)U\exp(-\frac{i}{2}g'\beta(x)\tau_3),\label{zetatrafo}
\end{eqnarray}
i.e.,
the physical scalar $\rho$ is a singlet.
Consequently, the first term in (\ref{twoextra}) is
gauge invariant while the second is not. However, this is no serious problem
since gauge invariance is not really
destroyed, it is just not completely obvious due
to the nonlinear parametrization\footnote{This can easily be
visualized with the help of an
example from quantum mechanics: If one studies a
translational invariant
Lagrangian and transforms the functional intergration
measure to polar
coordinates, one finds a translational noninvariant
extra term to the Lagrangian
although physics is still translational invariant.}.

In order to remove the unphysical scalar fields $\zeta_i$
from the theory  one could
apply one of the methods described in the previous two sections.
I do not explain this in detail here but only mention the main features.
If one imposes the gauge-fixing conditions \eref{ugf} (U-gauge)
or \eref{rgf} (\rx-gauge) with $\vp_i$ replaced by $\zeta_i$ one
finds expressions for the ghost term analogous to those obtained in sections
2.2 and 2.3, except that there are
additional interactions of more than one pseudo-Golstone boson with the ghost
fields (which do not affect the discussion) and that there are no couplings
of the Higgs boson to the ghosts, because from \eref{zetatrafo} it is obvious
that the changes of the $\zeta_i$ \eref{zetatrafo}, and therefore also those
of the $F_a$ \eref{ugf} or \eref{rgf}, under infinitesimal
gauge transformations do not depend on
$\rho$. Thus no ghost loops as in Figure~\ref{ghostloop} can be constructed.
Remembering the discussion of the previous sections, one can see
that the ghost term vanishes completely after integrating out
$\int{\cal D}\zeta_i\,\delta^3(\zeta_i)$ in the first case and
after taking the limit $\xi\to\infty$ in the second case.

Here I choose another possibility and apply
one further field transformation, which
affects the vector and the fermion fields. It is just a reversed
(non-Abelian) Stueckelberg transformation \cite{kugo,lezj,clt,dtt}:
\begin{eqnarray} w_\mu&=&U^\dagger W_\mu U -\frac{i}{g}
U^\dagger\partial_\mu U,\nonumber\\
b_\mu&=&B_\mu,\nonumber\\
\psi_L&=&U^\dagger\Psi_L,\nonumber\\
\psi_R&=&\Psi_R .\label{stue}\end{eqnarray}
The Jacobian of this transformation is  independent of the
physical fields
since (\ref{stue}) is
linear in them. With (\ref{det}) it yields the
(gauge noninvariant) extra term
\begin{equation}
{\cal L}^{\prime\prime}=-4i\delta^4(0)\ln(\sin^6\tilde{\zeta}+
\cos^6\tilde{\zeta})\label{sinsix}\end{equation}
to the Lagrangian.

Applying (\ref{nonlin}) and (\ref{stue}) to the SM Lagrangian
(\ref{linv}), one can easily convince
oneself that the $\zeta_i$ decouple from
the physical fields. The resulting Lagrangian turns out to
be the one obtained by removing all pseudo-Goldstone fields
from (\ref{linv}) and by replacing the fields
($W_{\mu}^i,B_\mu,\Psi,\rho$) by ($w_{\mu}^i,
b_\mu,\psi,\rho$).
Therefore one can integrate out the pseudo-Goldstone fields
in the PI, which yields a constant factor
\begin{equation}
\int{\cal D}\,\zeta_i\exp\left(i\int d^4x\,\tilde{{\cal L}}\right)
\label{fieldenlarge}\end{equation}
where $\tilde{{\cal L}}$ contains the self couplings of the
$\zeta_i$ in
(\ref{twoextra}) and (\ref{sinsix}):
\be \tilde{\lag}=-i\delta^4(0)\left(\ln\left(\frac{\sin^2\tilde\zeta}
{\tilde{\zeta}^2}\right)+4\ln(\sin^6\tilde{\zeta}+\cos^6\tilde{\zeta})\right).
\label{qwe}\ee
The expression \eref{fieldenlarge}
can be removed by multiplying the PI with the compensating
factor.

In summary one can  see that the unphysical fields $\zeta_i$
have decoupled and one
obtains the same result for the quantized U-gauge Lagrangian as
in the previous two
sections; in particular the extra term (\ref{extra}) is again
recovered.

Next I study the behaviour of the new fields under gauge
transformations.
With (\ref{gaugetrafo}), (\ref{zetatrafo}), and (\ref{stue})
one finds that all
physical fields are invariant under the action of SU(2)
and transform under
U(1) as
\begin{eqnarray}
w_\mu&\to&\exp\left(\frac{i}{2}g'\beta(x)\tau_3\right) w_\mu
\exp\left(-\frac{i}{2}g'\beta(x)\tau_3\right)-
\frac{1}{2}\frac{g'}{g}
\partial_\mu \beta(x)\tau_3,\nonumber\\
b_\mu&\to&b_\mu-\partial_\mu\beta(x),\nonumber\\
\rho&\to&\rho,\nonumber\\
\psi_{L,R}&\to&\exp\left(\frac{i}{2}g'(\tau_3+B-L)
\beta(x)\right)\psi_{L,R}.\end{eqnarray}
Introducing, in analogy to \eref{mix},
the mass and charge eigenstates
$w^\pm_\mu$, $z_\mu$ and $a_\mu$
and rescaling the gauge parameter $\beta(x)$ as
\begin{equation} e\kappa(x)\equiv g'\beta(x)\label{rescale}\end{equation}
(with $e$ being the electromagnetic coupling constant $e=g\sin\theta_W$)
one finds
\begin{eqnarray} w_\mu^\pm&\to&\exp(\pm ie\kappa(x))w_\mu^\pm,
\nonumber\\
z_\mu&\to&z_\mu,
\nonumber\\ a_\mu&\to& a_\mu-\partial_\mu \kappa(x),\nonumber\\
\rho&\to&\rho,\nonumber\\ \psi&\to& \exp(ieQ_f\kappa(x))\psi,
\label{emgauge}\end{eqnarray}
where $Q_f$ is the fermion charge matrix
\begin{equation} Q_f=\left(
\begin{array}{cc} q_u&0\\0&q_d\end{array}\right)=\frac{1}{2} (\tau_3+B-L).
\end{equation}
This is just an electromagnetic gauge
transformation\footnote{On the first
look this result seems surprising since I  made a $\rm U(1)_Y$
transformation to derive (\ref{emgauge}). However, due to the field
transformation (\ref{stue}), this acts differently
on the transformed fields
than on the original ones; so it becomes a
$\rm U(1)_{em}$ transformation.}.
Thus, after the fields have been parametrized
such that all pseudo-Goldstone bosons decouple,
the action of
the whole gauge group on the physical fields reduces to
a gauge transformation belonging to the unbroken subgroup.
The remaining
gauge freedom is only connected to the unphysical scalars and
has been ``removed'' by dropping (\ref{fieldenlarge}).
Finally, the $\rm
U(1)_{em}$ gauge freedom has to be fixed by adding the g.f. term
\begin{equation}
{\cal L}_{g.f.}=-\frac{1}{2\xi_\gamma}\partial_\mu a^\mu ,
\label{gfuem}\end{equation}
while the corresponding ghost field decouples.
Thus, one finally finds the same
quantized Lagrangian as in the previous two sections.

This derivation of the U-gauge is similar
to the classical procedure, i.e. the
removal of the pseudo-Goldstone scalars by a
means of a gauge transformation. In fact the
Stueckelberg transformation (\ref{stue}) {\em formally\/}
acts as an SU(2) gauge transformation on the vector
and fermion fields with the gauge {parameters}
$\alpha_i$ being replaced by the
the pseudo-Goldstone {fields} $-\zeta_i/M_W$.
But there remains a principal difference
between a gauge transformation and
a Stueckelberg transformation; namely the
gauge transformation which maps the $\zeta_i$ identically
to zero depends on the
numerical values of these fields at the various space-time
points and thus it is not
the same transformation for different functions $\zeta_i$.
The Stue\discretionary{k-}{k}{ck}elberg
transformation (\ref{stue}), however, decouples the
pseudo-Goldstone fields independently of their functional form
from the physical fields and thus it can be
applied to the PI, where an integration over
the $\zeta_i(x)$ at each space-time point $x$ is carried out. The only
effect of quantum physics on the above
discussion is the need of transforming
the functional integration measure,
too. The corresponding Jacobian
determinant gives rise to the extra term
(\ref{extra}).


\section{The Stueckelberg Formalism}
The procedure of the last section can be reversed:
one can start from the
U-gauge Lagrangian
and construct the corresponding SBGT by subsequent field
transformations. Although it is clear from the previous
section what do to, I
explain this procedure a bit more in detail, since it is a
``derivation'' of an SBGT (e.g.\ of the SM), which illustrates some
features of such a model
quite well. Furthermore, the U-gauge Lagrangian of an SBGT can be
motivated on a rather intuitive basis: it only
involves ``physical'' fields
(i.e.\ all the fields corresponding to observable particles)
and it is the
most general Lagrangian without unphysical fields
which guarantees tree-unitarity (i.e., $N$-particle $S$-matrix
elements
calculated at the tree level decrease at least as $E^{4-N}$ for high
energy $E$)
\cite{clt,jogl,llsm}. Therefore, it seems interesting to
look whether and how
the general (gauge invariant)
structure of an SBGT can be reconstructed from its
U-gauge Lagrangian. Tree
unitarity implies, in particular, the need of
physical scalars (Higgs bosons)
with appropriate couplings to the other particles and to itself
in order to ensure good
high-energy behaviour.  It is clear that the
extra term (\ref{extra}),
which
genuinely reflects quantum effects, is not obtained from tree-level
arguments,
but it is inferred from the requirement of vanishing quartically
divergent $N$-Higgs
self-couplings (which was shown for $N=3,4$ at the one-loop level in
\cite{apqu,jogl}).
In the following I take the term (\ref{extra}) as a given part of
the U-gauge Lagrangian.

Starting from this full
U-gauge Lagrangian (containing the fields $w_{\mu}^i,b_\mu,
\rho$ and $\psi$ as in the previous section)
one first recognizes the local
$\rm U(1)_{em}$ symmetry (\ref{emgauge}).
Applying the FP procedure reversely to the unbroken
subgroup,
one can remove the corresponding
g.~f.\ term (\ref{gfuem}) (and the ghost term if the
unbroken subgroup is non-Abelian). Next, one can see that the
Lagrangian is gauge invariant
also under the larger group $\rm SU(2)\times U(1)$ except
for the mass terms of the vector bosons and the fermions and the
couplings of
the physical scalar to these fields. In order to obtain a gauge invariant
Lagrangian, one can now use the Stueckelberg formalism.
The purpose of the Stueckelberg formalism is
to introduce, by field-enlarging transformations, unphysical
degrees of freedom
with an appropriate behaviour under $\rm SU(2)\times U(1)$
 gauge transformations that
compensate the effect of the transformations of the physical
fields in
the gauge noninvariant terms of the U-gauge
Lagrangian \cite{stue,kugo,clt,sots,dtt}.

In the Lagrangian PI formalism, the
field-enlarging transformation is constructed
as follows: First one introduces the completely decoupled fields
$\zeta_i$ by formally multiplying
the (infinite) constant (\ref{fieldenlarge}) to the PI.
This contains the
functional integration over the unphysical fields $\zeta_i$
(Stueckelberg
scalars) and an exponential function, which is needed
to remove the Jacobian determinant of the subsequent
field transformations.
Then the $\zeta_i$, parametrized
in terms of the unitary matrix $U$
(\ref{u}), are coupled to the physical
fields by the transformations (\ref{stue}).
The resulting Lagrangian is invariant under \suu\ gauge transformations,
\eref{gaugetrafo} and \eref{zetatrafo}, (except for the
extra term (\ref{qwe})),
because the fields in the U-gauge Lagrangian have been replaced
by the combinations
(\ref{stue}) and the effect of an arbitrary gauge transformation
on these is just an electromagnetic one
(\ref{emgauge}) which leaves the U-gauge Lagrangian unchanged.
Since the Stueckelberg transformation
(\ref{stue}) has the form of an SU(2) gauge
transformation\footnote{In
\cite{sots} one can find an alternative formulation of the Stueckelberg
formalism for the case of \suu\ symmetry.
There the Stueckelberg transformation does not look like an SU(2) gauge
transformation,
as in my treatment, but like a full $\rm SU(2)\times U(1)$
transformation with  the gauge parameters being
replaced by unphysical scalar fields; thus {\em four\/}
unphysical fields are introduced.
However, one can show that
there exists a reparametrization of the scalar fields (i.e.\ one more
field transformation) that decouples one of these, so
that one finally
obtains the same result as I do.},  its effect on the kinetic terms, the
vector-boson self-interaction terms and the vector-boson--fermion
interaction terms is just to replace $w_{\mu}^i $
by $ W_{\mu}^i$, etc. (because
these terms are already gauge invariant), while the
mass and Higgs coupling terms give rise to a kinetic
term of the $\zeta_i$
(stemming from the gauge-boson mass-term) and to
nonpolynomial couplings of the
Stueckelberg scalars to the physical particles. Thereby these terms
have become replaced by gauge invariant expressions.

Two important points should be mentioned here.
\begin{itemize}
\item
In the Stueckelberg formalism spontaneous
symmetry breaking (SSB) is implied by
the parametrization of the unphysical
scalar fields in terms of the unitary matrix $U$
\eref{u}. This yields the constraint
\begin{equation} UU^\dagger={\bf 1}
\label{constraint}\end{equation}
in the scalar sector, which implies the nonvanishing VEV
\be\langle U\rangle_0={\bf 1}. \ee
This VEV does not follow from a
scalar self-interaction potential with a minimum for nonvanishing fields
as in the
Higgs mechanism.
\item
In the Stueckelberg formalism the
origin of the (seemingly) bad high-energy
behaviour of the U-gauge Lagrangian is shifted from the
massive-vector-boson propagator, which can now be expressed
within to the $\rm R_\xi$-gauge as (\ref{mgbp}),
to the nonpolynomial interactions of the unphysical scalars.
\end{itemize}

In the original Stueckelberg formalism this is the end of the
story because
field transformations like (\ref{stue}) are not
applied to the U-gauge of a (linearly realized)
SBGT but to a Yang--Mills theory with mass terms added by hand. The
Stueckelberg formalism transforms such a model to a
gauged nonlinear $\sigma$-model, which is
in fact nonrenormalizable \cite{apbe,shiz},
since it is not possible to
reparametrize the unphycal scalars (\ref{u}) in a linear
form in order to avoid
nonpolynomial interactions \cite{burn}.
However, in this case the primordial Lagrangian contains
an extra Higgs field (or several Higgs fields in the case of more
extended theories). This  has been introduced in order to ensure good
high-energy
behaviour at the tree level and, in fact, it makes the model
renormalizable, too,
since {\em physical and unphysical scalar fields together\/}
can be rewritten as a linear
expression by means of the field transformation (\ref{nonlin}).
(The Jacobian determinant of (\ref{nonlin}) removes the extra
term (\ref{extra}).)
Thus, in the linear parametrization (\ref{higgslin}) with (\ref{higgsshift})
the nonpolynomial interactions have been removed.
It was shown in \cite{clt}
that for each tree unitary theory such a linearizing transformation exists,
although this transformation has not explicitly been
constructed there.

Let me briefly discuss the origin of the scalar self-interaction term
\begin{equation}
 -V(\Phi)=-\frac{1}{2}\mu^2{\,\rm tr\,}(\Phi\Phi^\dagger)-
\frac{1}{4}\lambda\,({\rm tr\,}(\Phi\Phi^\dagger))^2
\label{higgsself}\end{equation}
in the SM Lagrangian \eref{linv}
(which implies the nonvanishing VEV and thus the SSB)
within this derivation of the SM.
The self interactions of the physical Higgs field $h$
are already present in the U-gauge Lagrangian, they are necessary to ensure
tree unitarity \cite{llsm}. However, the VEV of $h$ is zero.
The nonvanishing VEV and the unphysical scalar fields become introduced in
the Stueckelberg formalism (i.e.\ by rewriting the theory
in a gauge invariant form which implies SSB through the constraint
\eref{constraint}).
In the linear parametrization of the scalar sector,
the Higgs scalar becomes ``coupled'' to the VEV and to the Stueckelberg
scalars, so that one finally obtains $V(\Phi)$ \eref{higgsself}.
One can see that is this derivation of a Higgs model, the Higgs boson with its
self-interactions and the SSB become introduced at different
stages of the procedure.

So finally one has ``recovered'' the SM from its unitary
gauge by applying
appropriate transformations to the physical fields.
At the level of the classical Lagrangian, this has originally
been done in \cite{clt}. At the quantum level
one has to consider the PI \eref{pi}, so that two new features arise:
\begin{itemize}
\item Field enlarging is easily achieved by multiplying
an infinite constant
(as (\ref{fieldenlarge})) to the PI.
\item The (quantized) U-gauge Lagrangian has to contain the extra term
(\ref{extra}). This cancels against the Jacobian
determinant which arises as a
consequence of the transformation
of the functional integration measure.
\end{itemize}
As the result of this section one can see that a Higgs model
can be derived from
the requirement of good high-energy behaviour of tree-level
amplitudes and of loops without explicitly using the
Higgs mechanism (although
SSB is implicitly included in this derivation). Thus, the
physical sector of an
SBGT ``contains'' the entire model. Non-Abelian
Stueckelberg models are
renormalizable if, in addition to the unphysical Stueckelberg
scalars, also physical
scalars with appropriate couplings to the other
particles and themselves are present.


\section{The Gauged Nonlinear $\bf \sigma$-Model}
As mentioned before, the original
purpose of the Stueckelberg formalism was to
construct gauge theories with massive gauge bosons but without physical scalars
\cite{stue,kugo,sots,dtt,burn}.
To construct a (non-Abelian) Stueckelberg model one
starts from a Yang-Mills theory, adds a mass term by hand, thereby
breaking gauge invariance explicitly, and then makes a
field-enlarging
Stueckelberg transformation as (\ref{stue}) in order
to restore gauge invariance.
As pointed out in the previous sections, the Stueckelberg
transformation only
affects the mass term in a nontrivial way thereby
yielding, besides a
kinetic term for the Stueckelberg scalars, nonpolynomial
interactions of the unphysical fields with the physical
fields. In the Yang--Mills
part of the Lagrangian the
original physical fields are simply replaced by the
transformed fields.
Thus, the addition of unphysical scalar fields with suitable
nonpolynomial couplings
to the physical fields (in the non-Abelian case)
embeds the original massive
Yang--Mills theory in an
equivalent gauge theory. The resulting model
is a gauged nonlinear $\sigma$-model. As it is well known,
such a model can be constructed in three alternative ways
(explained here for the case of \suu\ symmetry):
\begin{itemize}
\item By applying a Stueckelberg transformation (\ref{stue})
to a massive Yang--Mills theory
\cite{kugo} as explained above.
\item By gauging a nonlinear $\sigma$-model \cite{bash}, i.e. a model
which contains scalar fields parametrized in terms of a
unitary matrix $U$ \eref{u} and is symmetric with respect to {\em global\/}
transformation such as \eref{zetatrafo}. This global symmetry is spontaneously
broken due to the constraint \eref{constraint}. The gauged nonlinear
$\sigma$-model is obtained by coupling $U$ minimally to gauge fields so that
the global symmetry becomes a {\em local\/} one.
\item By formally taking the limit $M_H\to\infty$
of a Higgs model \cite{apbe}.
In this limit the linearly realized scalar fields $\Phi$
become substituted by
\begin{equation} \Phi\to \frac{v}{\sqrt{2}}U\label{sigmatrafo}
\end{equation}
(with $\Phi$ and $U$ given by (\ref{higgslin}).
(\ref{higgsshift}) and (\ref{u})).
This automatically removes the scalar self-interaction term \eref{higgsself}
and implies the constraint (\ref{constraint}).
\end{itemize}
Although the gauged nonlinear $\sigma$-model
is nonrenormalizable due to the nonpolynomial interactions, the
gauge freedom enables perturbative calculations in the
$\rm R_\xi$-gauge and so one
finds that the loops do not diverge as severely as one would
expect from naive
power counting. In fact, the one-loop divergences of the
gauged nonlinear
$\sigma$-model are only logarithmically cut-off dependent
\cite{apbe}.


\section{Effective Lagrangians}
During the last years effective electroweak Lagrangians
describing non--Yang--Mills self-interactions of massive vector bosons
have been studied in order
to parametrize possible deviations of the electroweak interaction
from the SM (e.g.\ \cite{nongauge}).
These are also gauge noninvariant,
because the mass terms
{\em and the nonstandard interaction terms\/}
violate gauge invariance. However,
repeating the reasoning of the previous two sections, one can easily
see, that {\em each\/} effective Lagrangian with arbitrary
vector-boson self-interactions (which is only invariant under
$\rm U(1)_{em}$ transformations) can be rewritten as an SBGT
with nonlinearly realized symmetry by applying the
Stueckelberg transformation (\ref{stue}).
The original effective Lagrangian is
the U-gauge Lagrangian of this SBGT.
Again, the behaviour of the gauge noninvariant
terms under gauge transformations is compensated by
the addition of appropriate couplings to
unphysical scalars but, in contrary to the case of a simple
massive Yang--Mills theory, not only the mass terms
but also the nonstandard interaction terms
give rise to new nonpolynomial couplings.
The resulting model is a generalized gauged nonlinear $\sigma$-model
(also called chiral Lagrangian)
with additional nonstandard (but gauge invariant) effective interaction terms.
Because of the gauge freedom, one can do loop
calculations within such a (nonrenormalizable)
model in the $\rm R_\xi$-gauge and thus has better opportunities
to subdue the divergences. Furthermore one can
avoid the ambiguities  \cite{fls,jawe} in calculations on the basis of
the original effective Lagrangian.
This result has already
been obtained in \cite{bulo}\footnote{For a more general and formal
treatment of the relations between
gauge noninvariant Lagrangians and SBGTs see \cite{alda}.},
however without identifying the
corresponding transformations as Stueckelberg transformtions.

A method of deriving nonstandard self-couplings of electroweak
vector bosons from a gauge
invariant Lagrangian with {\em linearly} realized symmetry
(Higgs model)
is to add to the the SM Lagrangian extra
$\rm SU(2)\times U(1)$ invariant interaction terms, which contain the
nonstandard couplings \cite{gauge}.
Although there are no nonpolynomial
interactions, these models are nonrenormalizable, too,
because the additional interaction terms have a (mass) dimension
higher than four (while in the renormalizable SM all terms have dimension
four). However, it has
recently been shown that, within models with extra dimension-six terms,
one-loop diagrams depend (after renormalization)
only logarithmically on the cut-off due to the linearly
realized gauge invariance
\cite{ruj,haze}. In fact, most of the additional terms contain extra
interactions of the Higgs boson to the gauge bosons and the
Higgs boson
contribution cancels the quadratic loop divergences \cite{haze}.
Thus the models discussed above,
which are obtained by applying a Stueckelberg transformation to an
effective Lagrangian and therefore contain no physical Higgs field,
will yield more severely divergent loop corrections.

It should be noted that the two applied methods to
obtain nonstandard interactions from a gauge invariant model are in
principle different. In \cite{gauge} terms
are added to the SM Lagrangian, most of them contain not only
vector-boson
self-interaction
terms but also couplings to the physical Higgs boson.
Thus the original
effective Lagrangian is {\em extended\/} to a gauge invariant model
which is not equivalent to it, since there
are additional Higgs
couplings.
In contrary, the generalized Stueckelberg formalism,
i.e.\ the introduction of nonlinearly realized symmetry enables one to
express an effective Lagrangian in terms of an {\em equivalent\/}%
\footnote{The proof that Lagrangians which are related to each other by
a Stueckelberg transformation are physically
equivalent will be given in the sections
4.2 and 5.3.2.} gauge theory. On the other hand,
from the above discussion it is clear how these two methods are
connected. Given an effective electroweak
theory with arbitrary vector-boson self-interactions,
one first makes the Stueckelberg
transformation (\ref{stue}) and finds the equivalent generalized
gauged
nonlinear $\sigma$-model. This can, like the usual gauged nonlinear
$\sigma$-model, be understood as the limit $M_H\to\infty$ of a
Higgs model
with linearly realized symmetry \cite{apbe}. To recover this Higgs
model one has to substitute\footnote{It should
be clear that (\ref{recov}) cannot be understood
as a field-enlarging transformation like (\ref{stue}), since the matrix
$\Phi$ (\ref{higgslin}) cannot be
expressed  as a unitary matrix $U$ (\ref{u}). This shows that the step
from the generalized gauged nonlinear $\sigma$-model
to the generalized Higgs model is indeed an {\em extension} of the theory.},
reversing (\ref{sigmatrafo}),
\begin{equation} U\to\frac{\sqrt{2}}{v}\Phi\label{recov}
\end{equation}
(with (\ref{higgslin}), (\ref{higgsshift}))
and to add the Higgs self-interaction potential (\ref{higgsself}),
which implies a nonvanishing VEV of $\Phi$ (and replaces the constraint
(\ref{constraint}) in the gauged nonlinear $\sigma$-model).
As in section~2.5 the addition of a physical Higgs
boson enables a linear parametrization of the scalar sector and
removes the nonpolynomial interactions.
Applying this simple formalism one can embed {\em each}
effective electroweak theory with arbitrary vector-boson self-interactions
in an
$\rm SU(2)\times U(1)$ gauge invariant theory with linearly realized symmetry
which is expected
to yield a more decent loop behaviour.
The SBGT which contains the most general trilinear vector-boson
self-interactions within a gauge invariant framework is
given in \cite{gore}, however,  it has
not explicitly been constructed there by applying this procedure.
Here, I have given the general formalism to
understand and to carry out this extension of an
effective Lagrangian to a linear Higgs model for all types of vector-boson
self-interactions\footnote{The investigation of this section
concerning arbitrary vector-boson self-interactions can also be
applied to obtain arbitrary fermionic interactions from a
(linearly or
nonlinearly realized) SBGT. However, I do not
stress this point here because this is of less phenomenological
interest since the SM fermionic interactions are very well
confirmed in experiments now.}.

\chapter{Constrained Hamiltonian Systems}
It is well known that quantization (both
canonical quantization and path integral
quantization) has to be based on the {\em Hamiltonian\/}
formalism (and not on the Lagrangian
formalism) \cite{fad,bedu,sen,const,gity,dirac} because the Hamiltonian is
the generator of the time evolution of a physical system \cite{cm}.
Besides, when deriving the path integral
formalism from the canonical formalism
one finds a Hamiltonian path integral \cite{books}.

Thus, to justify the use of
the convenient Lagrangian PI \eref{lpi} with \eref{eq},
one has to derive it within the Hamiltonian PI formalism. This will be
the subject of the following chapters. This chapter contains a review of
Hamiltonian dynamics and PI quantization, on which the subsequent
investigations will be based.

For the physically most interesting field theories, namely those which contain
vector or fermion fields, the Hamiltonian formalism becomes more involved
than for simple mechanical systems
because these theories are singular, i.e. they are subject to
{\em constraints\/}. Constraints are well known from classical mechanics
\cite{cm}; there they usually are realized by wires or surfaces which restrict
the motion of the particles. In the most important cases
these external circumstances result in holonomic constraints, i.e. in
constraints of the type
\be
\phi_a(q_i)=0.
\label{hol}\ee
In the usual mechanical formalism, one uses these relations
in order to reduce the number of coordinates to the number of physical
degrees of freedom; then the remaining coordinates
are independent of each other and in the subsequent Lagrangian and
Hamiltonian procedure one does not have to care for the constraints
anymore \cite{cm}.
In field theory\footnote{In fact, the formalism
explained in this chapter can be developed within a mechanical
framework; however its physically most
important applications are in field theory.}
the concept of constraints becomes more abstract, because there
the constraints are not induced by
wires and surfaces. Instead, one often introduces unphysical
degrees of freedom in order to obtain a manifestly Lorentz or gauge invariant
formulation of a theory \cite{books}.
For example, a spinor (fermion) field contains
eight real components at each space-time point, although it has
only four physical degrees of freedom; a real vector field has four components
but only two (in the massless case) or three (in the massive case) physical
degrees of freedom; furthermore SBGTs contain unphysical scalar fields.
In fact, the
presence of these unphysical degrees of freedom leads to constraints;
in most cases, however,
these are not holonomic constraints of the type
\eref{hol}, but they have the more general form
\be
\phi_a(q_i,p_i)=0.
\label{hamconst}\ee
This form of the constraints, which also involves the momenta,
strongly suggests a Hamiltonian treatment.
Besides, even in this case it is possible to eliminate the constraints by
reducing the number of variables to the number of
physical degrees of freedom (as it is usually done
when dealing with holonomic constraints \eref{hol}),
however in general such a procedure
is not desired because it would make it more difficult to find a
manifestly Lorentz or gauge invariant formulation of the theory.
Instead, one considers the constraints
within the Hamiltonian formalism instead
of eliminating them from the beginning.

Another feature not known from classical mechanics arises: in some
cases (but not
in all) constraints are connected with a gauge freedom. For example,
a massive vector field involves constraints, but it has
no gauge freedom (except if
it is embedded in an SBGT), while a massless vector field implies constraints
and a gauge freedom. Actually, these different features are also
reflected within the Hamiltonian formalism
because there are two types of constraints, namely:
first-class constraints, which are connected with a gauge freedom,
and second-class constraints, which are not. Actually, the
first-class constraints are the generators of the gauge transformations.
The gauge freedom can be removed from a
first-class constrained system by introducing
gauge-fixing conditions; constraints and g.f. conditons together form a
set of second-class constraints and the resulting second-class constrained
system is physically equivalent to the original first-class constrained system.

Another classification stems from the distinction between
primary, secondary, etc.\ constraints. It
refers to the stage of the formalism at which these constraints appear.
The requirement that the primary constraints have to be consistent
with the EOM yields the secondary constraints, and so on.

The Hamiltonian PI for a constrained system can easily
be derived because one can find a canonical transformation, such that
one part of the resulting pairs of canonical variables are completely
unconstrained, while the remaining ones are completely fixed
by the constraints and the g.f.\ conditions. The
Hamiltonian PI is then essentially given by the PI for the unconstrained
system corresponding to the free variables. Then, by doing the above
canonical transformation inversly, one finds the general expression for
the Hamiltonian PI.

In this chapter I will first explain the classical dynamics and then the
PI quantization of constrained Hamiltonian
systems. I will restrict myself to the case of a finite number of
degrees of freedom;
the generaliztion to field theory works as usual.
This chapter is a pure review; the reader who is interested in more details
is referred to the original literature \cite{fad,sen,dirac} or to textbooks
on this subject\footnote{In particular, I will omit all involved
proofs here because they can be found in the literature.}
\cite{const,gity}. The basic concepts of the
Hamiltonian formalism can be found in textbooks on classical mechanics
(e.g. \cite{cm}).


\section{Dynamics of Constrained Systems}
Consider a physical system given by the Lagrangian $L$ which is a function
of the coordinates $q_i$ ($i=1,\ldots,I$) and their first time derivatives.
(Lagrangians with higher time derivatives will be considered in section~5.1.)
This Lagrangian is {\em singular\/} if
\be
\det\left(\frac{\pa^2 L}{\pa\dot{q}_i\pa\dot{q}_j}\right)=0.
\ee
In this case not all of the equations that define the momenta $p_i$,
\be
p_i=\frac{\pa L}{\pa \dot{q}_i},
\label{pqd}\ee
can be solved for the velocities $\dot{q}_i$. Instead, some of these
relations yield the {\em primary constraints\/}
\be
\phi^{(1)}_a(q_i,p_i)=0.
\label{pcf}\ee
Now one can construct the Hamiltonian\footnote{Note that $H$ is not unique
because any expression proportional the constraints can be added to $H$.}
\be
H=\dot{q}_ip_i-L
\label{hamf}
\ee
where the $\dot{q}_i$ have to be expressed in terms of the $q_i$ and $p_i$
by applying \eref{pqd}.
Although \eref{pqd} cannot be solved for all
$\dot{q}_i$, one can easily show that, due to
the presence of the constraints \eref{pcf},
nevertheless all the $\dot{q}_i$ can be eliminated
from $H$, i.e. $H$ only depends on $q_i$ and $p_i$.
(See the analogous argument for Lagrangians with higher derivatives
in section~5.1.)

As in the unconstrained case,
the Hamiltonian equations of motion follow from Hamilton's priciple
\be
\delta\int[\dot{q}_ip_i-H]\,dt=0,
\ee
however in this case,
the variations $\delta q_i$ and $\delta p_i$ are not independent
of each other but they are restricted by the constraints \eref{pcf}.
This yields the EOM
\be
\dot{f}=\{f,H^{(1)}\}\big\pcs,\label{eomp}
\ee
($f$ stands for $q_i$, $p_i$ or
any other function of these variables) with
\be
H^{(1)}\equiv H+\lambda_a\phi_a^{(1)},
\label{hprim}\ee
where the $\lambda_a$ are (a priori undetermined) Lagrange multipliers.

The primary constraints have to be consistent with the EOM, i.e., the
time derivative of \eref{pcf} also has to vanish:
\be
\dot{\phi}_a^{(1)}=\{\phi^{(1)}_a,H^{(1)}\}\big\pcs=\{\phi^{(1)}_a,H\}\big\pcs
+\lambda_b\{\phi_a^{(1)},\phi_b^{(1)}\}\big\pcs=0.\label{eomconsp}
\ee
For a (physically meaningful) theory, the equations \eref{eomconsp}
can take three possible forms, viz.
\begin{itemize}
\item They are fulfilled automatically;
\item They serve to determine the Lagrange multipliers $\lambda_a$;
\item They can be written in the form
\be
\phi^{(2)}_a(q_i,p_i)=0.\label{scf}
\ee
In this case, the constraints \eref{scf}
also have to be satisfied in order to ensure consistency
with the EOM. They are called {\em secondary constraints\/}.
\end{itemize}
This procedure has to be iterated, i.e., the demand that the
time derivatives of the secondary constraints
have to vanish may imply tertiary constraints, and so on,
until one finally obtains a set of constraints which is consistent with the
EOM. Although the various constraints are obtained at different stages of
the procedure, there is no principle difference between them; actually they
can be treated on the same level. It has been shown in \cite{gity} that
one obtains an equivalent physical formulation of a constrained theory
if one rewrites \eref{eomp}
and \eref{hprim} as
\be
\dot{f}=\{f,H_T\}\Big\cs,\label{eom}
\ee
with the total Hamiltonian
\be
H_T\equiv H+\lambda_a\phi_a,
\label{htot}\ee
where $\phi_a$ denotes {\em all\/} constraints.

Next one wants to determine the Lagrange multipliers in \eref{htot}. Actually,
if the matrix
\be
\{\phi_a,\phi_b\}\Big\cs\label{pbc}
\ee
in nonsingular, the constraints are called {\em second-class\/}; in this case
the relations (analogous to \eref{eomconsp})
\be
\dot{\phi}_a=\{\phi_a,H_T\}\Big\cs=\{\phi_a,H\}\Big\cs
+\lambda_b\{\phi_a,\phi_b\}\Big\cs=0.\label{eomcons}
\ee
can be solved for the $\lambda_a$:
\be
\lambda_a=-\{\phi_a,\phi_b\}^{-1}\{\phi_b,H\}\Big\cs.
\label{lambda}
\ee
Inserting this into \eref{eom} with \eref{htot}, the EOM can be written
in the simple form
\be
\dot{f}=\{f,H\}_{DB},\label{eomdb}
\ee
where the {\em Dirac bracket\/} \cite{dirac} $\{\, ,\}_{DB}$ is defined as
\be
\{f,g\}_{DB}\equiv\{f,g\}\Big\cs-\{f,\phi_a\}\{\phi_a,\phi_b\}^{-1}
\{\phi_b,g\}\Big\cs.\label{db}
\ee
By replacing the Poisson brackets by Dirac brackets
the dynamics of a second-class constrained system can now be formulated
analogously to the dynamics of an unconstrained system.

A different situation arises if the matrix \eref{pbc} is singular.
Assuming that \eref{pbc} has the rank $R$, one orders the constraints such
that the upper left $R\times R$ submatrix of \eref{pbc} has a nonvanishing
determinant.
Then only the first $R$ constraints are second-class and the remaining ones
are {\em first-class\/}. The Lagrange multipliers corresponding to
the first-class constraints cannot be determined from
\eref{eomcons}\footnote{In fact, they cannot be determined from the complete
EOM, \eref{eom} with \eref{htot}, either \cite{gity}.}. Thus the equations
of motion \eref{eom} with \eref{htot} contain undetermined Lagrange
multipliers and therefore their solution (for given initial
conditions) is  not unique. \eref{eom} with \eref{htot} imply that
two solutions $f$ and $f'$ of the EOM with the same initial
condition at $t=0$ (but with distinct choices of the
Lagrange multipliers corresponding to the first-class constraints)
differ after an infinitesimal time interval $dt$ by
\be
\Delta f(dt)=dt(\lambda_a-\lambda^\prime_a)\{f,\phi_{1st}^a\}
\label{igt}
\ee
(where $\phi^a_{1st}$ denotes only the first-class constraints).
A transformation of the canonical variables,
which relates different solutions of the EOM is called
a {\em gauge transformation\/}. Equation \eref{igt} implies that the
first-class  constraints are the generators of (infinitesimal) gauge
transformations\footnote{In distinction from Dirac's original statement,
not only
the primary first-class constraints but all first-class constraints generate
gauge transformations because $H^{(1)}$ \eref{hprim}
in the EOM can be replaced by
$H_T$ \eref{htot} \cite{gity}.}. Thus, a first-class constrained theory has
a gauge freedom; it is called {\em degenerate}.
All solutions of the EOM with the same initial conditions describe the
same physical process; in other words,
all points in the phase space, which are related by gauge transformations
describe the same physical state of the system.

Given a degenerate theory, one can find a physically
equivalent nondegenerate theory. In that theory,
the unique solution of the EOM for given initial conditions turns out to be
one of the various solutions of the EOM in the degenerate theory. The
corresponding nondegenerate theory is constructed
by imposing a {\em gauge\/} on the original theory, i.e.\
by introducing additional {\em gauge-fixing
conditions\/}
\be\chi_a(q_i,p_i)=0\label{gff}\ee
such that the number of g.f. conditions is equal to the
number of first-class constraints and that the g.f. condtions and the
constraints together form a set of second-class constraints which
is consistent with the EOM\footnote{Faddeev also required that
$\{\chi_a,\chi_b\}=0$ \cite{fad}.
This condition is unnecessary \cite{const,gity,kadi}. Actually,
later in this thesis I will
use g.f.\ conditions that do not satisfy this relation.}.
In fact, the relations analogous to \eref{eomcons} with $\phi_a$ replaced by
$\chi_a$ determine the Lagrange multipliers corresponding to the first-class
constraints and the ambiguity in the solution of the EOM is removed; thus
the gauged first-class constrained system is a physically equivalent
second-class constrained system.
A convenient way to construct g.f.\ conditions is given in \cite{gity}:
One starts with primary g.f.\ conditions $\chi_a^{(1)}$ and constructs
secondary g.f. conditions $\chi_a^{(2)}$ by demanding
\be
\{\chi_a^{(1)},H\}=0\label{gfcons}
\ee
which ensures the consistency with the EOM.
Finally, if one considers a gauged first-class constrained system,
$H_T$ \eref{htot}
in the EOM \eref{eom} can be replaced by
\be
H_T\equiv H+\lambda_a\phi_a+\tilde{\lambda}_a\chi_a.
\label{htotgf}\ee
{}From \eref{htotgf}
it is obvious that the constraints and the g.f.\ conditions can be applied
in order to convert the Hamiltonian and each other because this corresponds
to a redefinition of the Lagrange multipliers.

The standard examples of a first- and a
second-class constrained system are the
massless and the massive vector field, which will
be discussed later in this thesis (within
the treatment of effective Lagrangians).
It turns out that a massive vector field is
subject to two second-class constraints. This means that among
the four field components and the four conjugate fields there are only six
independent degrees of freedom (three fields and three generalized momenta).
In the Hamiltonian treatment of a massless vector field, two first-class
constraints arise and therefore two gauge-fixing conditions have to be
introduced. Thus there are only two physical field components and two physical
momenta. Besides, the first-class constraints are related to the
gauge freedom of a massless vector field.

All the features discussed above become very obvious if one applies a theorem
derived in \cite{gity,gtp}, which states that one can find canonical
variables (written in pairs as
$(\bar{q}_i,\bar{p}_i), \, (Q_i,P_i),\,
(\Q_i,\P_i)$) for an arbitrary constrained system
such that the second-class constraints are
\be
Q_i=0,\qquad P_i=0\label{scccc}
\ee
and the first-class constraints are
\be
\P_i=0.\label{fcccc}
\ee
It should be noted that the single constraints $\phi_a$ do not
correspond to the
$Q_i$, $P_i$ and $\P_i$ (because, in general, the Poisson brackets among them
are not canonical ones) but all the constraints $\phi_a$
together can be expressed
as \eref{scccc} and \eref{fcccc}.
With this choice of canonical variables
the fundamental Poisson brackets immediately imply
that the Lagrange multipliers corresponding to the second-class constraints
\eref{scccc} can be determined from \eref{eomcons} and that those
corresponding to the first-class constraints are undetermined.
Thus the solutions of the EOM \eref{eom} with \eref{htot} for the
$\bar{q}_i$ and $\bar{p}_i$
are unique, while the $\Q_i$ are completely arbitrary.
The gauge transformations are
shifts of the $\Q_i$ which are generated
by the first-class constraints $\P_i$, while the second-class
constraints do not
give rise to gauge freedom. In order to construct a gauge of this theory,
one has to fix the $\Q_i$, i.e.,
one has to impose g.f.\ conditions
\be
\Q_i-f_i=0,\label{gfccc}
\ee
where the $f_i$ are functions of the remaining variables.
Equations \eref{scccc}, \eref{fcccc} and \eref{gfccc} imply that the
physical sector of this constrained theory is parametrized in terms of
the canonical variables $\bar{q}_i$ and $\bar{p}_i$ alone. Thus, in a reduced
phase space, which consists only of these variables, one can treat
this physical system as an unconstrained one. However,
as mentioned above, in general one
does not use this choice of unconstrained parameters because in the
primordial constrained parametrization it is easier to find a manifestly
Lorentz or gauge invariant formulation of the theory. Besides, it is not always
possible to find the explicit form of the abovementioned canonical
transformation \cite{gity}.


\section{The Hamiltonian Path Integral for Constrained Systems}
Now one can quantize a constrained physical system within the Hamiltonian
PI formalism\footnote{The derivation of the Hamiltonian PI that is briefly
reviewed in this section is the one given in \cite{gity}, which
is different from the
original derivation in \cite{fad,sen}.}.
To derive the Hamiltonian PI it is sufficient to consider first only
second-class constrained systems.
As explained at the end of the preceding section, such a system
is equivalent to an unconstrained one with a reduced phase space
consisting of the variables
$\bar{q}_i$  and $\bar{p}_i$. The Hamiltonian PI for this unconstrained
system has the well-known simple form \cite{books}
\be
Z=\int\Df\bar{q}_i\Df\bar{p}_i\,\exp\left\{
i\int dt\,\Big[\dot{\bq}_i\bp_i-H|_{\scriptstyle Q_i=P_i=0}\Big]
\right\}
\ee
In the enlarged phase space that also contains the variables
$Q_i$ and $P_i$ which are zero due to the second-class constraints
\eref{scccc}, this can
be written as
\be
Z=\int\Df\bar{q}_i\Df\bar{p}_i\Df Q_i\Df P_i\,\exp\left\{
i\int dt\,\left[\dot{\bq}_i\bp_i+\dot{Q}_iP_i-H\right]
\right\}\delta(Q_i)\delta(P_i).\label{zw1}
\ee
Now one can pass to the original second-class constraints
$\phi_a$ which are independent functions of the $Q_i$ and $P_i$.
This affects the $\delta$-functions in \eref{zw1}. One finds
\be
\delta(Q_i)\delta(P_i)=\delta(\phi_a) \mbox{Det}^{\frac{1}{2}}
(\{\phi_a,\phi_b\}\delta(x^0-y^0)).\label{deter}
\ee
The $\delta$-function in the argument of the determinant is missing in
\cite{fad,sen,const,gity}. However it necessarily has to be present,
because the factor $\mbox{Det}^{\frac{1}{2}}\{\phi_a,\phi_b\}$ (where $\phi_a$
and $\phi_b$ are taken at equal times)
has to be introduced for {\em all\/}
times and $\mbox{Det}^{\frac{1}{2}}(\{\phi_a,\phi_b\}\delta(x^0-y^0))
$ is the ``product'' of this expression over all times.

Inserting \eref{deter}
into \eref{zw1} one obtains an expression which is invariant
under canonical transformations. Thus, one can go back from the
canonical variables $(\bar{q}_i,\bar{p}_i)$, $(Q_i,P_i)$
to the primordial ones, $(q_i,p_i)$, and finds
\be
Z=\int\Df q_i\Df p_i\,\exp\left\{i\int dt\,\left[\dot{q}_ip_i
-H\right]\right\}\delta(\phi_a)\,\mbox{Det}^{\frac{1}{2}}\,(
\{\phi_a,\phi_b\}\delta(x^0-y^0))
.\label{zw2}
\ee
This is the general form of the Hamiltonian PI for a second-class
constrained system. This result can easily be extended to the case of
a system with first- and second-class constraints because that becomes
a second-class constrained system after the introduction of g.f. conditions.
Thus, \eref{zw2} can be applied to this case; then
the $\phi_a$ denote the
second-class constraints $\phi_{2nd}^a$, the first-class constraints
$\phi_{1st}^a$ and the g.f.\ conditions $\chi^a$.
One finally finds\footnote{Remember
that I will add the source terms in the generating functional
after all manipulations will
have been done, as pointed out in the introduction.}
\cite{fad,sen,const,gity}:
\bea
Z&=&\int\Df q_i\Df p_i\,\exp\left\{i\int dt\,\left[\dot{q}_ip_i
-H\right]\right\}\delta(\phi_{2nd}^a)\delta(\phi_{1st}^a)
\delta(\chi^a)\nn&&\times\mbox{Det}^{\frac{1}{2}}\,(
\{\phi_{2nd}^a,\phi_{2nd}^b\}\delta(x^0-y^0))\,\mbox{Det}\,
(\{\phi_{1st}^a,\chi^b\}\delta(x^0-y^0)).\label{hpi}
\eea
This is the Hamiltonian path integral for an arbitrary constrained system.
It has the following properties \cite{fad,sen,const,gity}:
\begin{itemize}
\item It is invariant under canonical transformations;
\item It is invariant with respect to the choice of an equivalent set of
constraints;
\item It is independent of the choice of the gauge-fixing conditions.
\end{itemize}
The investigations in the subsequent chapters will be based on this Hamiltonian
path integral (strictly speaking on its field theoretical generalization).

\chapter{Effective Lagrangians with Massive Vector Fields}
Now I come to the main point of this thesis, namely the proof of
the HLE theorem,
i.e. the derivation of the simple Lagrangian path integral \eref{lpi}
with \eref{eq} or \eref{fp}
from the correct but more involved Hamiltonian path integral \eref{hpi},
for effective interactions of scalars, fermions, massless and massive vector
bosons. This will be the subject of this and the following two chapters.

In this chapter I will restrict myself to effective (non--Yang--Mills)
self interactions of massive vector
fields, which is the most interesting case, first because it is
phenomenologically very important
with respect to the investigation of effective electroweak
Lagrangians \cite{nongauge,nonlin,gauge} and second because it is interesting
for theoretical reasons since massive vector fields can be embedded in
three types of theories  (as discussed in chapter~2), namely in
gauge noninvariant models \cite{nongauge}, in gauged nonlinear $\sigma$-models
\cite{nonlin}
(i.e., SBGTs with a nonlinearly realized scalar sector) and in Higgs models
\cite{gauge} (i.e., SBGTs with a linearly realized scalar sector).
I will first prove the HLE theorem for gauge noninvariant models and then
extend it to SBGTs by applying the Stueckelberg formalism
introduced in the chapter~2, which relates gauge noninvariant
Lagrangians to SBGTs. To this end I will
reformulate the Stueckelberg formalism within the Hamiltonian formalism
as a transition from a second-class constrained theory to a first-class
constrained theory, which is realized as follows:
One first enlarges the phase space by introducing new
(unphysical) variables; in addition one has extra constraints
so that the number of physical degrees of freedom remains unchanged.
Next, one uses these additional constraints in order to
convert the Hamiltonian and the primordial constraints.
Finally,
one half of the second-class constraints are considered to be first-class
constraints and the other half to be gauge-fixing conditions.
This procedure implies the equivalence of Lagrangians which
are related to each other by a Stueckelberg transformation.

The proof of the HLE theorem for (nonlinearly realized)
SBGTs goes then as follows:
Using the Stueckelberg formalism I will show
that the generating functional corresponding to an SBGT can be
written as a Lagrangian PI with the quantized Lagrangian being
identical to the U-gauge Lagrangian of this SBGT
(i.e., the Lagrangian which is
obtained by removing all unphysical scalar fields from the gauge
invariant one). In chapter~2 I have shown that
this generating functional is the
result of the FP procedure if the g.f.\
conditions that all unphysical scalar fields become equal to zero
are imposed. Then I will use the
equivalence of all gauges \cite{lezj,able} in order
to generalize this result to any other gauge.

Finally, I will prove the HLE theorem
for Higgs models, i.e.\ for SBGTs
with linearly realized
scalar fields. Because each Higgs model is related to a
nonlinear Stueckelberg model by a simple point transformation
such as \eref{nonlin}
that delinearizes the scalar sector (as discussed in chapter~2),
and since within the Hamiltonian formalism
a point transformation becomes a canonical transformation which leaves the
Hamiltonian PI invariant, the result for
nonlinearly realized SBGTs can easily be
extended to the case of linearly realized SBGTs. As in the nonlinear case,
the HLE theorem will first be derived for the special case of
the U-gauge and then be generalized to any other gauge.

As mentioned in the introduction, Hamiltonian PI quantization in general
yields additional $\delta^4(0)$-terms to the quantized Lagrangian. It is known
that in renormalizable theories (i.e.\ theories without effective interaction
terms) these $\delta^4(0)$-terms cancel against quartic divergent loops
\cite{apqu,jogl}; however, it is an open question how to interpret divergences
higher than logarithmic within an effective (nonrenormalizable)
field theory \cite{bl}. In order to prove the HLE theorem I will,
according to \cite{bedu}, neglect these terms because they become zero in
dimensional regularization \cite{bedu,leib}.
However, to give an example of such a
$\delta^4(0)$-term, I will derive the quartically divergent Higgs
self-interaction term in the quantized
U-gauge Lagrangian of a minimal Higgs model by applying
the Hamiltonian PI formalism. This term has already been
obtained within the Lagrangian PI formalism in chapter~2.

Furthermore, I will expand all quantities in powers of the effective coupling
constant\footnote{A different approach to the Hamiltonian
PI quantization of effective
Lagrangians has recently been made in \cite{moul}. There no expansion
in powers of the effective coupling constant is done, but instead an
expansion in powers of the momenta.}
$\ep$ and neglect $O(\ep^2)$-terms because I assume
$\ep \ll 1$. The restriction to the first order of $\ep$ is only a
technical simplification;
in their proof of the HLE theorem for effective interactions of
scalar fields Bernard and Duncan considered also $O(\ep^2)$-terms
and neglected $O(\ep^3)$-terms \cite{bedu}. They supposed that the HLE
theorem can in principle be proven in any finite order $N$ of $\ep$,
but that with increasing $N$ the proof
becomes technically more and more involved. However this proof of the HLE
theorem cannot be generalized
to {\em all\/} orders of $\ep$ because then it is
impossible to find a closed expression for the velocities and the
Hamiltonian in terms of the momenta. Thus, this proof of the
HLE theorem can only be applied if the effective interactions are treated
perturbatively.

The investigations of this chapter will be restricted to the case of
effective Lagrangians which do not depend on higher
time derivatives of the fields and which depend on
first time derivatives of the vector fields only
through the non-Abelian field strength tensor. (The latter requirement ensures
that the SBGTs corresponding
to such effective Lagrangians also do not involve higher
time derivatives.) The case of effective Lagrangians with
higher derivatives will be treated in the next chapter.

For the clearness of the representation,
I will only consider massive Yang--Mills
fields (of course with extra non--Yang--Mills interactions)
in which all vector bosons have equal masses together with
the corresponding SBGTs.
The results can easily be generalized to any other effective Lagrangian with
massive vector fields, e.g., to electroweak models. In these cases the
derivation becomes formally more complicated but the physically important
features remain the same.

The results of this chapter have first been published in \cite{gk1}.


\section[The HLE Theorem for Gauge Noninvariant Models]
{The Hamilton--Lagrange Equivalence Theorem for Gauge Noninvariant Models}
In this section I will quantize a massive (and thus gauge noninvariant)
Yang--Mills theory with
additional arbitrary non--Yang--Mills interactions \cite{nongauge}
(which are proportional to a parameter $\epsilon$
with $\epsilon\ll 1$) in the Hamiltonian PI formalism in order to
derive the simple Lagrangian form (\ref{lpi})
with (\ref{eq}) of the generating functional.
Some of the techniques used in the following derivation originate from
the quantization of a massive Yang--Mills theory without effective
interaction terms in \cite{sen} and some others from the quantization
of effective interactions of scalar fields in \cite{bedu}.

The effective Lagrangian has the form
\begin{equation}{\cal L}_{}= {\cal L}_{0}+\epsilon{\cal L}_{I}=
-\frac{1}{4}F^{\mu\nu}_aF_{\mu\nu}^a+\frac{1}{2}M^2A^\mu_aA_\mu^a
+\epsilon{\cal L}_{I}(A_\mu^a,F_{\mu\nu}^a)\label{leff1}
\end{equation}
($a=1,\ldots ,N$) with the field strength tensor
\begin{equation} F_{\mu\nu}^a=\partial_\mu
A_\nu^a-\partial_\nu A_\mu^a -g f_{abc}A_\mu^b
A_\nu^c.\label{fst1}\end{equation}
For the non--Yang--Mills part of the interactions,
denoted as ${\cal L}_{I}$, I make the following assumptions:
\begin{itemize}\label{req}
\item ${\cal L}_{I}$ does not depend on higher time derivatives;
\item ${\cal L}_{I}$ depends on first
time derivatives of $A_\mu^a$ only through the
non-Abelian field strength tensor\footnote{Actually,
this requirement is equivalent to the demand that ${\cal L}_I$ does not depend
on $\dot{A}_0^a$.} $F_{i0}^a$
(\ref{fst1}).
\end{itemize}
These conditions are fulfilled by the phenomenologically most important
effective interactions, especially by all nonstandard $P$, $C$ and $CP$
invariant trilinear interactions of electroweak vector bosons \cite{nongauge}.
Effective Lagrangians with higher derivatives will be treated in the next
chapter.

{}From (\ref{leff1}) one finds the momenta
\begin{eqnarray} \pi_0^a&=&\frac{\partial{\cal L}_{}}
{\partial \dot{A}^0_a}=0,\label{pn}\\
\pi_i^a&=&\frac{\partial{\cal L}_{}}{\partial
\dot{A}^i_a}=F_{i0}^a+\epsilon
\frac{\partial{\cal L}_{I}}{\partial \dot{A}^i_a}
=\dot{A}^i_a+\partial_iA^a_0
-gf_{abc}A^b_iA^c_0+\epsilon
\frac{\partial{\cal L}_{I}}{\partial \dot{A}^i_a}.
\label{pi1}\end{eqnarray}
Equation (\ref{pi1}) can be solved for the $\dot{A}_a^i$
(in first order\footnote{Remember that without neglecting higher powers of
$\ep$ it would, in general,
be impossible to find closed expressions for the $\dot{A}_i^a$
in terms of the $A_\mu^a$ and $\pi_i^a$ from \eref{pi1};
however the restriction to the
first order of $\ep$ is a purely technical simplification.} of $\epsilon$)
\begin{equation} \dot{A}^i_a=\left.\pi_i^a-\partial_i
A_0^a+gf_{abc}A^b_iA^c_0-
\epsilon\frac{\partial{\cal L}_{I}}
{\partial \dot{A}^i_a}\right|_{
\scriptstyle F_{i0}^a\to\pi_i^a}+O(\epsilon^2). \label{dai}
\end{equation}
The Hamiltonian is
\begin{eqnarray}
{\cal H}_{}&=&\pi_\mu^a\dot{A}^\mu_a-{\cal L}_{}\nonumber\\
&=&\frac{1}{2}\pi_i^a\pi_i^a-\pi_i^a
\partial_iA^a_0+gf_{abc}\pi^a_i
A^b_iA^c_0+\frac{1}{4}F_{ij}^aF_{ij}^a-\frac{1}{2}M^2 (A_0^aA_0^a-
A_i^aA_i^a)\nonumber\\
&&-\epsilon \bar{\cal L}_{I}+O(\epsilon^2),\label{heff}
\end{eqnarray}
where $\bar{\cal L}_{I}$ is given by
\begin{equation} \bar{\cal L}_{I}\equiv{\cal L}_{I}|_{
\scriptstyle F_{i0}^a\to\pi_i^a}.\label{lib} \end{equation}
Equation (\ref{pn}) yields the primary constraints
\begin{equation}\phi_1^a=\pi_0^a=0.\label{pc}\end{equation}
As discussed in section~3.1, the requirement \eref{eomconsp} that the primary
constraints have to be consistent
with the equations of motion, in general,
implies secondary constraints. In this case one finds
\begin{equation} \phi_2^a=\partial_i\pi_i^a+gf_{abc}\pi^b_i A^c_i-
M^2 A_0^a-\epsilon\frac{\partial\bar{\cal L}_{I}}
{\partial A_0^a} +O(\epsilon^2)=0.\label{sc}\end{equation}
There are no further constraints. The Poisson
brackets of the primary and the
secondary constraints (which are taken at equal times) are
\begin{equation}\{\phi_1^a({\bf x}),\phi_2^b({\bf y})\}=
\left(M^2\delta^{ab}
+\epsilon \frac{\partial^2
\bar{\cal L}_{I}}{\partial  A_0^a\partial  A_0^b} +O(\epsilon^2)\right)
\delta^3({\bf  x}-{\bf y})
.\label{pb}\end{equation}
Since $\{\phi_1^a({\bf x}),\phi_1^b({\bf y})\}=0$, one obtains
\begin{equation}
{\rm Det}\,^{\frac{1}{2}}\{\phi^a,\phi^b\}=(-1)^{N+1}
{\rm Det}\,\{\phi^a_1,\phi^b_2\}\ne 0 \label{det1}\end{equation}
(where the $\phi^a=(\phi^a_1,\phi^a_2)$ denote all constraints). Thus, the
constraints are second-class. This is
due to the fact that ${\cal L}_{}$ is gauge
noninvariant, since the mass term and
(in general) the non--Yang--Mills
interactions in ${\cal L}_{I}$ break gauge invariance
explicitly.

The generating functional for a second-class constrained system is
generally given by \eref{zw2} \cite{sen,const,gity}.
In this case it has the form\footnote{It is obvious that in \eref{hpi1}
one has to take the product of the integration measure and of the
$\delta$-functions over all Lorentz and gauge-group indices and that the
constraints in the Poisson brackets have to be taken at equal times, although
this is not explicitly written down there. I will use this simplified notation
throughout this thesis.}
\begin{eqnarray} Z&=&\int
{\cal D} A_\mu^a{\cal D}
\pi_\mu^a\,\exp\left\{i\int d^4x\,[\pi_\mu^a
\dot{A}^\mu_a-{\cal H}_{}
]\right\}\nonumber\\ &&
\qquad\times
\delta(\phi_1^a)\delta(\phi_2^a)
\,{\rm Det}\,^{\frac{1}{2}}(
\{\phi^a({\bf x}),\phi^b({\bf y})\}\delta(x^0-y^0))
.\label{hpi1}\end{eqnarray}
The determinant in (\ref{hpi1}) only yields a $\delta^4(0)$-term which is
neglected here. This can be seen as follows: With $\lib^{Q}$
being the part of
$\lib$ which is quadratic in the fields,
and $\lib^{NQ}$ being the part,
which contains the cubic, quartic, etc.\ terms,
equations (\ref{pb}) and (\ref{det1}) imply that the determinant has the form
\bea\mbox{Det}^\frac{1}{2}(
\{\phi^a({\bf x}),\phi^b({\bf y})\}\dnn)&=&\Det\left[\left(
M^2\delta^{ab}+\epsilon\frac{\partial^2
\bar{\cal L}_{I}^{NQ}}{\partial  A_0^a\partial  A_0^b}
+O(\epsilon^2)\right)
\delta^4(x-y)\right]\nn&&\times
\Det\left[\left(
\delta^{ab}+\epsilon\frac{1}{M^2}\frac{\partial^2
\bar{\cal L}_{I}^Q}{\partial  A_0^a\partial  A_0^b}
\right)\delta^4(x-y)\right].\label{pb2}\eea
Due to the definition of $\lib^{Q}$, the second of these determinants
is constant and can thus be neglected.
The first one can be rewritten as a
functional integral over Grassmann variables (in analogy to
\eref{ghosts}) which yields a ghost term
\be {\cal L}_{ghost}=-M^2\eta_a^\ast\eta_a-
\epsilon\eta^\ast_a\frac{\partial^2\bar{\cal L}_{I}^{NQ}}
{\partial A_0^a \partial  A_0^b}\eta_b+O(\epsilon^2).
\label{egh}\end{equation}
The ghost fields are static, i.e.\ \eref{egh} contains no kinetic terms for
them, only mass terms and couplings to the $A^\mu_a$-fields.
This means, all ghost propagators are simply inverse masses and thus
all ghost loops are (at least) quartically
divergent.  Therefore,
the ghost term can be replaced by a $\delta^4(0)$-term which
yields the same contribution to $S$-matrix elements as the ghost loops.
(See the similar reasoning in section~2.2.)

Dropping the determinant, integrating
out the $\pi_0^a$ and using the relation
\begin{equation} \delta(\phi_2^a)\propto\int
{\cal D}\lambda^a\,
\exp\left\{-i\int d^4x\,\lambda^a\phi_2^a\right\}\label{exp}
\end{equation}
one finds
\begin{eqnarray}
Z&=&\int{\cal D} A_\mu^a
{\cal D}\pi_i^a{\cal D}\lambda^a\,
\exp\Bigg\{ i\int d^4x\,\Bigg[
-\frac{1}{2}\pi_i^a\pi_i^a  \nn&&+\pi_i^a
(\dot{A}^i_a+\partial_i(A_0^a+\lambda^a)-
gf_{abc}A_i^b(A_0^c+\lambda^c))-\frac{1}{4}F_{ij}^a
F_{ij}^a\nn&&+\frac{1}{2}M^2((A_0^a+\lambda^a)
(A_0^a+\lambda^a)-\lambda^a
\lambda^a
-A_i^aA_i^a)\nn&&+\epsilon\left(\bar{\cal L}_{I}+\lambda^a
\frac{\partial\bar{\cal L}_{I}}{\partial A_0^a}\right)
+O(\epsilon^2)\Bigg]\Bigg\}.\end{eqnarray}
The substitution
\begin{equation} A_0^a\to A_0^a-\lambda^a,
\label{subs}\end{equation}
which obviously leaves the functional integration measure
invariant, yields (after neglecting $O(\ep^2)$-terms)
\begin{eqnarray} Z&=&\int{\cal D}
 A_\mu^a{\cal D}\pi_i^a
{\cal D}\lambda^a\,
\exp\bigg\{ i\int d^4x\,\bigg[
-\frac{1}{2}\pi_i^a\pi_i^a+\pi_i^a F^a_{i0}-
\frac{1}{4}F_{ij}^a
F_{ij}^a\nonumber\\
&&+\frac{1}{2}M^2( A_0^aA_0^a-\lambda^a\lambda^a
-A_i^aA_i^a)-\ep\tilde{\cal H}_{I}(A_\mu^a,\partial_i
A_\mu^a,\pi_i^a,\lambda^a)
\bigg]\bigg\}.\label{step1}\end{eqnarray}
with
\begin{equation} \tilde{\cal H}_{I}
(A_\mu^a,\partial_iA_\mu^a,\pi_i^a,\lambda^a)
\equiv-\left(\bar{\cal L}_{I}+\lambda^a
\frac{\partial\bar{\cal L}_{I}}{\partial A_0^a}\right)\Bigg|_{
\scriptstyle A_0^a\to A_0^a-\lambda^a}.
\label{thi}\end{equation}

Now one can generalize the proof of the HLE theorem for effective
interactions of scalar fields \cite{bedu} to the present case.
Introducing sources $K_i^a$ and $K_\lambda^a$ coupled to $\pi_i^a$
and $\lambda^a$, one can rewrite (\ref{step1}) in a form in
which $\pi_i^a$ and $\lambda^a$ occur at most quadratically:
\begin{eqnarray}
Z&=&\int{\cal D} A_\mu^a\,\exp\left\{-i\int
d^4x\,\ep\tilde{\cal H}_{I}\left
(A_\mu^a,\partial_iA_\mu^a,
\frac{\delta}{i\delta K_i^a},\frac{\delta}{i\delta
K_\lambda^a}\right)\right\}
\nonumber\\&&\times\int{\cal D}\pi_i^a{\cal D}
\lambda^a\,
\exp\bigg\{i\int d^4x\,\bigg[
-\frac{1}{2}\pi_i^a\pi_i^a+  \pi_i^aF_{i0}^a-\frac{1}{4}F_{ij}^a
F_{ij}^a\nonumber\\
 &&\quad+\frac{1}{2}M^2( A_0^aA_0^a-\lambda^a\lambda^a
-A_i^aA_i^a)+\pi_i
^aK_i^a+\lambda^aK^a_\lambda)\bigg]\bigg\}\bigg|_{
\scriptstyle K_i^a=K_\lambda^a=0}.\end{eqnarray}
Now one can carry out the Gaussian integrations over $\pi_i^a$ and
$\lambda^a$ and finds
\begin{eqnarray} Z&=&\int{\cal D}
 A_\mu^a\,\exp\left\{
i\int d^4x\, {\cal L}_{0}\right\}\nonumber\\
&&\times\exp\left\{-i\int d^4x\,\ep\tilde{\cal H}_{I}
\left(A_\mu^a,\partial_iA_\mu^a,
\frac{\delta}{i\delta K_i^a},\frac{\delta}
{i\delta K_\lambda^a}\right)\right\}
\nonumber\\&&\times\exp\left\{
i\int d^4x\,\left[\frac{1}{2}K_i^aK_i^a+\frac{1}{2}K_\lambda^aK_
\lambda^a
+K_i^aF_{i0}^a\right]\right\}\bigg|_{
\scriptstyle K_i^a=K_\lambda^a=0}\end{eqnarray}
where ${\cal L}_{0}$ is the massive Yang--Mills part of the
effective Lagrangian
(\ref{leff1}). The use of the functional identity \cite{col}
\begin{equation}
 F\left[\frac{\delta}{i\delta K}\right]G[K]\Bigg|_{K=0}=
G\left[\frac{\delta}{i\delta \rho}\right]F[\rho]\Bigg|_{\rho=0}
\end{equation}
yields
\begin{eqnarray}
Z&=&\int{\cal D} A_\mu^a\,\exp\left\{
i\int d^4x\,{\cal L}_{0}\right\}\nonumber\\
&&\times\exp\left\{\int d^4x\,\left[{}-\frac{i}{2}
\sum_{i,a}\left(\frac{\delta}
{\delta\rho_i^a}\right)^2-\frac{i}{2}
\sum_a\left(\frac{\delta}
{\delta\rho_\lambda^a}\right)^2+F^a_{i0}
\left(\frac{\delta}{\delta\rho_i^a}\right)\right]\right\}\nonumber\\
&&\times
\exp\left\{-i\int d^4x\,
\ep\tilde{\cal H}_{I}\left (A_\mu^a,\partial_iA_\mu^a,
\rho_i^a,\rho_\lambda^a\right)\right\}\bigg|_{
\scriptstyle \rho_i^a=\rho_\lambda^a=0}.
\label{form}\end{eqnarray}
The third exponential in (\ref{form})
can be expanded in powers of $\epsilon$:
\begin{equation} \exp\left\{-i\int d^4x\,
\ep\tilde{\cal H}_{I}\left (A_\mu^a,\partial_iA_\mu^a,
\rho_i^a,\rho_\lambda^a\right)\right\}=1-i\int d^4x\,
\ep\tilde{\cal H}_{I}\left
(A_\mu^a,\partial_iA_\mu^a,\rho_i^a,\rho_\lambda^a\right)+
O(\epsilon^2).
\label{hexp}\end{equation}
Obviously, second functional derivatives with respect to the $\rho$'s
acting on this expression yield terms which are proportional to
$\epsilon^2$ or to $\delta^4(0)$ and
thus neglected here. Therefore,
the second and the
third exponential in (\ref{form}) together reduce to
\begin{eqnarray} &&\exp\left\{\int d^4x\, F^a_{i0}
\left(\frac{\delta}{\delta\rho_i^a}\right)\right\}
\exp\left\{-i\int d^4x\,
\ep\tilde{\cal H}_{I}\left (A_\mu^a,\partial_iA_\mu^a,
\rho_i^a,\rho_\lambda^a\right)\right\}\Bigg|_{
\scriptstyle \rho_i^a=\rho_\lambda^a=0}\nonumber\\&&
\qquad\qquad\qquad=
\exp\left\{-i\int d^4x\,
\ep\tilde{\cal H}_{I}\left (A_\mu^a,\partial_iA_\mu^a,
\rho_i^a,\rho_\lambda^a\right)\right\}\bigg|_{
\begin{array}{l}
\scriptstyle \rho_i^a=F_{i0}^a\\
\scriptstyle \rho_\lambda^a=0 \end{array}}.\label{tl}
\end{eqnarray}
With the definitions of $
\tilde{\cal H}_{I}$ (\ref{thi}) and of $\bar{\cal L}_{I}$
(\ref{lib})
one finds
\begin{equation}
\tilde{\cal H}_{I}\left (A_\mu^a,\partial_iA_\mu^a,
\rho_i^a,\rho_\lambda^a\right)\bigg|_{
\begin{array}{l}
\scriptstyle \rho_i^a=F_{i0}^a\\
\scriptstyle \rho_\lambda^a=0 \end{array}} =-
{\cal L}_{I}(A_a^\mu, F^{\mu\nu}_a) \label{simp}.
\end{equation}
Inserting (\ref{tl}) with
(\ref{simp}) into (\ref{form}) and introducing source terms one finally
obtains
(apart from $O(\epsilon^2)$- and $\delta^4(0)$-terms)
\begin{equation} Z[J]=\int{\cal D}
A_\mu^a\,\exp\left\{i\int d^4x\,
[{\cal L}_{0}+
\epsilon{\cal L}_{I}+
J_\mu^aA^\mu_a]\right\},\label{result}\end{equation}
which is the expected result, namely
the simple Lagrangian path integral (\ref{lpi}) with (\ref{eq}).
Thus, the HLE theorem is proven for effective self
interactions of massive vector fields (within gauge noninvariant models).

This proof can easily be extended to Lagrangians which
also contain effective interactions of massive vector
fields with other
fields (scalar, fermion or additional vector fields). In order to
derive this result, one adds
the kinetic and the mass terms of the extra fields as well as their
standard interactions to ${\cal L}_{0}$ in (\ref{leff1}) and the
nonstandard couplings to ${\cal L}_{I}$
and then goes through the same procedure as above (see,
for comparison, the
treatment of effective gauge theories in section 6.1). Thus, the HLE
theorem also holds for effective vector--fermion and vector--scalar
interactions. An application of this result,
which will become important
in section 4.3, is to consider ${\cal L}_{0}$ as the U-gauge
Lagrangian of a (minimal) Higgs model,
while ${\cal L}_{I}$ contains additional effective interactions
of the vector and Higgs fields.


\section{The Stueckelberg Formalism}
In this section I will prove the HLE theorem for SBGTs with nonlinearly
realized symmetry which contain arbitrary gauge-boson self-interactions within
a gauge invariant framework \cite{nonlin}. It has been
shown in \cite{bulo} and in section~2.7 that each theory given by an effective
Lagrangian of the type (\ref{leff1}) can be rewritten as a
(nonlinearly realized)
SBGT by applying a Stueckelberg transformation \cite{stue,kugo}.
On the other hand, each nonlinear SBGT
can be obtained by applying a Stueckelberg transformation to
a gauge noninvariant Lagrangian such as (\ref{leff1}). Thus, I will
reformulate the Stueckelberg formalism within the Hamiltonian
formalism in order to prove that effective
Lagrangians which are related to each other by a Stueckelberg transformation
are equivalent (at the classical and at the quantum level).

As in chapter~2, I parametrize the Yang--Mills fields and the Stueckelberg
fields in the matrix notation: With $t_a$ being the generators of the gauge
group, which are
orthonormalized according to
\begin{equation}
{{\rm tr}\,}(t_at_b)=\frac{1}{2}\delta_{ab},
\label{tr}\end{equation}
one defines the matrix-valued fields
\begin{eqnarray} A_\mu&\equiv&A_\mu^a t_a,\label{amu}\\
\varphi&\equiv&i\frac{g}{M}\varphi^at_a,\label{mphi}\\
U&\equiv&\exp\varphi.\label{mu}\end{eqnarray}
The $\varphi^a$ are the unphysical pseudo-Goldstone scalars.

Let me consider a massive Yang--Mills theory with non--Yang--Mills
vector-boson self-interactions given by the Lagrangian $\lag$ \eref{leff1}.
I apply the Stueckelberg transformation
\begin{equation} A_\mu\to -\frac{i}{g}U^\dagger D_\mu U=
U^\dagger A_\mu U-\frac{i}{g}U^\dagger\partial_\mu U
=U^\dagger A_\mu U+\frac{1}{M}\partial_\mu\varphi^{a}
U^\dagger Q_a\label{stue1}\end{equation}
where $D_\mu U$ is the covariant derivative of $U$
\be
D_\mu U\equiv\pa_\mu U+igA_\mu U
\label{DU1}\ee
and $Q_a$ is given by
\begin{equation} Q_a\equiv \left(t_a +\frac{1}{2!}(
\varphi t_a+t_a\varphi) +\frac{1}{3!}(\varphi^2t_a+\varphi
t_a\varphi+t_a\varphi^2)+\ldots \right). \end{equation}
As pointed out in chapter~2,
the Stueckelberg transformation (\ref{stue1}) {\em formally}
acts like a gauge transformation, however, with the gauge parameters being
replaced by the pseudo-Goldstone fields. Thus, it only changes the mass term
and the effective
interaction term ${\cal L}_{I}$ in (\ref{leff1}) in a nontrivial way
but not the gauge
invariant Yang--Mills
term $-\frac{1}{4}F^{\mu\nu}_aF_{\mu\nu}^a$.
(\ref{stue1}) can be written in components by
multiplying with
$2t_a$ and taking the trace. With (\ref{tr}) and (\ref{amu})
one finds
\begin{equation}
A_\mu^a\to X_{ab} A_\mu^b+\frac{1}{M}Y_{ab}\partial_\mu\varphi^b
\label{st}\end{equation}
where the matrices $X_{ab}$ and $Y_{ab}$ are defined as
\begin{eqnarray} X_{ab}&\equiv& 2{{\rm tr}\,}(U^\dagger t_bUt_a),
\label{x}\\
Y_{ab}&\equiv&2{{\rm tr}\,}(U^\dagger Q_b t_a).\label{y}
\end{eqnarray}
$X_{ab}$ and $Y_{ab}$ are nonpolynomial expressions in the
pseudo-Goldstone fields
$\varphi^a$. They do not depend on
the derivatives $\partial_\mu \varphi^a$ and, due to
(\ref{tr}), they become unity
matrices for vanishing $\varphi^a$:
\begin{equation} X_{ab}(\varphi^a=0)=Y_{ab}(
\varphi^a=0)=\delta_{ab} \label{van}.\end{equation}

The SBGT corresponding to the effective Lagrangian $\lag$ (\ref{leff1}) is
given by the Lagrangian
\begin{equation} {\cal L}_{}^S\equiv{\cal L}_{}\big|_{
\scriptstyle A_\mu\to U^\dagger A_\mu U-\frac{i}{g}U^\dagger
\partial_\mu U
}.\label{leffs}\end{equation}
${\cal L}_{}$ can
be recovered from ${\cal L}^{S}$ simply by removing all
unphysical scalar fields
\begin{equation} {\cal L}_{}={\cal L}_{}^S\big|_{\varphi_a=0}.
\label{pz}\end{equation}
${\cal L}_{}^S$
describes a generalized gauged nonlinear $\sigma$-model (or chiral Lagrangian)
with extra non-Yang--Mills vector-boson self-interactions
contained in the gauge
invariant term
${\cal L}_{I}^S$, which is obtained by applying $(\ref{stue1})$ to
${\cal L}_{I}$
(see section~2.7). I will show in subsection~5.3.2 (for the more general case
that also higher derivatives occur in $\lag^S$) that
each nonlinearly realized effective SBGT given by
a Lagrangian ${\cal L}_{}^S$ can be constructed by
applying the Stueckelberg transformation (\ref{stue1}) to an
effective Lagrangian ${\cal L}_{}$ (\ref{leff1}), which is obtained
from $\lag^S$ by removing the
pseudo-Goldstone fields.
In this section I prove that the Lagrangians
${\cal L}_{}$ and ${\cal L}_{}^S$ are equivalent within the
Hamiltonian formalism\footnote{For a simple gauged nonlinear
$\sigma$-model {\em without\/}
effective interactions this equivalence was shown
in \cite{pape} (in a very formal and abstract way).}
and that ${\cal L}_{}$ is the U-gauge of ${\cal L}_{}^S$;
i.e. the U-gauge of a (nonlinear) effective SBGT can simply be constructed
by dropping all unphysical scalar fields (as in the Lagrangain
treatment of chapter~2). This is not obvious because the
Stueckelberg transformation (\ref{stue1}) involves derivatives of the
pseudo-Goldstone fields and thus it is not a point
transformation which corresponds to a canonical
transformation within the Hamiltonian framework\footnote{Actually,
within the Hamiltonian treatment of
Lagrangians with higher derivatives (Ostrogradsky formalism \cite{ost})
a Stueckelberg transformation becomes a canonical transformation. This
will be discussed in subsection 5.3.2. In this chapter, however, I do not
use the Ostrogradsky formalism.}.

One can easily see that,
if ${\cal L}_{}$ satisfies the conditions listed
at the beginning of section~4.1,
${\cal L}_{S}$ also fulfils these requirements, since the
field strength tensor transforms under a
Stue\discretionary{k-}{k}{ck}elberg transformation as
\begin{equation} F_{\mu\nu}\to U^\dagger F_{\mu\nu}U \label{fs}
\end{equation}
(with $F_{\mu\nu}=F^a_{\mu\nu}t_a$). Written in components this becomes
\begin{equation} F_{\mu\nu}^a\to X_{ab}F_{\mu\nu}^b.\end{equation}
For the subsequent treatment it is convenient to rewrite
(\ref{leffs}) as
\begin{equation} {\cal L}_{}^S={\cal L}_{}\bigg|_{
\begin{array}{l}
\scriptstyle A_\mu\to U^\dagger A_\mu
U-\frac{i}{g}U^\dagger\partial_\mu U\\
\scriptstyle F_{\mu\nu}\to U^\dagger F_{\mu\nu}U\end{array}}=
{\cal L}_{}\bigg|_{\begin{array}{l}
\scriptstyle A_0^a\to X_{ab} A_0^b+\frac{1}{M}Y_{ab}
\dot{\varphi}^b
\\\scriptstyle A_i^a\to
X_{ab} A_i^b+\frac{1}{M}Y_{ab}\partial_i\varphi^b\\
\scriptstyle F_{i0}^a\to X_{ab}F_{i0}^b
\\ \scriptstyle F_{ij}^a\to X_{ab}F_{ij}^b
\end{array}}, \label{ls}\end{equation}
where the following convention is used: While
in (\ref{leffs}) the
Stueckelberg transformation is applied to $A_\mu$
{\em everywhere\/} in ${\cal L}_{}$
(which automatically implies
the transformation of $F_{\mu\nu}$ (\ref{fs})),
in (\ref{ls}) it is applied to the $A_\mu$-field
only where this does not occur as a part of
the field strength tensor $F_{\mu\nu}$, and $F_{\mu\nu}$
becomes then transformed
seperately. I will use this convention throughout this section.

The momenta conjugate to the fields in ${\cal L}_{}^S$ are
\begin{eqnarray} \pi_0^a&=&\frac{\partial{\cal L}_{}^S}{
\partial \dot{A}^0_a}=0,\label{pns}\\
\pi_i^a&=&\frac{\partial{\cal L}_{}^S}
{\partial \dot{A}^i_a}=F_{i0}^a+\epsilon
\frac{\partial{\cal L}_{I}^S}
{\partial \dot{A}^i_a}
=\dot{A}^i_a+\partial_iA^a_0-gf_{abc}A^b_iA^c_0+\epsilon
\frac{\partial{\cal L}_{I}^S}{\partial \dot{A}^i_a},
\label{pis}\\
\pi_\varphi^a&=&\frac{\partial{\cal L}_{}^S}{\partial
\dot{\varphi}^a}=MY_{ca}\left(X_{cb}A_0^b
+\frac{1}{M}Y_{cb}\dot{\varphi}^b\right)+
\epsilon\frac{\partial{\cal L}_{I}^S}{\partial\dot{\varphi}^a}.
\label{pps}\end{eqnarray}
In first order of  $\epsilon$ one finds the velocities
\begin{eqnarray} \!\!\!\dot{A}^i_a&
=&\left.\pi_i^a-\partial_i A_0^a+gf_{abc}A^b_iA^c_0-
\epsilon\frac{\partial
{\cal L}_{I}^S}{\partial \dot{A}^i_a}\right|_{\begin{array}{l}
\scriptstyle F_{i0}^a\to\pi_i^a\\
\scriptstyle
\dot{\varphi}^a\to Y^{-1}_{ab}\left(Y^{-1}_{cb}\pi_\varphi^c-
MX_{bc}A^c_0\right)\end{array}}+
O(\epsilon^2),\\
\!\!\!\dot{\varphi}^a&=&Y^{-1}_{ab}\left(
Y^{-1}_{cb}\left(\pi_\varphi^c-\left.
\epsilon\frac{\partial{\cal L}_{I}^S}{\partial\dot{\varphi}^c}
\right|_{\begin{array}{l}\scriptstyle F_{i0}^a\to\pi_i^a\\
\scriptstyle \dot{\varphi}^a
\to Y^{-1}_{ab}\left(Y^{-1}_{cb}\pi_\varphi^c-
MX_{bc}A^c_0\right)\end{array}}\right)-MX_{bc}A^c_0\right)+
O(\epsilon^2)\end{eqnarray}
and the Hamiltonian
\begin{eqnarray}
{\cal H}_{}^S&=&\pi_\mu^a\dot{A}^\mu_a
+\pi_\varphi^a\dot{\varphi}^a-{\cal L}_{}^S
\nonumber\\
&=&\frac{1}{2}\pi_i^a\pi_i^a-\pi_i^a\partial_iA^a_0+gf_{abc}\pi^a_i
A^b_iA^c_0+\frac{1}{4}F_{ij}^aF_{ij}^a\nonumber\\
&&-\frac{1}{2}M^2
\left(\sum_a\left(X_{ab}A_0^b\right)^2-\sum_a\left(X_{ab}A_i^b+
\frac{1}{M}Y_{ab}\partial_i\varphi^b
\right)^2\right)\nonumber\\&&+\frac{1}{2}\sum_a
(Y^{-1}_{ba}\pi_\varphi^b-M X_{ab}A_0^b)^2
-\epsilon\bar{\cal L}_{I}^S+O(\epsilon^2)\label{hs}\end{eqnarray}
where $\bar{\cal L}_{I}^S$ is given by
\begin{equation}
\bar{\cal L}_{I}^S\equiv{\cal L}_{I}^S\bigg|_{\begin{array}{l}
\scriptstyle F_{i0}^a\to\pi_i^a\\
\scriptstyle \dot{\varphi}^a\to Y^{-1}_{ab}\left(Y^{-1}_{cb}\pi_
\varphi^c-
MX_{bc}A^c_0\right)\end{array}} =
{\cal L}_{I}\Bigg|_{
\begin{array}{l}\scriptstyle A_0^a \to \frac{1}{M}
Y_{ba}^{-1}\pi_\varphi^b \\
\scriptstyle A_i^a\to X_{ab} A_i^b+\frac{1}{M}Y_{ab}
\partial_i\varphi^b\\
\scriptstyle F_{i0}^a\to X_{ab}\pi_i^b
\\\scriptstyle F_{ij}^a\to X_{ab}F_{ij}^b\end{array}}. \label{con}
\end{equation}
The primary constraints are
\begin{equation}\phi_1^a=\pi_0^a=0\label{pcs}\end{equation}
and the secondary constraints, obtained from \eref{eomconsp}, are
\begin{equation} \phi^a_2=\partial_i\pi_i^a+gf_{abc}\pi_i^bA_i^c
-MX_{ba}Y^{-1}_{cb}\pi^c_\varphi +O(\ep^2)=0.\label{scs}\end{equation}
There are no terms proportional to $\epsilon$
in (\ref{scs}) (in first order of $\ep$)
since,
due to (\ref{con}),
$\bar{\cal L}_{I}^S$
does not depend on $A_0^a$ (neither directly nor through
$F_{i0}^a$). There are no further constraints.
The constraints are first-class due to the gauge freedom of
${\cal L}_{}^S$
(\ref{leffs}).

As discussed in section~3.1, such a first-class constrained sytem has to be
gauged in order to remove the ambiguities in the solutions of the EOM.
To prove the equivalence of ${\cal L}_{}$ (\ref{leff1}) and
${\cal L}_{}^S$ (\ref{leffs}) I construct the U-gauge by imposing
the primary g.f.\ conditions
\begin{equation} \chi_1^a=\varphi^a =0.\label{pgf}\end{equation}
The demand (\ref{gfcons}) yields the
secondary g.f.\ conditions
\begin{equation}\chi_2^a=Y^{-1}_{ab}\left(
Y^{-1}_{cb}\pi^c_\varphi-MX_{bc}A_0^c\right)-
\epsilon\frac{\partial\bar{\cal L}_{I}^S}{\partial
\pi_\varphi^a}+O(\epsilon^2)=0.
\label{sgf}\end{equation}
Using the primary g.f.\
conditions\footnote{Remember that the constraints and the g.f. condition may
be inserted into the Hamiltonian and each other as pointed out in sect.~3.1.}
(\ref{pgf}),  the relation (\ref{van}) and the defintions of
$\bar{\cal L}_{I}^S$ (\ref{con}) and $\bar{\cal L}_{I}$ (\ref{lib}) one can
express the Hamiltonian (\ref{hs}), the secondary constraints (\ref{scs})
and g.f.\ conditions (\ref{sgf}) as
\begin{eqnarray}{\cal H}_{}^S&=&\frac{1}{2}\pi_i^a
\pi_i^a-\pi_i^a\partial_iA^a_0+gf_{abc}\pi^a_i
A^b_iA^c_0+\frac{1}{4}F_{ij}^aF_{ij}^a
\nonumber\\&& -\frac{1}{2}M^2(A_0^aA_0^a-A_i^a
A_i^a)+\frac{1}{2}\sum_a(\pi_\varphi^a-
MA_0^a)^2-\epsilon\tilde{\cal L}_I
+O(\epsilon^2),\label{hs2}\\
\phi^a_2&=&\partial_i\pi_i^a+gf_{abc}\pi_i^bA_i^c
-M\pi^a_\varphi +O(\ep^2)=0,\label{scs2}\\\chi_2^a&=&
\pi^a_\varphi-MA_0^a-
\epsilon \frac{\partial\tilde{\cal L}_I}
{\partial \pi_\varphi^a}
+O(\epsilon^2) =0\label{zus}\end{eqnarray}
with
\begin{equation}
\tilde{\cal L}_I\equiv
\bar{\cal L}_I^S\Big|_{\scriptstyle
\varphi_a=0}=
\bar{\cal L}_{I}\Big|_{
\scriptstyle A_0^a\to\frac{1}{M}\pi_\varphi^a}.
\end{equation}
The $\ep$-dependent term in \eref{zus} can be converted by applying \eref{zus}
itself. One obtains
\begin{equation}
\chi_2^a =\pi^a_\varphi-MA_0^a-
\epsilon \frac{1}{M}\frac{\partial
\bar{\cal L}_{I}}{\partial A_0^a}+O(\epsilon^2)
=0.\label{sgf2}\end{equation}
Applying the secondary g.f.\ condition (\ref{sgf2}), one can
rewrite the
Hamiltonian  (\ref{hs2}) as (\ref{heff}) and the constraints
(\ref{pcs}), (\ref{scs2}) as (\ref{pc}), (\ref{sc}) (in the first
order of $\epsilon$),
i.e., as the Hamiltonian and
the constraints corresponding
to the gauge noninvariant Lagrangian ${\cal L}_{}$
(\ref{leff1}).
Finally, the g.f.\ conditions (\ref{pgf})
and (\ref{sgf2}) can be
omitted, since they involve the fields $\varphi^a$ and $\pi_
\varphi^a$ and
neither the Hamiltonian nor the constraints depend on
these fields anymore.
Thus, the Lagrangians ${\cal L}_{}$ and
${\cal L}_{}^S$ in (\ref{leffs})
describe equivalent physical
sytems, ${\cal L}_{}$ being the U-gauge of ${\cal L}_{}^S$.

As the Hamiltonians and the constraints
corresponding to ${\cal L}_{}^S$ and to ${\cal L}$ are equal,
the generating functional
obtained when quantizing $\lag^S$ within the Hamiltonian PI formalism
is the same as the one found in the last section by quantizing $\lag$, namely
\eref{result}. This, however,
is identical to the generating functional obtained in the
(Lagrangian) Faddeev--Popov PI formalism if one imposes the
g.f.\ conditions (\ref{pgf}), as one can see in analogy to the treatment
of section 2.2.
Due to the equivalence of all gauges \cite{lezj,able},
(\ref{result}) yields the same $S$-matrix elements
as the Faddeev--Popov PI corresponding to ${\cal L}^S$ in any other
gauge (e.g.\ $\rm R_\xi$ gauge, Lorentz gauge, etc.).
Thus, after the introduction of a source term,
the result \eref{lpi} with \eref{fp} is obtained in an arbitrary gauge.

As mentioned in the introduction, this result cannot be {\em directly\/}
derived in the $\rm R_\xi$-gauge
(which is the
most adequate for loop calculations within an SBGT), because the corresponding
g.f. conditions
\begin{equation}
\chi^a=\partial^\mu A_\mu^a-\xi M\varphi^a-C^a=0
\label{rx}\end{equation}
are {\em not\/} g.f. conditions within the Hamiltonian formalism.
This is because (\ref{rx})
cannot be written as relations among the
fields and the conjugate fields alone (such as \eref{gff}),
due to the presence of the
velocities $\dot{A}^a_0$ in \eref{rx} that are not expressable through
the momenta (remember the constraint (\ref{pcs})). Therefore, when
quantizing ${\cal L}^S$ within the $\rm R_\xi$-gauge, one has to
proceed {\em indirectly} as described above by first
constructing the U-gauge and then using the equivalence of all gauges
(i.e., the invariance of the $S$-matrix elements under a change of the
gauge in the FP formalism\footnote{Actually, loop calculations within
the U-gauge suffer from ambiguities in the determination of the
finite part of an $S$-matrix element \cite{fls,jawe}; they
yield the same $S$-matrix elements as loop calculations
within the $\rm R_\xi$-gauge, but only if the
correct renormalization prescription is used
\cite{fls}; other renormalization prescriptions yield distinct
results. In the $\rm R_\xi$-gauge these ambiguities are
absent.}) \cite{lezj,able} in order to rewrite this
result in the $\rm R_\xi$-gauge.

The above procedure demonstrates
how to interpret the Stueckelberg formalism at the
Hamiltonian level. While the gauge noninvariant
Lagrangian ${\cal L}_{}$ is related
to the gauge invariant Lagrangian
${\cal L}_{}^S$ by a Stueckelberg
transformation (\ref{stue1}), one can pass from the
second-class constrained
Hamiltonian sytem corresponding to
${\cal L}_{}$ to the first-class constrained
Hamiltonian system corresponding to ${\cal L}_{}^S$ by the
following procedure: One enlarges the phase space by
introducing the unphysical
variables $\varphi^a$ and $\pi_\varphi^a$
and the extra constraints $(\ref{pgf})$ and
(\ref{sgf}), which make the new variables dependent on the
others so that
the number of physical degrees of freedom remains unchanged.
Next, one applies the additional constraints (\ref{pgf}) and (\ref{sgf}),
in order to rewrite the Hamiltonian as (\ref{hs}) and
the primordial constraints as (\ref{pcs}) and
(\ref{scs}). Finally, half of the constraints, namely the new
ones, are considered to be g.f.\ conditions.

A similar transition from a second- to a first-class
constrained system has recently been investigated in several
works \cite{mira}. However, in \cite{mira}
no phase space enlargement is made
with the outcome, that the resulting model contains only half as many
first-class constraints as the original model has second-class constraints
(the other half has become g.f. conditions.)
In my treatment the number of constraints remains unchanged, since, due to the
phase space enlargement, new constraints are introduced. The method of
connecting first- and second-class constrained systems by a phase
space enlargement was first considered in \cite{fash}
(however without referring to the Stueckelberg formalism).


\section{Higgs Models}
Finally the HLE theorem has to be proven for SBGTs with
linearly realized symmetry (Higgs models)
which contain effective (non--Yang--Mills) gauge-boson
self-interactions \cite{gauge}. This result will  simply be obtained
by showing the
equivalence of a  linear Higgs model to a nonlinear
Stueckelberg model (with
(an) additional physical scalar(s)).

Unlike in the case of a gauged nonlinear $\sigma$-model
discussed in the previous
section it is not possible to write down the Lagrangian of a Higgs model
in a general form without specifying the gauge group. Therefore
I consider the case of SU(2) symmetry (i.e. $t_a =\frac{1}{2}\tau_a$,
$a=1,2,3$). The extension to other gauge groups is straightforward.

As shown in section~2.7, any effective Lagrangian (\ref{leff1}) can be
extended to a Higgs model by constructing the
corresponding Stueckelberg Lagrangian
(\ref{leffs}) and then introducing a physical scalar field
$h$ and linearizing the scalar sector of the theory by means of
the replacement
\begin{equation} \frac{v}{\sqrt{2}}U\equiv\frac{v}{\sqrt{2}}
\exp\left(\frac{i\varphi_a\tau_a}{v}\right)\quad\to\quad
\Phi\equiv\frac{1}{\sqrt{2}} ((v+h){\rm \bf 1} +
i\varphi_a\tau_a)\label{phi}\end{equation}
with $\frac{v}{\sqrt{2}}$ being the VEV of $\Phi$,
$v=\frac{2M}{g}$.
The Lagrangian of the Higgs model corresponding to (\ref{leff1})
becomes
\begin{equation}
{\cal L}_{}^H={\cal L}_{}^S
\big|_{\scriptstyle \frac{v}{\sqrt{2}}U\to\Phi}
-V(\Phi)\label{leffh}\end{equation}
with the scalar self-interaction potential $V(\Phi)$
which implies the nonvanishing vacuum expectation value. Remember that,
in distinction from ${\cal L}_{}^S$ (\ref{leffs}),
${\cal L}_{}^H$ is {\em
not} equivalent to the effective
Lagrangian ${\cal L}_{}$, since there is an additional
physical degree of freedom but that ${\cal L}_{}^H$ contains the
same effective
vector-boson self-interactions as ${\cal L}_{}$ and
${\cal L}_{}^S$.
Each effective Higgs model can
be constructed this way from a Lagrangian of type (\ref{leff1}).

In order
to extend the results of the previous two sections to Lagrangians
of the type (\ref{leffh}), one uses the fact that even within a linaerly
realized SBGT the scalar fields can be parametrized nonlinearly
by means of the point transformation
\begin{equation}  \Phi \to \frac{v+h}{\sqrt{2}}U \label{trafo}
\end{equation}
with $U$ and $\Phi$ given by (\ref{phi}) (see \cite{lezj,clt,higgs}
and sect.~2.4).
The Lagrangian of a Higgs model
in which the scalar sector is nonlinearly realized,
\begin{equation} {\cal L}_{}^{H,S} \equiv {\cal L}_{}^{H}\big|_{
\scriptstyle \Phi \to \frac{v+h}{\sqrt{2}}U}, \label{leffhs}
\end{equation}
describes a Stueckelberg model with an additional physical scalar $h$.
The equivalence between ${\cal L}_{}^{H}$ and ${\cal L}_{}^{H,S}$
can easily be established
since a point transformation such as (\ref{trafo}) becomes a canonical
transformation within the Hamiltonian formalism, i.e.\ the Hamiltonians
and also the constraints corresponding
to ${\cal L}_{}^{H}$ and to ${\cal L}_{}^{H,S}$
are related by a canonical transformation\footnote{For the Hamiltonian
and the primary constraints this statement is obvious and the secondary
constraints are obtained from the Poisson brackets (\ref{eomconsp}) which are
invariant under canonical transformations.}.
Using the result of the previous section
one finds that the U-gauge Lagrangian of $\lag^{H,S}$ (and thus also
of $\lag^H$) is
\begin{equation}  {\cal L}_{U}^{H}\equiv {\cal L}_{}^{H,S}\big|_{
\scriptstyle \varphi_a=0} =
{\cal L}_{}^{H}\big|_{\scriptstyle \varphi_a=0};\label{lu}
\end{equation}
i.e., also for a linearly realized Higgs model the U-gauge is obtained simply
by removing all unphysical scalar fields.

Remembering the last paragraph of section~4.1, the HLE theorem for Stueckelberg
models (section 4.2) can
be applied in order to quantize the nonlinear Lagrangian
${\cal L}_{}^{H,S}$;
the resulting generating functional has the form (\ref{lpi})
with the quantized Lagrangian
\begin{equation} {\cal L}_{quant}= {\cal L}_{U}^{H}.\end{equation}
Due to the invariance of the Hamiltonian PI under
canonical transformations
\cite{fad,sen,const,gity}, the generating
functional obtained when quantizing the
linear Lagrangian ${\cal L}_{}^H$  is the same.
As shown in sect.~2.2, this is identical
to the result of the Faddeev--Popov procedure (apart from a
$\delta^4(0)$-term) if the
g.f.\ conditions (\ref{pgf}) are imposed.
As in the previous section, this result can be generalized to
any other gauge,  which yields \eref{lpi} with \eref{fp}. This
completes the proof of the HLE theorem for any
effective Lagrangian with arbitrary interactions of massive
vector fields which
fulfils the requirements listed at the beginning of section~4.1.

The treatment of this
section shows that the Stueckelberg formalism, which was
originally introduced
in order to construct Higgs-less SBGTs \cite{stue,kugo,bulo}, also
represents a powerful tool when dealing with Higgs models.


\section{The Quartically Divergent Higgs Self-Interaction}
When proving the HLE theorem
in the previous sections of this chapter, I have neglected quartically
divergent $\delta^4(0)$-terms.
In this section I want to present a simple example of such a term. Therefore
I will quantize the SU(2) Higgs model
(without effective non--Yang--Mills interactions) in
the U-gauge
{\em while taking fully into account $\delta^4(0)$-terms\/} in order to derive
the well-known quartically divergent
Higgs self-coupling term \cite{lezj,wein2,leya,wein1,wein3,giro}.
In chapter~2 this term has been derived within the {Lagrangian} PI formalism;
in this section I will find the same term within the Hamiltonian PI
formalism.

{}From the discussion of the previous two sections it is
clear that it makes
no difference whether one quantizes the gauge invariant Lagrangian
of an SBGT or the corresponding U-gauge
Lagrangian\footnote{When establishing
this equivalence, no $\delta^4(0)$-terms have
been neglected, thus, even concerning the quartically
divergent extra terms,
quantization of both Lagrangians yields the
same result.}. Thus, for simplicity, I start
from the U-gauge Lagrangian
\begin{equation}{\cal L}_{}= -\frac{1}{4}F^{\mu\nu}_aF_{\mu\nu}^a+
\frac{1}{2}(\partial^\mu h)(\partial_\mu h)+
\frac{1}{8}g^2(v+h)^2A^\mu_aA_\mu^a-V(h,\varphi_a=0)
.\label{lsu2}\end{equation}
The momenta are given by
\begin{eqnarray} \pi_0^a&=&\frac{\partial
{\cal L}_{}}{\partial \dot{A}^0_a}=0,\label{pnh}\\
\pi_i^a&=&\frac{\partial{\cal L}_{}}
{\partial \dot{A}^i_a}=F_{i0}^a=
\dot{A}^i_a+\partial_iA_0^a-gf_{abc}A^b_iA^c_0,
\label{pih}\\
\pi_h&=&\frac{\partial
{\cal L}_{}}{\partial \dot{h}}=\dot{h},\label{ph}\end{eqnarray}
and the Hamiltonian is
\begin{eqnarray} {\cal H}_{}&=&\pi_\mu^a\dot{A}^\mu_a+\pi_h\dot{h}-
{\cal L}_{}\nonumber\\
&=&\frac{1}{2}\pi_i^a\pi_i^a+\frac{1}{2}\pi_h^2-
\pi_i^a\partial_iA^0_a+gf_{abc}\pi^a_i
A^b_iA^c_0+\frac{1}{4}F_{ij}^aF_{ij}^a+
\frac{1}{2}(\partial_ih)(\partial_ih)\nonumber\\&&-
\frac{1}{8}g^2(v+h)^2 (A_0^aA_0^a-
A_i^aA_i^a)+V(h,\varphi_a=0).\label{hh}\end{eqnarray}
The constraints turn out to be
\begin{eqnarray}\phi_1^a&=&\pi_0^a=0,\label{pch}\\
\phi_2^a&=&\partial_i\pi_i^a+gf_{abc}\pi^b_i A^c_i-
\frac{1}{4}g^2(v+h)^2 A_0^a=0.
\label{sch}\end{eqnarray}
The Poisson brackets of the primary and the
secondary constraints are
\begin{equation} \{\phi_1^a({\bf x}),\phi_2^b({\bf y})\}=
\frac{1}{4}g^2(v+h)^2\delta^{ab}\delta^3({\bf x}-\bf{ y}).
\label{pbh}\end{equation}
There are no further constraints. The constraints are second-class.

To quantize this model, one starts from the Hamiltonian PI (\ref{hpi1}),
integrates out the $\pi_0^a$, uses
(\ref{exp}) in order to rewrite
$\delta(\phi_2^a)$, substitutes \eref{subs} and
rewrites the deteminant using
(\ref{det1}) and (\ref{pbh}). The generating functional becomes
\begin{eqnarray}
 Z&=&\int{\cal D} A_\mu^a{\cal D} h
{\cal D}\pi_i^a{\cal D}\pi_h{\cal D}\lambda^a\,
\exp\bigg\{ i\int d^4x\,\bigg[-\frac{1}{2}\pi_i^a\pi_i^a
-\frac{1}{2}\pi_h^2+  \pi_i^a
F_{i0}^a+\pi_h\dot{h}\nonumber\\&&-\frac{1}{4}F_{ij}^a
F_{ij}^a-\frac{1}{2}(\partial_i h)(\partial_i h)+
\frac{1}{8}g^2(v+h)^2( A_0^aA_0^a-\lambda^a\lambda^a
-A_i^aA_i^a)\nonumber\\
&&-V(h,\varphi_a=0)+\bigg]
\bigg\}\,{\rm Det^3}\left(\frac{1}{4}g^2(v+h)^2\delta^4(x-y)
\right).\end{eqnarray}
Now one can carry out the Gaussian integrations over
$\pi_i^a$, $\pi_h$ and $\lambda^a$. Integrating out
$\lambda^a$ yields an extra factor
${\rm Det^{-\frac{3}{2}}}\left(\frac{1}{4}g^2(v+h)^2
\delta^4(x-y)\right)$.
One finds \begin{equation}
 Z=\int{\cal D} A_\mu^a{\cal D} h\,\exp\left\{i\int d^4x\,{\cal L}\right\}
\Det\left(\frac{1}{8}g^3(v+h)^3\delta^4(x-y)
\right).\label{aa}\end{equation}
Finally one applies
(\ref{det}) in order to rewrite the determinant as an exponential function
and introduces
the source terms; $Z$ becomes then a
Lagrangian PI
(\ref{lpi}) with the quantized U-gauge Lagrangian
\begin{equation} {\cal L}_{quant}=
{\cal L}_{}-3i\delta^4(0)\ln\left(1+\frac{h}{v}\right)
={\cal L}_{}
-3i\delta^4(0)\ln\left(1+\frac{g}{2M}h\right)\label{extra1}
\end{equation}
(after dropping a constant). Thus, the quantized U-gauge Lagrangian
contains, in addition to the classical Lagrangian \eref{lsu2},
an extra quartically divergent Higgs self-interaction term\footnote{From
the proof in section 4.1 it is obvious, that the Hamiltonian PI
quantization of Lagrangains with additional effective interaction terms
in general yields further $\delta^4(0)$-terms.
These may also contain other fields than the Higgs field.}
which is identical
to the one obtained in chapter~2 within the Lagrangian PI formalism
\eref{extra}.

\chapter{Effective Lagrangians with Higher Derivatives}
In this chapter I will generalize the results obtained in the last one
to effective Lagrangians which also depend on higher derivatives of the fields
\cite{phhide}.

Actually, the proof of the HLE theorem for first-order
Lagrangians can be extended to the higher-order case by applying the results
of \cite{eom,pol}, where it has been shown that each effective
higher-order Lagrangian can be reduced to a first-order one by
applying the equations of motion to the effective interaction term
(upon neglecting higher powers of the effective coupling constant $\ep$)
since one can find field transformations which effectively
result in applications of the EOM. However, the
treatment of \cite{eom,pol} is incomplete because it has not been shown
there that two Lagrangians which are related
to each other by such field transformations
are physically equivalent. In fact, this is not trivial since these
transformations, in general, involve derivatives of the fields.
I will show that within the Hamiltonian formalism for Lagrangians with higher
derivatives (Ostrogradsky formalism \cite{ost}) even a field
transformation
which involves derivatives becomes a canonical transformation;
this establishes the equivalence of Lagrangians that are related to each other
by an arbitrary local field transformation and thus justifies the use
the EOM in order to convert the effective interaction term.
Since the Ostrogradsky formalism is not very
well-known, I will first give a short introduction to it before
deriving these results.

Making use of the abovementioned procedure of reducing effective
higher-order Lagrangians to first-order ones and of the results of
the last chapter, I will then prove
the HLE theorem for the case of effective Lagrangians with higher
derivatives\footnote{The Lagrangian PI \eref{lpi} with \eref{eq} or \eref{fp}
has been derived for some special examples of higher-order Lagrangians in
\cite{bedu,gity,bnn}, but a general theorem has not been proven.}.
I will derive
this result first for the simple case of effective interactions
of scalar fields and then for effective
interactions of massive vector fields. In the latter case I will, as in the
previous chapter, first consider gauge noninvariant models and then extend
the results to SBGTs by aplying the Stueckelberg formalism.
Actually, within the Ostrogradsky formalism a
Stueckelberg transformation becomes a canonical transformation
(see above) and thus
the proof in section 4.2 that Lagrangians which are related
to each other by a Stueckelberg
transformation are physically equivalent can be simplified very much.

The results of this chapter have first been published in \cite{gk2}.


\section{The Ostrogradsky Formalism}
The Hamiltonian formalism for Lagrangians with higher
derivatives was formulated one and a half century ago by
Ostrogradsky \cite{ost}. Since this formalism
is not as well-known as the Hamiltonian
formalism for Lagrangians with at most first derivatives, I will first
briefly review it in this section taking also into account the case of
singular higher-order Lagrangians \cite{gity,shd,pons}, i.e.\ of
Lagrangians which involve constraints.
Furthermore, I will consider the quantization of higher-order
Lagrangians within the Hamiltonian PI formalism. Then,
using the Ostrogradsky
formalism, I will show that Lagrangians which related to each other by
field transformations involving derivatives
are equivalent (at the classical and at the quantum level).
For simplification I will restrict myself to the case of
finitely many degrees of freedom;
the generalization to field theory works as in the treatment of
Lagrangians without higher derivatives.

Consider a Lagrangian of order $N$, i.e. a Lagrangian which depends
on the coordinates $q_i$, (with $i=1,\ldots,I$) and their
derivatives up to the order $N$:
\be
L(q_i,q_i^{(1)},\ldots,q_i^{(N)}),\qquad\mbox{with}
\qquad q_i^{(n)}\equiv
\left(\frac{d}{dt}\right)^n q_i.
\label{hol2}\ee
Within in the Ostrogradsky formalism one defines the coordinates as
\be
Q_{i,n}\equiv q_i^{(n-1)},\qquad n=1,\ldots,N
\label{Q}\ee
($q_i^{(0)}\equiv q_i$),
i.e. the $q_i^{(n)}$ with $n\le N-1$ are formally treated as independent
coordinates and only $q_i^{(N)}$ is treated as a derivative.
The momenta are defined as
\be
P_{i,n}\equiv \sum_{k=n}^N\left(-\frac{d}{dt}\right)^{k-n}
\frac{\partial L}{\partial q_i^{(k)}},\qquad
n=1,\ldots,N.
\label{P}\ee
The Hamiltonian is given by
\be
H\equiv\sum_{i=1}^I\sum_{n=1}^NP_{i,n}\dot{Q}_{i,n}-L
=\sum_{i=1}^I
\left(\sum_{n=1}^{N-1}P_{i,n}Q_{i,n+1}+P_{i,N}q_i^{(N)}
\right)-L.
\label{ham}
\ee
In \eref{ham}, the $q_i^{(N)}$ ($i=1,\ldots ,I$)
have to be expressed in terms of the
$P_{i,N}$ by using \eref{P} with $n=N$:
\be
P_{i,N}=\frac{\partial L}{\partial q_i^{(N)}}.
\label{PN}\ee
In analogy to the treatment of first-order Lagrangians discussed in sect.~3.1,
the theory is called nonsingular if the $I\times I$-matrix
\be
M_{ij}\equiv\frac{\partial^2 L}{\partial q_i^{(N)}
\partial q_j^{(N)}}
\label{sing}\ee
is nonsingular. In this case \eref{PN} can be solved for all $q_i^{(N)}$.

For singular Lagrangians one finds
\be
\mbox{rank}\,M_{ij}=R<I.
\label{R}\ee
The indices $i,j$
can be ordered such that the upper left
$R\times R$-submatrix
of $M$ has the rank $R$. In this case, the first $R$ of the relations
\eref{PN} can be solved for $q_i^{(N)}$
with $i=1,\ldots,R$, while the remaining ones are the primary constraints.
Nevertheless, $H$ \eref{ham} does
not depend on $q_i^{(N)}$ with $i=R+1,\ldots,I$ because
\be
\frac{\partial H}{\partial q_i^{(N)}}=
P_{i,N}-\frac{\partial L}{\partial q_i^{(N)}}=0,
\qquad i=R+1,\ldots,I\, ,
\ee
where the constraints \eref{PN} with $i=R+1,\ldots,I$ have been used.
Furthermore, \eref{R} implies that the first $R$ of the relations
\eref{PN} can be used to rewrite the remaining ones in a form which
does not involve the $q_i^{(N)}$.
Thus the Hamiltonian and the primary constraints are functions of the
$Q_{i,n}$ and the $P_{i,n}$, $n=1,\ldots,N$, alone.
As in the first-order case, the Hamiltonian EOM have the form \eref{eom} with
\eref{htot}, where $f$ may be $Q_{i,n}$, $P_{i,n}$ or an arbitrary
function of these variables.

A higher-order Lagrangian can be quantized within the
Hamilitonian PI formalism \cite{fad,sen,const,gity}
in analogy to a first-order Lagrangian. The Hamiltonian  PI
\eref{hpi} turns out to be
\bea
Z&=&\int\prod_{i=1}^I\prod_{n=1}^N(\dfun Q_{i,n} \dfun P_{i,n})
\exp\left\{i\int d^4x\,\left[\sum_{i=1}^I \sum_{n=1}^N
P_{i,n}\dot{Q}_{i,n}-H\right]\right\}\nn&&\qquad\times
\delta(\phi_a)\,{\rm Det}^\frac{1}{2}\,(\{\phi_a,\phi_b\}\delta(x^0-y^0))
\label{hpi2}\eea
where the $\phi_a$ are
the primary constraints
\eref{PN} with $i=R+1,\ldots,I$, the secondary, tertiary, etc.\
constraints and the gauge-fixing
conditions if there are first-class constraints.

For the following investigations it is important, that
{the Ostrogradsky formalism is not affected by changing the
(formal) order $N$ of the Lagrangian\/} as long as
$N$ is greater or
equal to the order of the highest derivative actually occurring in $L$.
In other words, two physical systems, which are both given in terms of
the same Lagrangian $L$, but in the one $L$ is formally treated as
an $N$-th order Lagrangian (within the Ostrogradsky formalism) and in the other
as an $M$-th order one, are equivalent. This theorem was
proven in \cite{gity}. I repeat this short proof here and, in addition,
I show that also the Hamiltonian PIs corresponding to the two
systems are identical.

In order to prove this theorem,
it is sufficient to assume that $M=N+1$. (The result for an arbitrary $M$
follows then by induction.) Thus I treat the
Lagrangian \eref{hol2} (which does not depend on the $q_i^{(N+1)}$)
formally as an $(N+1)$st-order Lagrangian. One finds the
canonical variables
\bea
\tilde{Q}_{i,n}&=& q_i^{(n-1)},\qquad n=1,\ldots,N+1,
\label{Q2}\\
\tilde{P}_{i,n}&=& \sum_{k=n}^{N+1}
\left(-\frac{d}{dt}\right)^{k-n}
\frac{\partial L}{\partial q_i^{(k)}},\qquad
n=1,\ldots,N+1.
\label{P2}\eea
With $n=1,\ldots,N$, these equations become
\be
\tilde{Q}_{i,n}=Q_{i,n},\qquad
\tilde{P}_{i,n}=P_{i,n},\qquad n=1,\ldots,N,
\label{e1}\ee
i.e., these variables are identical to the corresponding ones
obtained in the $N$th-order formalism, \eref{Q} and \eref{P}.
With $n=N+1$ one finds
\bea
\tilde{Q}_{i,N+1}&=&q^{(N)}_i,\label{QN1}\\
\tilde{P}_{i,N+1}&=&\frac{\pa L}{\pa q_i^{(N+1)}}
=0.\label{pc2}
\eea
The Hamiltonian is given by
\be
\tilde{H}=\sum_{i=1}^I\sum_{n=1}^{N+1}\tilde{P}_{i,n}
\dot{\tilde{Q}}_{i,n}-L
=\sum_{i=1}^I\sum_{n=1}^{N}\tilde{P}_{i,n}
\tilde{Q}_{i,n+1}-L.
\label{H2}\ee
$\tilde{H}$ is identical to $H$ \eref{ham}, except that in $H$
the $q_i^{(N)}$ are
expressed in terms of the remaining variables by using \eref{PN},
but in $\tilde{H}$ they are still present
(because, due to \eref{QN1}, they are independent coordinates
in the $(N+1)$st-order formalism).
The $(N+1)$st-order system is singular; the relations \eref{pc2}
are the primary constaints.
The requirement \eref{eomconsp} that these constraints have to be
consistent with the
EOM yields the secondary constraints
\be
\dot{\tilde{P}}_{i,N+1}=-\frac{\partial \tilde{H}}
{\partial \tilde{Q}_{i,N+1}}
=\frac{\partial L}{\partial q_i^{(N)}}-\tilde{P}_{i,N}=0.
\label{sc2}
\ee
This is identical to the relation \eref{PN}. Applying the constaints \eref{sc2}
with $i=1,\ldots,R$  in order to eliminate all
$\tilde{Q}_{i,N+1}$ from the Hamiltonian $\tilde{H}$
and from the remaining constraints \eref{sc2}
(in analogy to the procedure outlined above) one finds
\be
\tilde{H}=H.
\label{e2}
\ee
Then the Lagrange
multipliers corresponding to the constraints \eref{sc2}
with $i=1,\ldots,R$,
which can be solved for $\tilde{Q}_{i,N+1}$,
i.e.\ that can be rewritten as
\be
\tilde{Q}_{i,N+1}-f_i(\tilde{Q}_{j,1},
\ldots ,\tilde{Q}_{j,N}, \tilde{P}_{j,N})=0
\qquad i=1,\ldots,R,
\label{solve}
\ee
become zero in order to ensure
$\dot{\tilde{P}}_{i,N+1}=0$
(because $\tilde{H}$ and the other constraints
do not depend on the $\tilde{Q}_{i,N+1}$ anymore).
The remaining constraints in the $(N+1)$st-order
formalism are identical
to those obtained in the $N$th-order formalism (because \eref{PN} and
\eref{sc2} are identical). Therefore, also the total Hamiltonians \eref{htot}
which imply the EOM are identical%
\footnote{Actually, $\tilde{H}_T$ contains an additional term
$\lambda^i P_{i,N}$ due to the presence of
the primary constraints \eref{pc2}. This however does
not affect the EOM \eref{eom} for the
variables $\tilde{Q}_{i,n}$ and $\tilde{P}_{i,n}$
with  $n=1,\ldots,N$.}
\be
\tilde{H}_T=H_T.
\label{e3}
\ee
{}From \eref{e1} and \eref{e3} follows the equivalence of
the $N$th- and the $(N+1)$st-order Ostrogradsky formalism, i.e. the
independence of the Ostrogradsky formalism from the choice of the (formal)
order $N$ of the Lagrangian $L$.

To extend this classical result to quantum physics,
one has to show that the Hamiltonian PIs
obtained within the $N$th and the $(N+1)$st-order
formalism are identical. In the
$(N+1)$st-order formalism the Hamiltonian PI has the form
\bea
\tilde{Z}&=&\int\prod_{i=1}^I\prod_{n=1}^{N+1}
(\dfun \tilde{Q}_{i,n}\dfun \tilde{P}_{i,n})
\exp\left\{i\int d^4x\,\left[\sum_{i=1}^I\sum_{n=1}^{N+1}
\tilde{P}_{i,n}\dot{\tilde{Q}}_{i,n}-\tilde{H}\right]\right\}\nn&&\qquad\times
\delta( \tilde{\phi}_a)
\,{\rm Det}^\frac{1}{2}\,(\{\tilde{\phi}_a,\tilde{\phi}_b\}\delta(x^0-y^0)).
\label{hpi22}\eea
The constraints and g.f.\ conditions $\tilde{\phi}_a$ include all
the constraints and g.f.\ conditions $\phi_a$
obtained in the $N$th-order formalism and, in addition, the constraints
\eref{pc2} and \eref{solve}. The fundamental Poisson brackets imply
that \eref{pc2} with $i=1,\ldots, R$ and \eref{solve} represent a
system of second-class constraints, while the constraints \eref{pc2}
with $i=R+1,\ldots,I$,
which have vanishing Poisson brackets with all
other constraints, are first-class.
Therefore one has to include
gauge-fixing conditions corresponding to these
first-class constraints
into the $\tilde{\phi}_a$. A possible and suitable choice is
\be
\tilde{Q}_{i,N+1}=0,\qquad i=R+1,\ldots,I.
\label{gf2}
\ee
Using the fundamental Poisson brackets and remembering that the
$\phi_a$ do not depend on the $\tilde{Q}_{i,N+1}$
and the $\tilde{P}_{i,N+1}$ , one finds
\be
{\rm Det}\, (\{\tilde{\phi}_a,\tilde{\phi}_b\}\delta(x^0-y^0))=
{\rm Det}\, (\{\phi_a,\phi_b\}\delta(x^0-y^0)).
\ee
Because of the presence of
the $\delta$-functions corresponding to the constraints
\eref{pc2} and \eref{solve} and to the g.f.\ conditions
\eref{gf2} in the PI
one can carry out the functional
integrations over the $\tilde{Q}_{i,N+1}$ and the
$\tilde{P}_{i,N+1}$.
Equations \eref{e1} and \eref{e2} imply then that the
Hamiltonian PI \eref{hpi22} is identical to the one
obtained in the $N$th-order formalism \eref{hpi2}.  Thus, {the
Hamiltonian PI is also independent of the formal order $N$.}

The conclusion of this section, which is most important for the
subsequent investigations is the following: In the $N$th-order
formalism all derivatives up to the order $N-1$ are treated
as independent coordinates and
not as derivatives. Furthermore, the order $N$ can be
chosen arbitrarily
high without affecting the physical content of the
theory. This implies that {each local coordinate transformation
which involves derivatives of the coordinates (up to a finite
order) can be considered to be a point transformation,
i.e. a transformation which formally only depends on
the coordinates but not on the derivatives.
If one wants to apply a coordinate
transformation involving derivatives up to the order $N$, one simply
has to treat the Lagrangian as an $(N+1)$st-order one (even if
no $(N+1)$st derivatives occur in $L$)
so that this transformation can be identified as
a point transformation.
Such a transformation
becomes a canonical transformation within the
Hamiltonian framework\/} because the fact that two Lagrangians
are related to each other
by a point transformation implies that the
corresponding Hamiltonians and constraints are related to each other by a
canonical transformation (if the order $N$ is chosen sufficiently high);
namely the canonical transformation corresponding to the point transformation
\be
Q_{i,n}\to f_{i,n}(Q_{j,m})\label{pointq},
\ee
is (in analogy to the first-order formalism \cite{cm})
given by\footnote{This can easily be seen
if the order $N$ is chosen so high that
the two Lagrangians related by the transformation
\eref{pointq} only depend
on $Q_{i,n}$ with $n<N$ and that the transformations
of the $Q_{i,n}$ which occur in the Lagrangians do not involve the $Q_{i,N}$.
In analogy to the treatment of first-order Lagrangians one
finds then that the Hamiltonians \eref{ham} and the
primary constraints ($P_{i,N}=0$ in this case) are related to each other
by the canonical transformation \eref{pointq}, \eref{pointp}.
The secondary, tertiary, etc.\ constraints follow
from the Poisson brackets \eref{eomconsp}
which are invariant under canonical transformations.}
\eref{pointq} and
\be
P_{i,n}\to \left(\frac{\pa f_{i,n}}{\pa Q_{j,m}}\right)^{-1}P_{j,m}
.\label{pointp}\ee
Since the physical content of a theory is not affected
by a canonical transformation, one finds
that {\em Lagrangians, which are related to each other by a local
coordinate transformation are physically
equivalent even if this
transformation involves  derivatives.}

The Hamiltonian PIs corresponding to Lagrangians
that are related by such a field transformation
are identical because of the invariance of the
Hamiltonian PI under canonical transformations\footnote{Within the
{\em canonical\/} quantization formalism the invariance of the $S$-Matrix
elements with respect to a point transformation of the Lagrangian was
shown in \cite{chis}.} \cite{fad,sen,const,gity} and
its independence of the (formal) order $N$ (see above). Therefore,
{\em this equivalence is also valid in quantum physics.}


\section[Reduction of Higher-Order Effective Lagrangians]
{Reduction of Higher-Order Effective Lagrangians and
the Hamilton--Lagrange Equivalence Theorem}
In this section I will show  that
a higher-order effective Lagrangian can be reduced
to a first-order one by applying the equations of
motion to the effective interaction term.
Using this reduction, I will prove the HLE theorem for
effective Lagrangians with higher derivatives.
For simplicity I will first only
consider effective interactions of a scalar field.

I consider a Lagrangian of the form
\be
\lag=\lag_0+\ep\lag_I=\frac{1}{2}(\pa^\mu\vp)(\pa_\mu\vp)-
\frac{1}{2}M^2\vp^2+\ep\lag_I(\vp,\pa^\mu\vp,\,\ldots,\,
\pa^{\mu_1}\cdots\pa^{\mu_N}\vp).
\label{leff2}
\ee
$\lag_I$ contains effective interactions
of the scalar field $\varphi$ which depend on
the derivatives of $\varphi$ up to the order $N$. These
interactions are governed
by the coupling constant $\ep$ with $\ep\ll 1$.

Now I want to remove all higher
time derivatives from the effective
interaction term by applying the
EOM (upon neglecting higher powers of
$\ep$). This procedure must be carefully justified
since, in general, the EOM must not be applied in order to convert
the Lagrangian. Therefore, I use the results of \cite{eom,pol}
where it was shown
that it is always possible to find
field transformations which effectively result
in applying the EOM following from
$\lag_0$ to $\lag_I$ (in first order of $\epsilon$).

This can be seen as follows:
Assume that $\ep\lag_I$ contains a term
\be
\ep T\ddot{\vp}
\label{term}\ee
(where $T$ is an arbitrary expression in $\vp$ and its derivatives),
i.e.\ that \eref{leff2} can be written as
\be
\lag=\lag_0+\ep T\ddot{\vp}+\ep\tilde{\lag}_I
\label{l1}
\ee
(with $\tilde{\lag}_I\equiv\lag_I-T\ddot{\vp}$).
Making the field transformation \cite{eom,pol}
\be
\vp\to\vp+\ep T
\label{trafo2} \ee
(and the corresponding transformations of the derivatives of $\vp$) one finds
(in first order of $\ep$)
\bea
\lag&=&\lag_0+\ep\frac{\pa\lag_0}{\pa\vp}T+\ep\frac{\pa\lag_0}{\pa(\pa^\mu\vp)}
(\pa^\mu T)+\ep T \ddot{\vp}+\ep\tilde{\lag}_I+ O(\ep^2)\nonumber\\
&=&\lag_0+\ep T\ddot{\vp}+\ep T\left(\frac{\pa\lag_0}{\pa\vp}-
\pa^\mu\frac{\pa\lag_0}{\pa(\pa^\mu\vp)}\right)+
\ep\tilde{\lag}_I+O(\ep^2)\nonumber\\
&=&\lag_0+\ep T (\Delta \vp-M^2 \vp)
+\ep\tilde{\lag}_I+O(\ep^2)
\eea
(after dropping a total derivative term\footnote{In general, only
total derivatives of expressions that
depend on nothing but the coordinates can be dropped.
However, since the
derivatives are treated as coordinates
within the Ostrogradsky formalism
if the order $N$ is chosen sufficiently high (as discussed in the
previous section),
all total derivative terms can be neglected
\cite{gity}.}). This means
that effectively the second time derivative has been removed
from the term \eref{term} by applying
the free EOM (i.e.\, the EOM
implied by $\lag_0$ alone)
\be
\ddot{\vp}=\Delta\vp-M^2\vp.
\label{kgeom}\ee
If there are terms with
higher than second time derivatives in $\ep\lag_I$,
they can be put into the form
\eref{term} by
applying product differentiation and dropping total derivative
terms.
Repeating the above procedure, one can remove all higher
time derivatives from the effective Lagrangian.

To prove that a higher-order effective Lagrangian and
the first-order Lagrangian obtained from it by applying the
above procedure are physically equivalent, I
use the results of the previous section. The Lagrangians are
connected by field transformations of the type \eref{trafo2} which,
in general, involve derivatives of $\vp$ (contained in $T$).
However, within
the Ostrogradsky formalism these transformations are point (i.e.\
canonical) transformations (if the formal order of $\lag$ is
choosen sufficiently high)
which establishes the equivalence of both
Lagrangians. To be strictly correct, one must remember that, within
the Ostrogradsky formalism, the time derivatives of $\vp$ are
fields that are formally independent  of $\vp$.
Therefore, the transformation \eref{trafo2} is completely specified by
\be
\pa_0^n\vp\to\pa_0^n(\vp+\ep T), \qquad n=0,\ldots, N,
\label{ttrafo2}\ee
where $N$ is the order of the highest time derivative occurring in $\lag$
\eref{leff2}. Now one can
easily see that, if $T$ contains at most $M$th time derivatives,
the Lagrangian formally has to be treated as an
$(N+M+2)$nd-order one so that \eref{ttrafo2} becomes
a canonical transformation.
The result of this procedure is that
{\em the equations of motion
imlied by $\lag_0$ may be applied to convert the effective interaction term
$\lag_I$ in order
to eliminate all higher time
derivatives\/} (by neglecting higher powers of
$\ep$).

Now it is easy to quantize the effective Lagrangian \eref{leff2}
and to prove the HLE theorem, i.e. to derive the Lagrangian PI
\eref{lpi} with \eref{eq}.
The proof goes as
follows:
\begin{enumerate}
\item Given an effective higher-order Lagrangian $\lag$ \eref{leff2},
it can be reduced to an equivalent
first-order\footnote{Formally, the reduced Lagrangian $\lag_{red}$
still has to be treated as
an $N$th-order one although it contains no
more higher derivatives. But, as shown in
section~5.1, it does not affect the physical content
of the theory to treat it as
a first-order Lagrangian.} Lagrangian $\lag_{red}$ by
applying field transformations such as \eref{trafo2}.
As discussed in section~5.1 this does not affect the Hamiltonian PI.
\item $\lag_{red}$ can be
quantized by applying the HLE theorem for
first-order Lagrangians \cite{bedu}. The resulting generating functional
can be written as a Lagrangian PI
\be
Z=\int\dfun\varphi\exp\left\{i\int\dx\lag_{red}\right\}
.\label{lred}
\ee
\item In the Lagrangian PI \eref {lred},
the field transformations \eref{trafo2}
are done inversely in order to reconstruct the primordial higher-order
Lagrangian $\lag$ \eref{leff2}. After the introduction of a source term
this finally yields \eref{lpi} with \eref{eq}.
\end{enumerate}
The last step needs some additional clarification because a field
transformation like \eref{trafo2}, if it is made in the
PI \eref{lred}, does not only affect the quantized
Lagrangian, but also the integration measure.
However, it was shown in \cite{pol} that the transformation of the
measure only yields an extra $\delta^4(0)$-term which is neglected here.
I briefly repeat here the derivation of this result in the second ref.\
in \cite{pol} because it is very similar to an argument that I have used
several times in chapters~2 and 4: The functional Jacobian determinant
corresponding to
the transformation \eref{trafo2} which arises due to the change of the
functional integration measure has the form
\be
\Det\left(\frac{\delta(\vp+\ep T)(x)}{\delta\vp(y)}\right)=
\Det\left(\delta^4(x-y)+\ep \frac{\delta T(x)}{\delta \vp(y)}\right).
\label{jade}\ee
With $T^L$ being the part of $T$ which is linear in $\vp$ and with $T^{NL}$
being the part, which contains the quadratic,
cubic, etc.\ terms, one can rewrite \eref{jade} as
\be
\Det\left(\delta^4(x-y)+\ep \frac{\delta T^{NL}(x)}
{\delta \vp(y)}+O(\ep^2)\right)
\Det\left(\delta^4(x-y)+\ep \frac{\delta T^L(x)}{\delta \vp(y)}\right).
\label{jade2}\ee
Due to the definition of $T^L$, the second determinant in \eref{jade2} is
constant and can thus be neglected. Applying \eref{ghosts},  the first one
yields a ghost term
\be
\lag_{ghost}=-\eta^\ast\eta-\ep\eta^\ast
\frac{\pa T^{NL}}{\pa \vp}\eta+O(\ep^2).\label{ggg}
\ee
$\lag_{ghost}$ contains a mass term for the ghosts and couplings to the scalar
field but no kinetic term. Therefore, using a similar
argumentation as in section~2.2 one
finds that all ghost loops are (at least)
quartically divergent and thus the term
\eref{ggg} can be replaced by a $\delta^4(0)$-term.

The HLE theorem implies, that an {\em effective\/} higher-order
Lagrangian can be quantized in the same way as
a first-order one without worrying about the
unphysical effects \cite{hide} that are
normally connected with higher-order Lagrangians.
In particular, the Feynman rules can be obtained in the standard manner
from the effective interaction term.

Closing this section, I want to add several remarks:
\begin{itemize}
\item
By repeating the above procedure, one can iteratively
eleminate the higher derivatives in
any finite order of $\ep$ \cite{eom,pol}. Thus, remembering the discussion
at the beginning of chapter~4, one can see that also in the higher-order case
the HLE theorem can in principle be proven in an arbitrary order of $\ep$.
However, by this iterative procedure one cannot eliminate the
higher derivatives in all orders of $\ep$.
\item
The procedure described in this section can only be applied
to {\em effective\/} Lagrangians, since then the supposed existence
of well-behaved ``new physics'' beyond the theory described by the
effective Lagrangian justifies the omission of all
unphysical effects. The above proof rests on the fact
that higher derivative terms proportional to higher powers of
$\ep$ are neglected (see the previous item). The assumption $\ep\ll 1$
{alone} is not sufficient for neglecting these terms since
theories with higher derivatives have no analytic limit
for $\ep\to 0$. Thus, the effects of a term with higher
derivatives are not small even if the coupling constant of this
term is extremely small \cite{hide}.
However, as mentioned in the introduction, the neglect of these terms
is justified when considering an effective Lagrangian
because effects implied by $O(\epsilon^n)$-terms with $n>1$
are assumed to be cancelled by other effects of (well-behaved) ``new physics''.
This means that the unphysical
features cannot be eliminated within models with higher
derivatives that are {\em not\/} considered to be effective ones.
\item In principle, the Ostrogradsky
formalism itself is a reduction of
a higher-order Lagrangian to a first-order one\footnote{For a formal proof of
this see \cite{pons}.} because the higher
derivatives are considered to be independent coordinates;
however, this means that new degrees of
freedom are introduced. These additional degrees of
freedom involve the unphysical effects \cite{hide}.
Here, an effective higher-order Lagrangian is reduced to a first-order one
{\em without\/} introducing extra degrees of freedom and thus the unphysical
effects are eliminated.
\item The use of the EOM \eref{kgeom} may, in general, yield
expressions  in $\lag_{red}$
which are not manifestly Lorentz invariant (see especially the
treatment of vector fields in the following section).
Also the Hamiltonian PI quantization
procedure implies such
terms (see \cite{fad,bedu,sen,const,gity} and chapter~4). However, these
expressions only occur in intermediate steps
of the derivation but not
in the resulting PI \eref{lpi} with \eref{eq}.
Actually, the HLE theorem enables calculations based
on the manifestly Lorentz invariant Lagrangian PI.
\item Since each effective Lagrangian with higher derivatives
can be reduced to a first-order one, it is in
principle sufficient to consider only effective Lagrangians with
at most first derivatives \cite{eom,pol,ruj,buwy}. However,
the reduction of a quite simple higher-order effective
interaction term to a first-order term
by applying the equations of motion, in general,
yields a lengthy and awkward expression, which is
a linear combination of several terms; each one alone of these terms
yields effects that are not implied by the primordial term
but among them complicated cancellations take place \cite{ruj,gkks}.
This means the physical effects of
such a higher-order term are quite untransparent after the application of the
EOM. Thus, for practical calculations
it is much more convenient to use the Feynman rules deriving from the
the primordial higher-order Lagrangian instead of those implied by
the reduced first-order Lagrangian. Therefore,
I have used this reduction only for technical purposes in order to apply
the HLE theorem for first-order Lagrangians, but in the final result
\eref{lpi} with \eref{eq} I have reconstructed
the original higher-order Lagrangian. Actually, this result
enables calculations based on a higher-order effective
Lagrangian without doing this reduction.
\end{itemize}


\section[Massive Vector Fields]{Higher-Order Effective Interactions of
Massive Vector Fields}
In this section I will extend the results of the preceding
one to higher-order effective (non Yang--Mills)
self-interactions of massive vector fields. I will again examine
the three different types of effective Lagrangians which are found in
the literature, namely gauge noninvariant Lagrangians,
gauged nonlinear $\sigma$-models and Higgs models. As in chapter~4
I will for simplicity only consider
massive Yang--Mills theories (with additional effective
interactions) in which all vector bosons have equal
masses and the corresponding SBGTs.


\subsection{Gauge Noninvariant Models}
I consider the effective Lagrangian
\be
\lag=\lag_0+\ep\lag_I=-\frac{1}{4}
F^{\mu\nu}_aF_{\mu\nu}^a+\frac{1}{2}M^2A_a^\mu
A_\mu^a+\ep\lag_I (A_a^\mu,\pa^\nu A_a^\mu,
\, \ldots\, , \pa^{\nu_1}\cdots\pa^{\nu_N}A_a^\mu).
\label{ymeff}
\ee
$\lag_0$ represents a massive Yang--Mills theory and  the effective
interaction term $\lag_I$ contains the deviations from the
Yang--Mills interactions which involve derivatives up to the order
$N$ and which are proportional to $\ep$ with $\ep\ll 1$.

By applying the procedure described in section~5.2 one
can now use the EOM following from
$\lag_0$  in order to eliminate the higher
time derivatives in
$\lag_I$. The EOM are:
\be
D_\mu F^{\mu\nu}_a=-M^2A_a^\nu
\label{ymeom}
\ee
with the covariant derivative
\be
D^\sigma F^{\mu\nu}_a\equiv \pa^\sigma F^{\mu\nu}_a
-gf_{abc}A^\sigma_b F^{\mu\nu}_c.
\label{df}\ee
For $\nu=i=1,2,3$ \eref{ymeom} can be written as
\be
\ddot{A}^i_a=-M^2A^i_a-D_jF^{ij}_a-gf_{abc}A_b^0F^{i0}_c
-\pa_i\dot{A}_a^0-gf_{abc}(\dot{A}_b^iA_c^0
+A^i_b\dot{A}_c^0).
\label{ddai}
\ee
This equation serves to eliminate all higher
time derivatives of the
$A^i_a$. Next one has to get rid of the time derivatives of the
$A^0_a$.
To be able to apply the HLE theorem for first-order Lagrangians
derived in section~4.1, one even has to remove the
first time derivatives
of the $A_a^0$ because in section~4.1 it is
assumed that $\lag$ does not depend on $\dot{A}_a^0$ (see
footnote~2 there).
For $\nu=0$ \eref{ymeom} becomes
\be
A_a^0=\frac{1}{M}[-\pa_i F^{i0}_a-gf_{abc}A_b^i F^{i0}_c].
\label{a0}
\ee
Differentiation yields
\be
\dot{A}_a^0=\frac{1}{M}[{}-\pa_i\dot{F}^{i0}_a-
gf_{abc}(\dot{A}_b^i F^{i0}_c+A_b^i \dot{F}^{i0}_c)],
\label{da0}
\ee
where $\dot{F}^{i0}_a$ can be written (using \eref{ddai}) as
\be
\dot{F}^{i0}_a=M^2 A^i_a+ D_jF^{ij} + gf_{abc} A^0_b F^{i0}_c.
\label{dfi0}\ee
By repeated application of \eref{ddai}, \eref{da0} and \eref{dfi0}
one can reduce the effective Lagrangian \eref{ymeff} to an
equivalent Lagrangian $\lag_{red}$ which contains neither higher
time derivatives of the fields
nor first time derivatives of the $A_a^0$.

Now the HLE theorem can be proven as in the previous section. The
effective higher-order Lagrangian becomes reduced to a first-order
Lagrangian $\lag_{red}$ as described above,
this gets quantized by applying the HLE
theorem for effective first-order interactions of massive vector
fields (sect.~4.1)  and finally one can reconstruct
the primordial Lagrangian $\lag$
by making the appropriate field transformations in the Lagrangian PI.


\subsection{Gauged Nonlinear $\bf \sigma$-Models}
Now I consider SBGTs with additional effective interaction terms.
First I restrict myself to gauged nonlinear $\sigma$-models;
in these the unphysical pseudo-Goldstone fields $\vp_a$
are nonlinearly parametrized as \eref{mu} with \eref{mphi}.

Each effective gauged nonlinear $\sigma$-model can be
rewritten in the gauge noninvariant form
\eref{ymeff} by applying
the inverse of the Stueckelberg transformation \eref{stue1}.
This can be seen as follows: Due to the gauge invariance
of the effective interaction term, the fields only occur there
in the gauge invariant combinations
\bea &&
U^\dagger(D^{\sigma_1}\cdots D^{\sigma_N}F^{\mu\nu})U,
\label{ddF}\\&&
U^\dagger D^{\sigma_1}\cdots D^{\sigma_N}U.
\label{ddu}\eea
(The higher covariant derivatives of $F^{\mu\nu}$ and $U$
are defined in analogy to the first-order ones, \eref{DU1} and
\eref{df}.) Each effective interaction term can be constructed
from the expressions \eref{ddF},
\eref{ddu} and constants like $t_a$, $g^{\mu\nu}$ and
$\ep^{\rho\sigma\mu\nu}$
by taking products, sums, derivatives and traces \cite{apbe}.
The term \eref{ddF} becomes
\be
D^{\sigma_1}\cdots D^{\sigma_N}F^{\mu\nu}
\ee
after an inverse Stueckelberg transformation.
By using the unitarity of $U$ (\eref{mu} with \eref{mphi})
and by product differentiation, the term \eref{ddu} can be
expressed through terms
\be
U^\dagger D^\mu U \label{DU}
\ee
and their derivatives.
E.g.\ for $N=2$ it can be written as
\be
U^\dagger D^{\sigma_1}D^{\sigma_2}U=
(U^\dagger D^{\sigma_1} U)(U^\dagger D^{\sigma_2} U)
+\pa^{\sigma_1}(U^\dagger D^{\sigma_2} U).
\ee
(Similar formulas can be found for $N>2$).
\eref{DU} becomes
\be
ig A^\mu
\ee
after an inverse Stueckelberg transformation.
This means that, by applying the inverse of the
Stueckelberg transformation \eref{stue1} (and of the corresponding
transformations of the derivatives of $A^\mu$),
an arbitrary nonlinear gauge invariant Lagrangian $\lag^S$ becomes
a gauge noninvariant Lagrangian $\lag$ (U-gauge Lagrangian)
which is obtained by simply dropping all unphysical scalar fields in
$\lag^S$, i.e., $\lag$ is given by \eref{pz}.

In section~4.2 I have proven that Lagrangians
(without higher derivatives)
which are related to each other by a Stueckelberg transformation
are equivalent within the Hamiltonian formalism.
Using the Ostrogradsky formalism
this result can be generalized to higher-order Lagrangians
and besides it can be derived more easily.
Since a Stueckelberg transformation
is a field transformation that depends on the derivatives of the
fields, it is a  canonical
transformation within this formalism\footnote{One may wonder
why a gauge invariant (i.e.\ first-class constrained) system
can be related to a gauge noninvariant (i.e.\
second-class constrained)
system by a canonical transformation, because the number of second-class
constraints is given by ${\rm rank}\,\{\phi_a,\phi_b\}|_
{\scriptstyle \phi_a=0}$ which is invariant under canonical
transformations. One should remember that these systems
are only related by a canonical transformation if their order is
artificially increased. This procedure yields additional constraints
(see section~5.1). In fact, $\lag^S$ and $\lag$ imply equal numbers
of first-class and of second-class constraints if their order is chosen
sufficiently high.}
(if the formal
order of $\lag^S$
is chosen sufficiently high) as discussed in section~5.1.
Since the time derivatives of $A^\mu$
are formally considered to be independent fields, the
inverse Stueckelberg transformation is completely specified by
\be
\pa^n_0A^\mu
\to\pa^n_0\left(U A^\mu U^\dagger
-\frac{i}{g}U\pa^\mu U^\dagger\right)
,\qquad n=0,\ldots,N
\label{stue2}
\ee
(with $N$ being the
order of the highest time derivative occurring in $\lag^S$).
The Lagrangian $\lag^S$ has thus formally to be treated as an
$(N+3)$rd-order one to establish the equivalence of
$\lag^S$ and $\lag$.

The HLE theorem for effective
gauged nonlinear $\sigma$-models with higher
derivatives can now be proven as follows:
Given a nonlinear gauge invariant Lagrangian $\lag^S$, it can be
converted into the  equivalent gauge noninvariant U-gauge Lagrangian $\lag$
\eref{pz} as described above. This does
not affect the Hamiltonian path integral (see section~5.1).
Since $\lag$ is of the type
\eref{ymeff}, the results of the previous subsection can be applied
to quantize it. In analogy to the treatment of section~4.2, this yields
a generating functional which is equal to the one
obtained in the (Lagrangian) Faddeev--Popov
formalism if the g.f. conditions \eref{pgf} are imposed.
Due to the equivalence of all gauges \cite{lezj,able}, this result
can be rewritten in any other gauge. This yields \eref{lpi} with \eref{fp}.


\subsection{Higgs Models}
Finally let me consider effective SBGTs with linearly realized symmetry.
As in sect.~4.3 I
restrict myself to the case of SU(2) symmetry (i.e.\ $t_a=\frac{1}{2}\tau_a$,
$a=1,2,3$).

With the same arguments as in sections~4.3 and 5.3.2 one finds that,
like in the first-order case, the Lagrangian $\lag^H$ of an effective
Higgs model is equivalent to its U-gauge Lagrangian $\lag_U^H$ given by
\eref{lu}.
$\lag^H_U$ is a gauge noninvariant Lagrangian such as \eref{ymeff} but it
contains the additional physical scalar field $h$. Therefore it
has the form
\bea
\lag^H_U&=&\lag_0+\ep\lag_I\nn&=&-\frac{1}{4}
F^{\mu\nu}_aF_{\mu\nu}^a+\frac{1}{2}(\pa^\mu h)(\pa_\mu h)
+\frac{1}{8}g^2 (v+h)^2A_a^\mu
A_\mu^a-V(h,\vp_a=0)
\nn&&{}
+\ep\lag_I (A_a^\mu,\pa^\nu A_a^\mu,
\, \ldots\, , \pa^{\nu_1}\cdots\pa^{\nu_N}A_a^\mu
,h,\pa^\mu h,\,\ldots\, ,\pa^{\mu_1}\cdots\pa^{\mu_N}h).
\label{aaa}\eea
($\lag_0$ is the U-gauge Lagrangian of the Higgs model without
effective interaction terms.)
The EOM corresponding to \eref{ddai}, \eref{da0}, \eref{dfi0}
and the EOM for the Higgs field implied by $\lag_0$ are
\bea
\!\!\!\!\!\!\!\!\!\!\!\!\!
\ddot{A}^i_a&=&{}-\frac{1}{4}g^2(v+h)^2
A^i_a-D_jF^{ij}_a-gf_{abc}A_b^0F^{i0}_c
-\pa_i\dot{A}_a^0-gf_{abc}(\dot{A}_b^iA_c^0
+A^i_b\dot{A}_c^0),\\\!\!\!\!\!\!\!\!\!\!\!\!\!
\dot{A}_a^0&=&\frac{4}{g^2(v+h)^2}\left[-\frac{1}{2}
g^2(v+h)\dot{h}A^0_a-\pa_i\dot{F}^{i0}_a-
gf_{abc}(\dot{A}_b^i F^{i0}_c+A_b^i
\dot{F}^{i0}_c)\right],\label{nonpol}\\\!\!\!\!
\!\!\!\!\!\!\!\!\!
\dot{F}^{i0}_a&=& \frac{1}{4}g^2(v+h)^2
A^i_a+ D_jF^{ij} + gf_{abc} A^0_b F^{i0}_c,
\\\!\!\!\!\!\!\!\!\!\!\!\!\!
\ddot{h}&=&\Delta h
+\frac{1}{4}g^2(v+h)A_a^\mu A^a_\mu-\frac{\partial}{\partial h}V(h,\vp_a=0)
\eea
These EOM can be used in order to eliminate all higher time
derivatives of the $A^\mu_a$ and of $h$
and also the first time derivatives of the $A^0_a$ in
$\lag^H_U$ \eref{aaa}.

On the basis of this result, one can easily
prove the HLE theorem for effective Higgs models by
using the same arguments as in sections~4.3 and 5.3.2.

It should be mentioned
that the transformation \eref{trafo} and
the EOM \eref{nonpol}
involve nonpolynomial interactions, which are not
present in the primordial Lagrangian. This, however, is no serious
problem, since these expressions only
occur in intermediate steps of the derivation
and not in the resulting Faddeev-Popov PI
\eref{lpi} with \eref{fp}. The
application of the HLE theorem for first-order Lagrangians is not
affected by these terms, since the treatment of chapter~4 does not
necessarily require polynomial interactions.

The derivation in this subsection illustrates that the
above proof can be extended to the case of Lagrangians with
matter fields which have been negelcted here for
simplicity (also see the treatment of chapter~6). The additional
couplings in $\lag_0$ alter the EOM, but
the general statement
that each effective higher-order Lagrangian can
be reduced to a first-order one by using the EOM remains unaffected.
Therefore, the HLE theorem holds for any
Lagrangian with arbitrary interactions of
massive vector fields.

\chapter{Effective Gauge Theories}
After the derivation of the HLE theorem for effective interactions of massive
vector fields in the preceding two chapters, I will now also consider massless
vector fields and fermion fields, which completes the proof of the HLE
theorem for the physically most important types of particles.

As pointed out in the introduction, massless vector fields necessarily have to
be understood as gauge fields; therefore I will give a general proof
of the HLE theorem for gauge theories with effective
(but still gauge invariant) gauge-boson
self-interactions and with effective couplings to fermion and scalar
fields. The result \eref{lpi} with \eref{fp}
will first be derived within the Coulomb gauge
and then be generalized to any other gauge by
using the equivalence of all gauges \cite{lezj,able}.
This proof is also valid for
SBGTs (both with linearly and with nonlinearly
realized symmetry) i.e., gauge theories in which
the gauge bosons are massive. Thus, this thesis contains two alternative
proofs of the HLE theorem for effective SBGTs, the one given in the
previous two chapters and the one presented in this chapter. Both are
conceptionally different; in the first proof an SBGT is related to a gauge
noninvariant Lagrangian (U-gauge Lagrangian) by applying the Stueckelberg
formalism (actually, the reformulation of the Stueckelberg formalism
and the construction of the U-gauge within the Hamiltonian framework
is an important result by itsself)
while the proof in this chapter is more direct and
does not use the Stueckelberg formalism. This proof also applies to SBGTs in
which not all gauge bosons are massive (like effective electroweak models
\cite{nonlin,gauge}) and, of course, to gauge theories without
spontaneous symmetry breaking (like theories with effective gluon
self-interactions \cite{gluons}).

As in the previous two chapters, I will first consider effective Lagrangians
without higher time derivatives and then I will generalize the HLE theorem
to the higher-order case by using the EOM in order to eleminate the higher time
derivatives
from the effective interaction term.
Actually, this simplifies the treatment of fermion fields very much, because,
by applying the EOM, one can eliminate not only higher but also first
time derivatives of these fields
from the effective interaction term and therefore it is sufficient to prove
the HLE theorem for effective interactions that do not involve
time derivatives of the fermion fields.

The results of this chapter have first been published in \cite{gk3}.


\section[The HLE Theorem for Effective Gauge Theories]
{The Hamilton--Lagrange Equivalence Theorem for Effective Gauge Theories}
In this section I quantize an effective gauge theory  (including fermion
and scalar fields) without higher derivatives in the
Hamiltonian PI formalism in order to derive the Faddeev--Popov PI
\eref{lpi} with (\ref{fp}). Some of the techniques used in the following
derivation originate from the quantization of a Yang--Mills theory
without effective interaction terms in \cite{gity} and some others from the
quantization of gauge noninvariant effective Lagrangians in \cite{bedu} and
in section~4.1.

The effective Lagrangian is
\begin{equation}
{\cal L}={\cal L}_0+\epsilon{\cal L}_I=-\frac{1}{4}
F_a^{\mu\nu}F^a_{\mu\nu}
+i{\bar{\psi}}_a\gamma^\mu D_\mu\psi_a+
(D^\mu\varphi_a^\dagger)(D_\mu\varphi_a)
-V(\psi_a,{\bar{\psi}}_a,\varphi_a,\varphi^\dagger_a)+
\epsilon{\cal L}_I.
\label{leff3}\end{equation}
The field strength tensor
$F_{\mu\nu}^a$ is given by \eref{fst1} and its covariant derivative
by \eref{df}. The covariant derivatives of the $\psi_a$, $\bar{\psi}_a$,
$\vp_a$ and $\vp^\dagger_a$ are:
\begin{eqnarray}
D_\mu\psi_a&\equiv&\partial_\mu\psi_a
+igA_\mu^c t^{ab}_c\psi_b,\label{Dpsi}\\
D_\mu{\bar{\psi}}_a&\equiv&\overline{(D_\mu\psi_a)},\\
D_\mu\varphi_a&\equiv&\partial_\mu\varphi_a
+igA_\mu^c \bar{t}^{ab}_c\varphi_b,\label{Dphi}\\
D_\mu\varphi^\dagger_a&\equiv&(D_\mu\varphi_a)^\dagger.
\end{eqnarray}
(Higher covariant derivatives are defined analogously.)
$g$ is the gauge coupling constant, $f_{abc}$ are the structure
constants and $t^{ab}_c$ and $\bar{t}^{ab}_c$ are the generators of
the gauge group  in its representation in the fermion sector
and in the
scalar sector respectively.
$V(\psi_a,{\bar{\psi}}_a,\varphi_a,\varphi^\dagger_a)$ contains
derivative-free interactions of the fermion and scalar fields,
viz. Yukawa couplings and scalar self-interactions.

The effective interaction term
$\epsilon{\cal L}_I$, which parametrizes the
deviations from the minimal gauge theory (given by $\lag_0$),
contains arbitrary
interactions of the fields which are governed by the effective
coupling constant
$\epsilon$ with $\epsilon\ll 1$. As pointed out in the
introduction, an effective Lagrangian like (\ref{leff3}) only has a
physical meaning if the effective interaction term is gauge
invariant. The gauge invariance implies
that {\em the gauge fields $A_\mu^a$ do not occur arbitrarily in
${\cal L}_I$ but only through the field strength
tensor and through covariant derivatives.\/} Furthermore,
in this section I assume that {\em ${\cal L}_I$ does neither
depend on higher time
derivatives of the fields nor on first time
derivatives of the $A_0^a$\footnote{\rm Actually, the absence
of $\dot{A}_0^a$ already follows from the gauge invariance
and the requirement that no
higher time derivatives occur in ${\cal L}_I$.}
and of the fermion fields $\psi_a$
and ${\bar{\psi}}_a$.\/} The case of
interactions with higher derivatives will be treated in
the next section.

{}From (\ref{leff3}) one finds the conjugate fields (generalized
momenta):
\begin{eqnarray}
\pi_0^a&=&\frac{\partial{\cal L}}{\partial\dot{A}^0_a}=0,\label{p0}\\
\pi_i^a&=&\frac{\partial{\cal L}}{\partial\dot{A}^i_a}
=F_{i0}^a+\epsilon
\frac{\partial{\cal L}_I}{\partial\dot{A}^i_a}
=\dot{A}^i_a+\partial_iA_0^a
-gf_{abc}A_i^bA_0^c+\epsilon
\frac{\partial{\cal L}_I}{\partial\dot{A}^i_a}
,\label{pi3}\\
\pi_\psi^a&=&\frac{\partial{\cal L}}
{\partial\dot{\psi}_a}=i{\bar{\psi}}_a\gamma^0
,\label{ppsi}\\
\pi_{\bar{\psi}}^a&=&\frac{\partial{\cal L}}
{\partial\dot{{\bar{\psi}}}_a}=0,\label{ppsib}\\
\pi_\varphi^a&=&\frac{\partial{\cal L}}{\partial\dot{\varphi}_a}=
D_0\varphi^\dagger_a+\epsilon
\frac{\partial{\cal L}_I}{\partial\dot{\varphi}_a}=
\dot{\varphi}_a^\dagger-igA_0^c\bar{t}_c^{ba}\varphi^\dagger_b
+\epsilon\frac{\partial{\cal L}_I}
{\partial\dot{\varphi}_a},\label{pphi}\\
\pi_{\varphi^\dagger}^a&=&
\frac{\partial{\cal L}}{\partial\dot{\varphi}_a^\dagger}=
D_0\varphi_a+\epsilon\frac{\partial{\cal L}_I}
{\partial\dot{\varphi}_a^\dagger}=
\dot{\varphi}_a+igA_0^c\bar{t}_c^{ab}\varphi_b+
\epsilon\frac{\partial{\cal L}_I}
{\partial\dot{\varphi}_a^\dagger}.\label{pphic}
\end{eqnarray}
The relations (\ref{p0}), (\ref{ppsi}) and (\ref{ppsib}) do not
contain $\epsilon$-terms due
to the assumption that ${\cal L}_I$ does not
depend on $\dot{A}_0^a$,
$\dot{\psi}_a$ and $\dot{{\bar{\psi}}}_a$. These
relations cannot be solved for the velocities; they are constraints.
The remaining of the above equations can be solved for the
velocities, they become (in first order of $\epsilon$):
\begin{eqnarray}
\dot{A}_a^i&=&\pi_i^a-\partial_i A_0^a+gf_{abc}A_i^bA_0^c
-\epsilon\frac{\partial{\cal L}_I}
{\partial\dot{A}^i_a}\Bigg|_{\begin{array}{l}
\scriptstyle F_{i0}^a\to\pi_i^a\\\scriptstyle D_0\varphi^\dagger_a
\to\pi_\varphi^a\\\scriptstyle D_0\varphi_a\to
\pi_{\varphi^\dagger}^a\end{array}}
+O(\epsilon^2),\label{da}\\
\dot{\varphi}_a^\dagger&=&\pi_\varphi^a
+igA_0^c\bar{t}^{ba}_c\varphi^\dagger_b
-\epsilon\frac{\partial{\cal L}_I}
{\partial\dot{\varphi}_a}\Bigg|_{\begin{array}{l}
\scriptstyle F_{i0}^a\to\pi_i^a\\\scriptstyle D_0\varphi^\dagger_a
\to\pi_\varphi^a\\\scriptstyle D_0\varphi_a\to
\pi_{\varphi^\dagger}^a\end{array}}
+O(\epsilon^2),\label{dphic}\\
\dot{\varphi}_a&=&\pi_{\varphi^\dagger}^a
-igA_0^c\bar{t}^{ab}_c\varphi_b
-\epsilon\frac{\partial{\cal L}_I}
{\partial\dot{\varphi}_a^\dagger}\Bigg|_{\begin{array}{l}
\scriptstyle F_{i0}^a\to\pi_i^a\\\scriptstyle D_0\varphi^\dagger_a
\to\pi_\varphi^a\\\scriptstyle D_0\varphi_a\to
\pi_{\varphi^\dagger}^a\end{array}}
+O(\epsilon^2).\label{dphi}
\end{eqnarray}
One finds the Hamiltonian
\begin{eqnarray}
{\cal H}&=&\pi_\mu^a\dot{A}_a^\mu+\pi_\psi^a
\dot{\psi}_a+\dot{{\bar{\psi}}}_a\pi_{\bar{\psi}}^a+
\pi_\varphi^a\dot{\varphi}_a+\pi_{\varphi^\dagger}^a\dot{\varphi}_a
^\dagger-{\cal L}\nonumber\\
&=&\frac{1}{2}\pi_i^a\pi_i^a-\pi_i^a\partial_i A_0^a
+gf_{abc}\pi_i^a A_i^b A_0^c+\frac{1}{4}F_{ij}^aF_{ij}^a
\nonumber\\&&
-igA_0^c t^{ab}_c(\pi_\psi^a \psi_b-{\bar{\psi}}_a\pi_{\bar{\psi}}^b)
+i{\bar{\psi}}_a\gamma_iD_i\psi_a\nonumber\\
&&+\pi_{\varphi^\dagger}^a\pi_
\varphi^a-igA_0^c\bar{t}_c^{ab}(\pi_\varphi^a\varphi_b
-\varphi^\dagger_a\pi_{\varphi^\dagger}^b)
+(D_i\varphi_a^\dagger)(D_i\varphi_a)+V\nonumber\\
&&-\epsilon\bar{{\cal L}}_I
(A_i^a,\psi_a,{\bar{\psi}}_a,\varphi_a,\varphi^\dagger_a,\pi_i^a,
\pi_\varphi^a,\pi_{\varphi^\dagger}^a)+O(\epsilon^2)
\label{ham3}\end{eqnarray}
with
\begin{equation} \bar{{\cal L}}_I
(A_i^a,\psi_a,{\bar{\psi}}_a,\varphi_a,\varphi^\dagger_a,\pi_i^a,
\pi_\varphi^a,\pi_{\varphi^\dagger}^a)
\equiv{\cal L}_i\Bigg|_{\begin{array}{l}
\scriptstyle F_{i0}^a\to\pi_i^a\\\scriptstyle D_0\varphi^\dagger_a
\to\pi_\varphi^a\\\scriptstyle D_0\varphi_a\to
\pi_{\varphi^\dagger}^a\end{array}}.
\label{lib3}\end{equation}

One can use the identities
\begin{eqnarray}
[D_\mu ,D_\nu ]\psi_a&=&igF^c_{\mu\nu}t^{ab}_c
\psi_b,\label{compsi}\\
{}[D_\mu, D_\nu ]\varphi_a&=&igF^c_{\mu\nu}\bar{t}^{ab}_c
\varphi_b,\label{comphi}\\
{}[D_\mu,D_\nu]F^a_{\kappa\lambda}&=&-
gf_{abc}F^b_{\mu\nu}F^c_{\kappa\lambda}
\label{com}\end{eqnarray}
(and the corresponding relations
for ${\bar{\psi}}_a$ and $\varphi^\dagger_a$) in order
to rewrite those expressions in ${\cal L}_I$,
in which time and spatial covariant derivatives
act on the fields,
such that the time derivatives are applied first.
Remembering the discussion of the paragraph preceding equation
(\ref{p0}) one can then easily
see that ${\cal L}_I$ depends on the fields $A_0^a$ only
through the expressions
\begin{equation}
F_{i0}^a,
\qquad D_0F_{ij}^a,\qquad D_0\varphi_a,\qquad D_0\varphi^\dagger_a
.\label{a0terms}\end{equation}
Using the Bianchi identity
\begin{equation}
D_\lambda F_{\mu\nu}^a+D_\mu F_{\nu\lambda}^a
+D_\nu F_{\lambda\mu}^a=0
\label{homeq}\end{equation}
in order to rewrite $D_0F_{ij}^a$ as
\begin{equation}
D_0F_{ij}^a=D_jF_{i0}^a-D_i F_{j0}^a
\label{d0fij}\end{equation}
and the definition (\ref{lib3})
one finds that $\bar{{\cal L}}_I$ does not depend
on the $A_0^a$. Thus, the gauge invariance and the absence of
higher time derivatives (and of first time derivatives
of $A_0^a$, $\psi_a$ and ${\bar{\psi}}_a$) in ${\cal L}_I$ yields
\begin{equation}
\frac{\partial\bar{{\cal L}}_I}{\partial A_0^a}=0.
\label{noa0}\end{equation}

As mentioned above, the relations (\ref{p0}), (\ref{ppsi}) and
(\ref{ppsib})
imply the primary constraints
\begin{eqnarray}
\phi_1^a&=&\pi_0^a=0,\label{pc3}\\
\phi_\psi^a&=&\pi_\psi^a-i{\bar{\psi}}_a\gamma^0=0,\label{pcpsi}\\
\phi_{\bar{\psi}}^a&=&\pi_{\bar{\psi}}^a=0.\label{pcpsib}
\end{eqnarray}
The requirement \eref{eomconsp} in general
yields secondary  constraints (see sect.~3.1). Actually, (\ref{pcpsi}) and
(\ref{pcpsib}) do not imply secondary constraints, since the demand
(\ref{eomconsp}) only determines
the Lagrange multipliers corresponding to
these constraints \cite{gity}. The constraints (\ref{pc3})
imply the secondary constraints
\begin{equation}
\phi_2^a=\partial_i\pi_i^a+gf_{abc}\pi_i^b A_i^c
-igt_a^{bc}(\pi_\psi^b\psi_c-{\bar{\psi}}_b\pi_{\bar{\psi}}^c)
-ig\bar{t}^{bc}_a(\pi_\varphi^b
\varphi_c-\varphi^\dagger_b\pi_{\varphi^\dagger}^c)=0.
\label{sc3}\end{equation}
Due to (\ref{noa0}),
these secondary constraints do not contain $O(\epsilon)$-terms,
i.e.\, they are independent of the form of
the effective interaction term ${\cal L}_I$
(in first order of $\epsilon$).
There are no tertiary, etc.\ constraints.
The Poisson brackets
\bea
\{\phi_\psi^a(\vx),\phi_{\psi}^b(\vy)\}&=&\{\phi_{\bar{\psi}}^a
(\vx),\phi_{\bar{\psi}}^b(\vy)\}=0\label{xxx12}\\
\{\phi_\psi^a(\vx),\phi_{\bar{\psi}}^b(\vy)\}&=&-i\gamma^0\delta^{ab}\dd
\label{scpb}\eea
imply that the constraints $\phi_\psi^a$ (\ref{pcpsi})
and $\phi_{\bar{\psi}}^a$ (\ref{pcpsib}) are second-class
and the Poisson brackets
\bea
\{\phi_1^a(\vx),\phi_1^b(\vy)\}&=&\{\phi_1^a(\vx),\phi_2^b(\vy)\}
=\{\phi_1^a(\vx),\phi_\psi^b(\vy)\}=
\{\phi_1^a(\vx),\phi_{\bar{\psi}}^b(\vy)\}=0,\\
\{\phi_2^a(\vx),\phi_2^b(\vy)\}&=&gf_{abc}\phi_2^c(\vx)\dd,\\
\{\phi_2^a(\vx),\phi_\psi^b(\vy)\}&=&-igt^{cb}_a\phi_\psi^c(\vx)\dd,\\
\{\phi_2^a(\vx),\phi_{\bar{\psi}}^b(\vy)\}&
=& igt^{cb}_a\phi_{\bar{\psi}}^c(\vx)\dd\eea
imply that
the constraints $\phi_1^a$ (\ref{pc3}) and $\phi_2^a$ (\ref{sc3})
are first-class (see sect.~3.1).

Due to the presence of the first-class constraints one has to introduce
an equal number of gauge-fixing conditions (as discussed in section~3.1).
Following the reasoning of section~4.2,  one finds that the usual
Lorentz g.f.\ conditions
\begin{equation}
\chi_1^a=\partial^\mu A^a_\mu-C^a=0
\label{lg}\end{equation}
(and also the $\rm R_\xi$-g.f. conditions for SBGTs) are not g.f.
conditions within the Hamiltonian formalism
\cite{fad,sen,const,gity}
because they are not
relations among the fields and the conjugate fields alone. Therefore
I quantize the effective gauge theory \eref{leff3}
within the Coulomb gauge,
i.e.\ by choosing the primary g.f. conditions
\begin{equation}
\chi_1^a=\partial^i A^a_i-C^a=0.
\label{pgf3}\end{equation}
(Instead of the
Coulomb gauge, one can alternatively choose the axial gauge or, for
SBGTs, the U-gauge as in section~4.2).
The demand \eref{gfcons} yields the secondary g.f.\ conditions
\begin{equation}
\chi_2^a=\Delta A_0^a-\partial_i\pi_i^a
-gf_{abc}\partial_i(A_i^bA_0^c)
+\epsilon\partial_i\frac{\partial\bar{{\cal L}}_I}
{\partial\pi_i^a}=0.
\label{sgf3}\end{equation}

The Hamiltonian path intergral \eref{hpi} for this system is
\begin{eqnarray}
Z& =&\int{\cal D} A_\mu^a{\cal D}\psi_a
{\cal D}{\bar{\psi}}_a{\cal D} \varphi_a{\cal D}\varphi^\dagger_a
{\cal D} \pi_\mu^a{\cal D}\pi_\psi^a
{\cal D}\pi_{\bar{\psi}}^a{\cal D}
\pi_\varphi^a{\cal D}\pi_{\varphi^\dagger}^a\,
\nonumber\\&&\times \exp\left\{i\int d^4x\,\left[\pi_\mu^a
\dot{A}_a^\mu+\pi_\psi^a\dot{\psi}_a+
\dot{{\bar{\psi}}}_a\pi_{\bar{\psi}}^a+
\pi_\varphi^a\dot{\varphi}_a+\pi_{\varphi^\dagger}^a
\dot{\varphi^\dagger}_a-{\cal H}\right]\right\}
\nonumber\\&&
\times\delta(\phi_\psi^a)\delta(\phi_{\bar{\psi}}^a)\delta(\phi_1^a)
\delta(\phi_2^a)\delta(\chi_1^a)\delta(\chi_2^a)\nonumber\\&&
\times{\rm Det^{\frac{1}{2}}}\,(\{\phi_{2nd}^a({\bf x}),
\phi_{2nd}^b({\bf y})\}\delta(x^0-y^0))
\, {\rm Det}\,(\{\phi_{1st}^a({\bf x}),
\chi^b({\bf y})\}\delta(x^0-y^0))
\label{hpi3}\end{eqnarray}
(where $\phi^a_{2nd}=(\phi_\psi^a,\phi_{\bar{\psi}}^a)$,
$\phi^a_{1st}=(\phi_1^a,\phi_2^a)$ and $\chi^a=(\chi_1^a,\chi_2^a)$ denote
{\em all\/} second-class constraints, first-class constraints
and g.f.\ conditions respectively).
First let me consider the determinants
occurring in (\ref{hpi3}). Equations \eref{xxx12} and \eref{scpb} imply
\begin{equation}
\mbox{Det}^{\frac{1}{2}}\,
(\{\phi_{2nd}^a({\bf x}),\phi_{2nd}^b({\bf y})\}\delta(x^0-y^0))
=\mbox{constant}.
\end{equation}
Therefore, this term can be neglected in the PI. Furthermore, one
finds
\begin{eqnarray}
\{\phi_1^a({\bf x}),\chi_1^b({\bf y})\}
&=&0,\\ \!\!\!\!\!\!\!\!\! - \{\phi_1^a({\bf x}),\chi_2^b({\bf y})\}
= \{\phi_2^a({\bf x}),\chi_1^b({\bf y})\}
&=&(\delta_{ab}\Delta+gf_{abc}(\partial_i A_i^c)+
gf_{abc}A_i^c\partial_i)\delta^3({\bf x}-{\bf y}).\label{det21}
\label{const}\end{eqnarray}
The absence of $O(\epsilon)$-terms in (\ref{det21})
is again a consequence of (\ref{noa0}). This yields
\begin{equation}
\, {\rm Det}\,(\{\phi_{1st}^a({\bf x}),
\chi^b({\bf y})\}\delta(x^0-y^0))
=\,\mbox{Det}^2\,[(\delta_{ab}\Delta+gf_{abc}(\partial_i A_i^c)+
gf_{abc}A_i^c\partial_i)\delta^4(x-y)].
\label{ghostdet}\end{equation}
Next, one observes that ${\cal H}$ (\ref{ham3})
contains a term $A_0^a\phi_2^a$. Due to the presence of
$\delta(\phi_2^a)$ in the PI
this term can be omitted. Then one integrates over
$\pi_0^a$, $\pi_\psi^a$ and $\pi_{\bar{\psi}}^a$ and finds
\begin{eqnarray}
Z& =&\int{\cal D} A_\mu^a{\cal D}\psi_a
{\cal D}{\bar{\psi}}_a{\cal D} \varphi_a{\cal D}\varphi^\dagger_a
{\cal D} \pi_i^a{\cal D}\pi_\varphi^a
{\cal D}\pi_{\varphi^\dagger}^a\,\nonumber\\&&
\times\exp\Bigg\{i\int d^4x\,\Bigg[-\frac{1}{2}\pi_i^a\pi_i^a
+\pi_i^a\dot{A}^i_a-\frac{1}{4}F_{ij}^aF_{ij}^a\nonumber\\&&\quad
+i{\bar{\psi}}_a\gamma^0\dot{\psi}_a-i\psi_a\gamma_i D_i\psi_a
-\pi_\varphi^a\pi_{\varphi^\dagger}^a+
\pi_\varphi^a\dot{\varphi}_a+\pi_{\varphi^\dagger}^a\dot{\varphi}_a
^\dagger-(D_i\varphi_a^\dagger)
(D_i\varphi_a)-V\nonumber\\&&\quad+\epsilon
\bar{{\cal L}}_I(A_i^a,\psi_a,
{\bar{\psi}}_a,\varphi_a,\varphi^\dagger_a,\pi_i^a,
\pi_\varphi^a,\pi_{\varphi^\dagger}^a)\Bigg]\Bigg\}
\nonumber\\&&\times
\delta(\tilde{\phi}_2^a)\delta(\chi_1^a)
\delta(\chi_2^a)
\, {\rm Det}\,(\{\phi_{1st}^a({\bf x}),
\chi^b({\bf y})\}\delta(x^0-y^0))\label{pih1}
\end{eqnarray}
with
\begin{equation}
\tilde{\phi}_2^a=\partial_i\pi_i^a+gf_{abc}\pi_i^b A_i^c +
g{\bar{\psi}}^b\gamma^0t_a^{bc}\psi_c-ig\bar{t}_a^{bc}
(\pi_\varphi^b\varphi_c-\varphi^\dagger_b\pi_{\varphi^\dagger}^c)=0.
\label{sct}\end{equation}
After rewriting
\begin{equation}
\delta(\chi_2^a)=\delta(A_0^a-\tilde{A}_0^a)\,\mbox{Det}^{-1}
[(\delta_{ab}\Delta+gf_{abc}(\partial_i A_i^c)+
gf_{abc}A_i^c\partial_i)\delta^4(x-y)]
\label{deti}\end{equation}
(where $\tilde{A}_0^a$ is the solution of the differential
equation (\ref{sgf3})\footnote{One should not be confused
by the occurrence of the nonlocal expression $\tilde{A}_0^a$. Since the
integrand of the PI \eref{pih1} is independent of $A_0^a$ except for
the factor $\delta(A_0^a-\tilde{A}_0^a)$, the $\tilde{A}_0^a$ become
eliminated from the PI after the integration over $A_0^a$.}
with the boundary condition
that $\tilde{A}_0^a$ vanishes for $|{\bf x}|\to\infty$)
one can also integrate over $A_0^a$. Due to (\ref{noa0}),
the argument of the determinant in (\ref{deti})
does not contain $O(\epsilon)$-terms and,
besides, the integration over
$A_0^a$ does not affect $\bar{{\cal L}}_I$.
Next one reintroduces the variables $A_0^a$ by using
\begin{equation}
\delta(\tilde{\phi}_2^a)=
\int{\cal D} A_0^a\,\exp\left\{-i\int d^4x\, A_0^a\tilde{\phi}_2^a
\right\}
\end{equation}
and gets
\begin{eqnarray}
Z& =&\int{\cal D} A_\mu^a{\cal D}\psi_a
{\cal D}{\bar{\psi}}_a{\cal D} \varphi_a{\cal D}\varphi^\dagger_a
{\cal D} \pi_i^a{\cal D}\pi_\varphi^a
{\cal D}\pi_{\varphi^\dagger}^a\,\nonumber\\&&
\times\exp\Bigg\{i\int d^4x\,\Bigg[-\frac{1}{2}\pi_i^a\pi_i^a
+\pi_i^a F_{i0}^a
-\frac{1}{4}F_{ij}^aF_{ij}^a\nonumber\\&&\quad
+i\psi_a\gamma^\mu D_\mu\psi_a
-\pi_\varphi^a\pi_{\varphi^\dagger}^a+\pi_\varphi^a D_0\varphi_a
+\pi_{\varphi^\dagger}^a D_0\varphi^\dagger_a
-(D_i\varphi_a^\dagger)(D_i\varphi_a)
-V\nonumber\\&&\quad+\epsilon
\bar{{\cal L}}_I(A_i^a,\psi_a,{\bar{\psi}}_a,
\varphi_a,\varphi^\dagger_a,\pi_i^a,
\pi_\varphi^a,\pi_{\varphi^\dagger}^a)\Bigg]\Bigg\}
\nonumber\\&&\times
\delta(\partial^iA_i^a-C^a)
\, {\rm Det}\,[(\delta_{ab}\Delta+gf_{abc}(\partial_i A_i^c)+
gf_{abc}A_i^c\partial_i)\delta^4(x-y)].
\end{eqnarray}
In order to obtain expressions quadratic in
the momenta, one rewrites
this as
\begin{eqnarray}
Z& =&\int{\cal D} A_\mu^a{\cal D}\psi_a
{\cal D}{\bar{\psi}}_a{\cal D} \varphi_a{\cal D}\varphi^\dagger_a
\nonumber\\&&
\times
\exp\left\{i\epsilon\int d^4x\,\bar{{\cal L}}_I\left(A_i^a,\psi_a,
{\bar{\psi}}_a,\varphi_a,\varphi^\dagger_a,
\frac{\delta}{i\delta K_i^a},\frac{\delta}{i\delta K_\varphi^a},
\frac{\delta}{i\delta K_{\varphi^\dagger}^a}\right)\right\}
\nonumber\\&&
\times\int{\cal D}\pi_i^a{\cal D}\pi_\varphi^a{\cal D}
\pi_{\varphi^\dagger}^a
\exp\Bigg\{i\int d^4x\,\Bigg[-\frac{1}{2}\pi_i^a\pi_i^a
+\pi_i^a F_{i0}^a
-\frac{1}{4}F_{ij}^aF_{ij}^a\nonumber\\&&\quad
+i\psi_a\gamma^\mu D_\mu\psi_a
-\pi_\varphi^a\pi_{\varphi^\dagger}^a+\pi_\varphi^a D_0\varphi_a
+\pi_{\varphi^\dagger}^a D_0\varphi^\dagger_a
-(D_i\varphi_a^\dagger)(D_i\varphi_a)
-V\nonumber\\&&\quad +K_i^a\pi_i^a+K_\varphi^a\pi_\varphi^a
+K_{\varphi^\dagger}^a\pi_{\varphi^\dagger}^a\Bigg]\Bigg\}
\Bigg|_{\scriptstyle K_i^a=K_\varphi^a=
K_{\varphi^\dagger}^a=0}\nonumber\\&&\times
\delta(\partial^iA_i^a-C^a)
\, {\rm Det}\,[(\delta_{ab}\Delta+gf_{abc}(\partial_i A_i^c)+
gf_{abc}A_i^c\partial_i)\delta^4(x-y)].
\end{eqnarray}
Now one can carry out the Gaussian integrations over the momenta. With
${\cal L}_0$ given in (\ref{leff3}) one finds
\begin{eqnarray}
Z& \!=\!&\int{\cal D} A_\mu^a{\cal D}\psi_a
{\cal D}{\bar{\psi}}_a{\cal D} \varphi_a{\cal D}\varphi^\dagger_a\,
\exp\left\{i\int d^4x\,{\cal L}_0\right\}
\nonumber\\&&\times
\exp\left\{i\epsilon\int d^4x\,
\bar{{\cal L}}_I\left(A_i^a,\psi_a,{\bar{\psi}}_a,
\varphi_a,\varphi^\dagger_a,
\frac{\delta}{i\delta K_i^a},\frac{\delta}{i\delta K_\varphi^a},
\frac{\delta}{i\delta K_{\varphi^\dagger}^a}\right)\right\}
\nonumber\\&&
\times\exp\left\{i\int d^4x\,\left[\frac{1}{2}K_i^aK_i^a
+K_\varphi^a K_{\varphi^\dagger}^a+
K_i^a F_{i0}^a+K_\varphi^aD_0\varphi^\dagger_a+
K_{\varphi^\dagger}^aD_0\varphi_a\right]\right\}
\Bigg|_{\scriptstyle K_i^a=K_\varphi^a
=K_{\varphi^\dagger}^a=0}\nonumber\\&&
\times\delta(\partial^iA_i^a-C^a)
\, {\rm Det}\,[(\delta_{ab}\Delta+gf_{abc}(\partial_i A_i^c)+
gf_{abc}A_i^c\partial_i)\delta^4(x-y)].
\end{eqnarray}
This expression can be simplified in complete analogy to the
procedure described in section~4.1.
One finds (by neglecting $O(\epsilon^2)$- and
$\delta^4(0)$-terms)
\begin{eqnarray}
Z& =&\int{\cal D} A_\mu^a{\cal D}\psi_a
{\cal D}{\bar{\psi}}_a{\cal D} \varphi_a{\cal D}\varphi^\dagger_a\,
\exp\left\{i\int d^4x\,[{\cal L}_0+
\epsilon\tilde{{\cal L}}_I]\right\}
\nonumber\\&&\times
\delta(\partial^iA_i^a-C^a)
\, {\rm Det}\,[(\delta_{ab}\Delta+gf_{abc}(\partial_i A_i^c)+
gf_{abc}A_i^c\partial_i)\delta^4(x-y)],
\label{fpcg}\end{eqnarray}
where $\tilde{{\cal L}}_I$ turns out to be
\begin{equation}
\tilde{{\cal L}}_I=\bar{{\cal L}}_I\Bigg|_
{\begin{array}{l}
\scriptstyle \pi_i^a\to F_{i0}^a\\\scriptstyle \pi_\varphi^a\to
D_0\varphi^\dagger_a\\\scriptstyle \pi_{\varphi^\dagger}^a
\to D_0\varphi_a\end{array}}
={\cal L}_I.
\label{a}\end{equation}
(\ref{fpcg}) with (\ref{a})
is identical to the result obtained within
the (Lagrangian) Faddeev--Popov formalism
by choosing the (Coulomb) g.f.\ conditions $\chi_1^a$
(\ref{pgf3}) because the change of $\chi_1^a$ under
variations of the gauge parameters $\alpha_a$ is
\begin{equation}
\frac{\delta\chi_1^a(x)}{\delta\alpha_b(y)}=
(\delta_{ab}\Delta+gf_{abc}(\partial_i A_i^c)+
gf_{abc}A_i^c\partial_i)\delta^4(x-y).
\end{equation}
To derive the form \eref{lpi} of the generating
functional (with \eref{fp}) one has to construct a
g.f.\ term by using the $\delta$-fuction and to write
the determinant as an exponential function
by introducing ghost fields, as discussed in section~2.1.
Finally the source terms have to be added.
It is essential for the derivation of this result
that, due to (\ref{noa0}), no $O(\epsilon)$-terms occur in
the argument of the determinant in (\ref{fpcg}).
Thus the ghost term is independent of the form of the effective
interaction term $\lag_I$
as in the Faddeev--Popov formalism (see section~2.1).
Because of the equivalence of all gauges \cite{lezj,able}
the result \eref{lpi} with (\ref{fp}) can be rewritten in any other gauge
which can be derived within the Faddeev--Popov formalism.

The gauge
theory given by (\ref{leff3}) is spontaneously broken if the vacuum
expectation value of the scalar fields (implied by the scalar
self-interactions contained in $V$) is nonzero; this does not affect the above
proof. Actually, this proof holds for both,
SBGTs with a {linearly}
realized scalar sector \cite{gauge}, and gauged nonlinear $\sigma$-models
\cite{nonlin}, because the latter
ones can be obtained from the first ones by making a point
transformation (as in \eref{leffhs}) in order to rewrite the scalar sector
nonlinearly (this does not effect the Hamiltonion PI and yields only
$\delta^4(0)$-terms in the Lagrangian PI (see sections~2.4, 2.7 and 4.3))
and then removing all Higgs contributions from the Lagrangian.
Thus for an arbitrary effective gauge theory (without
higher derivatives) the convenient
Faddeev--Popov PI can be derived within the
correct Hamiltonian PI formalism.


\section{Effective Gauge Theories with Higher Derivatives}
In this section I will generalize the HLE theorem to
effective gauge theories with higher time derivatives by applying the result
of section~5.2 that the equations of motion
following from ${\cal L}_0$ in (\ref{leff3}) can be applied
in order to convert the effective interaction term ${\cal L}_I$
(by neglecting higher powers of $\epsilon$) because the use of the
EOM corresponds to a field transformation \eref{trafo2}
(in this case $\vp$ may represent any field occurring in ${\cal L}$
\eref{leff3}) which becomes a canonical transformation within the Ostrogradsky
formalism and thus does not affect the Hamiltonian PI.

On the basis of this result,
an arbitrary effective gauge theory can be reduced
to one of the type considered in
the previous section as follows: Due to the gauge invariance,
derivatives of the fields occur in the effective interaction term
only as covariant derivatives or through the field strength tensor.
Using the identities (\ref{compsi}), (\ref{comphi}) and (\ref{com})
(and the corresponding relations
for ${\bar{\psi}}_a$ and $\varphi^\dagger_a$) one again
rewrites all expressions in ${\cal L}_I$
such that the covariant time derivatives are applied
to the fields before the
covariant spatial derivatives. Then,
higher time derivatives (and first time derivatives of
$A_0^a$, $\psi_a$ and ${\bar{\psi}}_a$) occur in ${\cal L}_I$ only
through the expressions
\begin{equation}
D_0 F^a_{i0}, \qquad D_0D_0 F_{ij}^a,\qquad
D_0\psi_a,\qquad D_0{\bar{\psi}}_a,\qquad
D_0D_0\varphi_a,\qquad D_0D_0\varphi^\dagger_a
\label{hdterms}\end{equation}
and even higher derivatives of these terms.
After using (\ref{com}) and (\ref{homeq}) in order to rewrite
$D_0D_0 F_{ij}^a$ as
\begin{equation}
D_0D_0 F_{ij}^a=D_jD_0 F_{i0}^a-D_iD_0 F_{j0}^a
-2gf_{abc}F_{i0}^b F_{j0}^c
\end{equation}
one can convert all the terms (\ref{hdterms}) to terms
without higher time derivatives (and without first time
derivatives of $A_0^a$, $\psi_a$
and ${\bar{\psi}}_a$) by using the EOM
following from ${\cal L}_0$, viz.\
\begin{eqnarray}
D_0F_{i0}^a&=&D_jF^a_{ij}-g{\bar{\psi}}_b\gamma_it^{bc}_a\psi_c
+ig\bar{t}^{bc}_a\left((D_i\varphi_b^\dagger)\varphi_c
-\varphi^\dagger_b(D_i\varphi_c)\right), \label{EOMA}\\
D_0\psi_a&=&\gamma^0\left(\gamma_iD_i\psi_a
-i\frac{\partial V}{\partial{\bar{\psi}}_a}\right),\label{EOMpsi}\\
D_0D_0\varphi_a&=&D_iD_i\varphi_a
-\frac{\partial V}{\partial\varphi^\dagger_a}\label{EOMphi}
\end{eqnarray}
(and the corresponding
equations for ${\bar{\psi}}_a$ and $\varphi^\dagger_a$).
By repeated application of the EOM one can eliminate all
higher time derivatives from ${\cal L}_I$.
The fact that the EOM do not contain second time derivatives of
$A_0^a$, $\psi_a$ and ${\bar{\psi}}_a$ makes it possible to
eliminate not only higher but also
first time derivatives of these fields.
The reduced first-order
Lagrangian obtained by this procedure is gauge
invariant, too, because the form of the EOM is invariant under gauge
transformations and thus the use of the EOM does not affect the gauge freedom.

Now the HLE theorem for effective gauge theories with higher
time derivatives can be
proven as follows:
\begin{enumerate}
\item Given an arbitrary gauge invariant
effective Lagrangian ${\cal L}$ \eref{leff3}, this can be reduced to
an equivalent gauge invariant
Lagrangian ${\cal L}_{red}$ without higher time
derivatives (and without
first time derivatives of $A_0^a$, $\psi_a$
and ${\bar{\psi}}_a$)
by applying the EOM, i.e.,  actually by making field
transformations such as (\ref{trafo2}). This does not affect the
Hamiltonian PI (see chapter~5).
\item ${\cal L}_{red}$ can be quantized within the Hamiltonian PI
formalism by applying the HLE theorem for first-order Lagrangians
derived in section~6.1. This yields the Lagrangian PI
\begin{equation}
Z=\int{\cal D}\Phi\,
\exp\left\{i\int d^4x\,[{\cal L}_{red}+{\cal L}_{g.f.}+
{\cal L}_{ghost}]\right\}
\label{lred3}\end{equation}
in an arbitrary gauge.
\item Going reversely through the Faddeev--Popov
procedure one can
rewrite (\ref{lred3}) (after
introducing an infinite constant into the PI) as
\begin{equation}
Z=\int{\cal D}\Phi\,\exp\left\{i\int d^4x\,{\cal L}_{red}
\right\}\label{lred2}.
\end{equation}
\item In the Lagrangian PI (\ref{lred2})
the field transformations (\ref{trafo2})
applied in step~1 are done inversely in order to
reconstruct the primordial effective
Lagrangian\footnote{Note that the use of the
transformations (\ref{trafo2}) in (\ref{lred3}) would result in an
application of the EOM following from ${\cal L}_0+{\cal L}_{g.f.}
+{\cal L}_{ghost}$ (and not from ${\cal L}_0$ alone)
which would not yield the desired result.}. This yields
\begin{equation} Z=\int{\cal D}\Phi\,\exp\left\{i\int d^4x\,{\cal L}
\right\}.\label{lpi3}
\end{equation}
The Jacobian determinant implied by
change of the functional integration measure
corresponding to these transformations only yields
extra $\delta^4(0)$-terms (see \cite{pol} and section~5.2)
which are neglected here.
\item Applying the Faddeev--Popov formalism\footnote{In
distiction from naive
Lagrangian PI quantization, (\ref{lpi3}) is not taken as an ansatz
here but it has been derived from the correct Hamiltonian PI.}
to (\ref{lpi3}) and adding the source terms
one finally finds \eref{lpi} with (\ref{fp}) in an arbitrary gauge.
\end{enumerate}
This completes the proof of the HLE theorem for effective gauge
theories.

\chapter{Summary}
In this thesis I have proven the Hamilton--Lagrange equivalence
theorem (Matthews's theorem) for effective Lagrangians with
arbitrary interactions of the physically most important types of particles,
namely scalars, fermions, massless and massive and vector bosons. This theorem
states that the convenient Lagrangian path
integral can be derived from the correct (but more involved) Hamiltonian
path intergral.
This means that the Feynman rules, which are the basis for
calculations of $S$-Matrix elements and cross sections, can directly be
obtained from the Lagrangian in the usual way.

In particular,
this theorem is valid for all types of gauge theories with arbitrary
(but gauge invariant)
interaction terms, namely for gauge theories
without spontaneous symmetry breaking, for spontaneously broken gauge
theories with a nonlinearly realized scalar sector (gauged nonlinear
$\sigma$-models, chiral Lagrangians)
and for sponteneously broken gauge theories
with a linearly realized scalar sector (Higgs models).
This means that an arbitrary gauge theory
can be quantized within the (Lagrangian)
Faddeev--Popov formalism.

I have paid specific attention to the investigation of effective
Lagrangians with massive vector fields (both with and without a spontaneously
broken gauge symmetry). Each
spontaneously broken gauge theory can be related to its unitary gauge
by applying
a Stueckelberg transformation (after making a point transformation that
delinearizes the scalar sector in the case of a Higgs model);
the U-gauge Lagrangian
is obtained by simply removing all unphysical scalar fields
from the gauge invariant Lagrangian. On the other
hand, by using the Stueckelberg formalism one can rewrite an arbitrary
gauge noninvariant
Lagrangian as a (nonlinearly realized) spontaneously broken gauge theory
and extend it, by introducing (a) physical Higgs boson(s), to a (linearly
realized) Higgs model. I have reformulated the
Stue\discretionary{k-}{k}{ck}elberg formalism
within the Hamiltonian formalism, thereby establishing the equivalence of
Lagrangians which are related to each other by a Stueckelberg transformation.

The Hamilton--Lagrange equivalence
theorem even applies to effective Lagrangians with higher
derivatives of the fields. Each effective higher-order Lagrangian
can be reduced to a first-order one by applying the equations of motion
to the effective interaction term in order to remove all higher time
derivatives. This use of the equations of motion is correct
because it can be realized
by making field transformations that involve derivatives of the fields.
I have shown that Lagrangians which are related
to each other by local field transformations
are physically equivalent (at the classical and at the quantum level)
even if these transformations depend on derivatives, because
they become canonical transformations within the Hamiltonian treatment of
higher-order Lagrangians (Ostro\-gradsky formalism). Thus, an effective
higher-order Lagrangian can formally be treated in the same way as a
first-order one; all unphysical effects, which normally occur when dealing
with higher-order Lagrangians, are absent because an effective
Lagrangian is
assumed to parametrize the low-energy effects of well-behaved ``new physics''.

Many of the above statements are obvious within the naive Lagrangian
path integral formalism.
However, one has to apply the more elaborate Hamiltonian
procedure of this thesis to derive them correctly. Indeed, the results
obtained in this thesis
justify the straightforward treatment of effective Lagrangians in the
phenomenological literature.

\chapter*{Acknowledgements}
\addcontentsline{toc}{chapter}{Acknowledgements}
I thank Prof.\ Dr.\ Dieter Schildknecht for giving me the opportunity
to do this work, for arising my interest in the
Stueckelberg formalism, for his tolerance of letting me develop own
ideas and concepts aside from his original intentions and for interesting
discussions. I am very grateful to Prof.\ Dr.\ Reinhart K\"ogerler for
his encouragement and advice, for many important, illuminating
and stimulating discussions
and for very good collaboration. I enjoyed useful discussions with
Dr.\ Mikhail Bilenky and with Dr. Jan S{\l}adkowski.


\begin{thebibliography}{00}
\addcontentsline{toc}{chapter}{Bibliography}
\bibitem{nongauge}K. Gaemers and
G. Gounaris, Z. Phys.\ {\bf C1}, 259 (1979);\\
K. Hagiwara, R. D. Peccei, D. Zeppenfeld and K. Hisaka,
Nucl. Phys. {\bf B282}, 253 (1987);\\
M. Kuroda, J. Maalampi, K. H. Schwarzer and D. Schildknecht, \nphb{284}, 271
(1987);\\
D. Zeppenfeld and S. Willenbrock,
Phys.\ Rev.\ {\bf D37}, 1775 (1988);\\
G. Gounaris, J. L. Kneur, J. Layssac, G. Moultaka, F. M. Renard
and D. Schildknecht, in
{\em $e^+e^-$ Collisions at 500 GeV, The Physics Potential\/},
ed.\ P. W. Zerwas, DESY Report DESY 92-123 (1992), p.~735;\\
G. Gounaris, J. Layssac, G. Moultaka, and F. M. Renard, \ijmpha{8},
3285 (1993);\\
M. Bilenky, J. L. Kneur, F. M. Renard and D. Schildknecht,
\nphb{409}, 22 (1993)
\bibitem{nonlin}T. Appelquist and C. Bernard,
Phys.\ Rev.\ {\bf D22}, 200 (1980);\\
A. C. Longhitano, Nucl.\ Phys.\ {\bf B188}, 118 (1981);\\
A. Dobado, D. Espriu and M. J. Herrero, Phys.\ Lett.\ {\bf B225},
405 (1991);\\
B. Holdom, Phys.\ Lett.\ {\bf B258}, 156 (1991);\\
A. Falk, M. Luke and E. Simmons,
Nucl.\ Phys.\ {\bf B365}, 523 (1991);\\
D. Espriu and M. J. Herrero, Nucl.\ Phys.\ {\bf B373}, 117 (1991);\\
T. Appelquist and G.-H. Wu, \phrd{48}, 3235 (1993);\\
M. J. Herrero and E. R. Morales, Madrid Preprint FTUAM 93/24 (1993),
hep-ph/9308276
\bibitem{gauge}C. N. Leung, S. T. Love and S. Rao,
Z. Phys. {\bf C31}, 433 (1986);\\
W. Buchm\"uller and D. Wyler, Nucl.\ Phys.\ {\bf B268},
621 (1986);\\
A. de R\'{u}jula, M. B. Gavela, P. Hern\'{a}ndez and E. Mass\'{o},
Nucl.\ Phys.\ {\bf B384}, 3 (1992);\\
K. Hagiwara, S. Ishihara, R. Szalapski and D. Zeppenfeld,
Phys.\ Lett.\ {\bf B283}, 353 (1992);\\
G. J. Gounaris and F. M. Renard,
Z. Phys.\ {\bf C59}, 133 (1993);\\
C. Grosse-Knetter and D. Schildknecht, \phlb{302}, 309 (1993);\\
P. Hern\'{a}ndez and F. J. Vegas, Phys.\ Lett.\
{\bf B307}, 116 (1993);\\
K. Hagiwara, S. Ishihara, R. Szalapski and D. Zeppenfeld,
\phrd{48}, 2182 (1993);\\
C. Grosse-Knetter, I. Kuss and D. Schildknecht,
\zphc{60}, 375 (1993)
\bibitem{gluons}C. N. Leung, S. T. Love and S. Rao,
ref.\ \cite{gauge};\\
W. Buchm\"uller and D. Wyler, ref.\ \cite{gauge};\\
E. H. Simmons, Phys.\ Lett.\ {\bf B226}, 132 (1989);
Phys.\ Lett.\ {\bf B246}, 471 (1990);\\
H. Dreiner, A. Duff and D. Zeppenfeld, Phys.\ Lett.\
{\bf B282}, 441 (1992)
\bibitem{mat}P. T. Matthews, Phys.\ Rev.\ {\bf 76}, 684 (1949)
\bibitem{feyn}R. P. Feynman, Rev.\ Mod.\ Phys.\ {\bf 20}, 367 (1948);\\
R. P. Feynman and A. R. Hibbs, {\em Quantum Mechanics and Path Integrals\/},
McGraw-Hill (1965)
\bibitem{fapo}L. D. Faddeev and V. N. Popov,
Phys.\ Lett.\ {\bf B25}, 29 (1967)
\bibitem{books}T.-P. Cheng and L.-F. Li, {\em Gauge theory of
elementary particle physics\/}, Oxford University Press (1984);\\
L. H. Ryder, {\em Quantum Field Theory\/}, Cambridge University Press
(1985);\\
D. Bailin and A. Love, {\em Introduction to
Gauge Field Theory\/}, Hilger (1986);\\
S. Pokorsky, {\em Gauge Field Theories\/}, Cambridge University Press
(1987);\\
P. H. Frampton, {\em Gauge Field Theories\/}, Addison-Wesley (1987);\\
P. Ramond, {\em Field Theory: A Modern Primer\/}, Addison-Wesley (1989)
\bibitem{fad}L. D. Faddeev, Teor.\ Mat.\ Fiz. {\bf 1},
3 (1969) [Transl.:
Theor.\ Math.\ Phys.\ {\bf 1}, 1 (1970)]
\bibitem{bedu}C. Bernard and A. Duncan, Phys.\ Rev.\ {\bf D11},
848 (1975)
\bibitem{sen}P. Senjanovic, Ann.\ Phys.\ {\bf 100}, 227 (1976)
\bibitem{const}A. Hanson, T. Regge and C. Teitelboim,
{\em Constrained Hamiltonian Systems\/}, Accad. Naz. dei Lincei (1976);\\
K. Sundermeyer, {\em Constrained Dynamics}, Springer (1982);\\
J. Govaerts, {\em Hamiltonian Quantization and Constrained Dynamics\/},
Leuven University Press (1991);\\
M. Henneaux and C. Teitelboim,
{\em Quantization of Gauge
Systems\/}, Princeton University Press (1992)
\bibitem{gity}D. M. Gitman and I. V. Tyutin,
{\em Quantization of Fields with
Constraints}, Springer (1990)
\bibitem{dirac}P. A. M. Dirac, Can.\ J. Math. {\bf 2}, 129 (1950);
{\em Lectures on Quantum Mechanics}, Belfar (1964)
\bibitem{bfvbv}A. Dresse, J. M. L. Fisch, P. Gregoire and M. Henneaux,
\nphb{354}, 191 (1991);\\
G. V. Grigorian, R. P. Grigorian and I. V. Tuytin, \nphb{379}, 304 (1992);\\
F. De Jonghe, \phlb{316}, 503 (1993)
\bibitem{bfv}E. S. Fradkin and G. A. Vilkovisky, \phlb{55}, 224 (1975);\\
I. A. Batalin and G. A. Vilkovisky, \phlb{69}, 309 (1977);\\
E. S. Fradkin and T. E. Fradkina, \phlb{72}, 343 (1978);\\
M. Henneaux, \phrp{126}, 1 (1985)
\bibitem{bv}I. A. Batalin and G. A. Vilkovisky,
\phlb{102}, 27, (1981); \phrd{28}, 2567 (1983);
\nphb{234}, 106 (1984); J. Math.\ Phys.\ {\bf 26}, 172 (1985)
\bibitem{stue}E. C. G. Stueckelberg,
Helv.\ Phys.\ Acta {\bf 11}, 299 (1938)
\bibitem{kugo}T. Kunimasa and T. Goto,
Prog.\ Theor.\ Phys.\ {\bf 37}, 452 (1967)
\bibitem{lezj}B. W. Lee and J. Zinn-Justin,
Phys.\ Rev.\ {\bf D5}, 3121, 3137, 3155 (1972),
{\bf D7}, 1049 (1973)
\bibitem{clt}J. M. Cornwall, D. N. Levin and G. Tiktopoulos,
Phys.\ Rev.\ {\bf D10}, 1145 (1974)
\bibitem{bulo}M. Chanowitz, M. Golden and H. Georgi,
Phys.\ Rev.\ {\bf D36}, 1490 (1987);\\
C. P. Burgess and D. London, McGill Preprint McGill-92/04 (1992),
hep-ph/9203215; Phys.\ Rev.\ Lett.\ {\bf 69}, 3428 (1992)
\bibitem{able}E. S. Abers and B. W. Lee, Phys.\ Rep.\ {\bf 9},
1 (1973)
\bibitem{phhide}T. Appelquist and C. Bernard, ref.\ \cite{nonlin};\\
A. C. Longhitano, ref.\ \cite{nonlin};\\
C. N. Leung, S. T. Love and S. Rao, ref.\ \cite{gauge};\\
B. Grinstein and M. B. Wise, ref.\ \cite{gauge};\\
E. H. Simmons, ref.\ \cite{gluons};\\
G. J. Gounaris and F. M. Renard, ref.\ \cite{gauge};\\
K. Hagiwara, S. Ishihara, R. Szalapski and D. Zeppenfeld, ref.\
\cite{gauge}
\bibitem{ost}M. V. Ostrogradsky, Mem.\ Acad.\ St.\ Petersbourg {\bf
6}, 385 (1850)
\bibitem{hide}A. Pais and G. Uhlenbeck, Phys.\ Rev.\ {\bf 79},
145 (1950);\\
C. Bernard and A. Duncan, ref.\ \cite{bedu};\\
S. W. Hawking, in {\em Quantum Field Theory and Quantum
Statistics}, ed.: I. A. Batalin, C. J. Isham and C. A. Vilkovisky,
Hilger (1987), p.~129;\\
D. A. Eliezer and R. P. Woodard, Nucl.\ Phys.\ {\bf B325},
389 (1989);\\
J. Z. Simon, Phys.\ Rev.\ {\bf D41}, 3720 (1990)
\bibitem{eom}D. Barua and S. N. Gupta, Phys.\ Rev.\ {\bf D16},
413 (1977);\\
G. Sch\"afer, Phys.\ Lett.\ {\bf A100}, 128 (1984);\\
T. Damour and G. Sch\"afer, J.\ Math.\ Phys.\ {\bf 32}, 127 (1991);\\
H. Georgi, Nucl.\ Phys.\ {\bf B361}, 339 (1991);\\
H. Leutwyler, Bern Preprint BUTP-91/26 (1991)
\bibitem{pol}H. D. Politzer, Nucl.\ Phys.\ {\bf B172}, 349 (1980);\\
C. Arzt, Michigan Preprint UM-TH-92-28 (1992),
hep-ph/9304230
\bibitem{leib}G. Leibbrandt, Rev.\ Mod.\ Phys.\ {\bf 47}, 849 (1975)
\bibitem{fls}K. Fujikawa, B. W. Lee and A. I. Sanda,
Phys.\ Rev.\ {\bf D6},
2923 (1972)
\bibitem{wein2}S. Weinberg,
Phys.\ Rev.\ {\bf D7}, 2887 (1973)
\bibitem{leya}T. D. Lee and C. N. Yang, Phys.\ Rev.\ {\bf
128}, 885 (1962)
\bibitem{sast}A. Salam and J. Strathdee,
Phys.\ Rev.\ {\bf D2}, 2869 (1970)
\bibitem{wein1}S. Weinberg,
Phys.\ Rev.\ {\bf D7}, 1068 (1973)
\bibitem{apqu}T. Appelquist and H. Quinn,
Phys.\ Lett.\ {\bf B39}, 229 (1972)
\bibitem{jogl}S. D. Joglekar, Ann.\ Phys.\ {\bf 83}, 427 (1974)
\bibitem{jawe}R. Jackiw and S. Weinberg, Phys.\ Rev.\ {\bf D5},
2396 (1972);\\
I. Bars and M. Yoshimura, Phys.\ Rev.\ {\bf D6}, 374 (1972)
\bibitem{sots}T. Sonoda and S. Y. Tsai,
Prog.\ Theor.\ Phys.\ {\bf 71}, 878 (1984)
\bibitem{dtt}R. Delburgo, S. Twisk
and G. Thompson, Int.\ J. Mod.\
Phys.\ {\bf A3}, 435 (1988)
\bibitem{bash}W. A. Bardeen and K. Shizuya,
Phys.\ Rev.\ {\bf D18}, 1969 (1978)
\bibitem{apbe}T. Appelquist and C. Bernard, ref.\ \cite{nonlin};\\
A. C. Longhitano, ref.\ \cite{nonlin}
\bibitem{sm}S. Weinberg, Phys.\ Rev.\ Lett.\
{\bf 19}, 1264 (1967);\\
A. Salam, in {\em Elementary Particle Theory: Relativistic Groups
and Analyticty (Nobel Symposiun No.\ 8)\/},
ed.\ N. Svartholm, Almquist
and Wiksells (1968), p.~367
\bibitem{gkko}C. Grosse-Knetter and R. K\"ogerler, \phrd{48}, 2865 (1993)
\bibitem{higgs}P. W. Higgs,
Phys.\ Rev.\ {\bf 145}, 1156 (1966);\\
T. W. B. Kibble, Phys.\ Rev.\ {\bf 155}, 1554 (1967)
\bibitem{llsm}J. M. Cornwall, D. N. Levin and G. Tiktopoulos,
\phrl{30}, 1268 (1973);\\
C. H. Llewellyn Smith,
Phys.\ Lett.\ {\bf B46}, 233 (1973)
\bibitem{shiz}K. Shizuya, Nucl.\ Phys.\ {\bf B121}, 125 (1977)
\bibitem{burn}A. Burnel,
Phys.\ Rev.\ {\bf D33}, 2985 (1986)
\bibitem{alda}J. Alfaro and P. H. Damgaard,
Ann.\ Phys.\ {\bf 202}, 398 (1990)
\bibitem{ruj}A. de R\'{u}jula, M. B. Gavela,
P. Hern\'{a}ndez and E. Mass\'{o}, ref. \cite{gauge}
\bibitem{haze}
K. Hagiwara, S. Ishihara, R. Szalapski and D. Zeppenfeld, ref.\ \cite{gauge};\\
P. Hern\'{a}ndez and F. J. Vegas, ref.\ \cite{gauge}
\bibitem{gore}G. J. Gounaris and F. M. Renard, ref.\ \cite{gauge}
\bibitem{cm}H. Goldstein, {\em Classical Mechanics\/}, Addison-Wesley,
second edition (1980);\\
L. D. Landau and E. M. Lifschitz, {\em Lehrbuch der theoretischen
Physik, Band I: Mechanik\/}, Akademie-Verlag (1976);\\
F. Kuypers, {\em Klassische Mechanik\/}, VCH, second edition (1989)
\bibitem{kadi}S. Kaptano\v{g}lu, Phys.\ Lett.\
{\bf B98}, 77 (1981);\\
P. Ditsas, Ann.\ Phys.\ {\bf 167}, 36 (1986)
\bibitem{gtp}D. M. Gitman, I. V. Tuytin and Y. S. Prager,
Sov.\ Phys.\ Journ.\ {\bf 26}, 760 (1983)
\bibitem{bl}C. P. Burgess and D. London, McGill Preprint McGill-92/05 (1992),
hep-ph/9203216; Phys.\ Rev.\ Lett.\ {\bf 69}, 3428 (1992)
\bibitem{moul}G. Moultaka, Montpellier Preprint PM/93-20 (1993),
hep-ph/9310282
\bibitem{gk1}C. Grosse-Knetter, \phrd{48}, 2854 (1993)
\bibitem{col}S. Coleman, Lectures given at the the 1973
International Summer
School of Physics ``Ettore Majorana'', Harvard report
(unpublished), p.~42
\bibitem{pape}N. K. Pak and R. Percacci,
J. Math.\ Phys.\ {\bf 30}, 2951 (1989)
\bibitem{mira}P. Mitra and R. Rajaraman, Ann.\ Phys.\ {\bf 203},
157 (1990);\\
K. Harada and H. Mukaida, Z. Phys. {\bf C48}, 151 (1990);\\
R. Anishetty and A. S. Vytheeswaran, Madras Preprint
imsc-92/10 (1992)
\bibitem{fash}L. D. Faddeev and S. L. Shatashvili,
Phys.\ Lett.\ {\bf B167}, 225 (1986)
\bibitem{wein3}S. Weinberg, Phys.\ Rev.\ Lett.\ {\bf 27}, 1688 (1971)
\bibitem{giro}H. O. Girotti and H. J. Rothe, Nuov.\ Cim.\
{\bf 75A}, 62 (1983)
\bibitem{bnn}J. Barcelos-Neto and C. P. Natividade, \zphc{51}, 313
(1991)
\bibitem{gk2}C. Grosse-Knetter, Bielefeld Preprint BI-TP 93/29 (1993),
hep-ph/9306310
\bibitem{shd}T. Kimura, Lett.\ Nuov.\ Cim. {\bf 5}, 81 (1972);\\
V. Tapai, Nuov.\ Cim.\ {\bf 90B}, 15 (1985);\\
Y. Saito, R. Sugano, T. Ohta and T. Kimura,
J. Math.\ Phys.\ {\bf 30}, 1122 (1989);\\
V. V. Nestorenko, J. Phys.\ {\bf A22}, 1673 (1989)
\bibitem{pons}J. M. Pons, Lett.\ Math.\ Phys.\ {\bf 17}, 181 (1989)
\bibitem{chis}J. S. R. Chisholm, \nph{26}, 469 (1961);\\
S. Kamefuchi, L. O'Raifeartaigh and A. Salam, \nph{28}, 529 (1961)
\bibitem{buwy}W. Buchm\"uller and D. Wyler, ref.\ \cite{gauge}
\bibitem{gkks}C. Grosse-Knetter, I. Kuss and D. Schildknecht,
ref.\ \cite{gauge}
\bibitem{gk3}C. Grosse-Knetter, Bielefeld Preprint BI-TP 93/40 (1993),
hep-ph/9308201, to be published in \phrd{}
\end{thebibliography}
\end{document}